\documentclass[12pt]{article}

\usepackage{color}
\usepackage{amsmath,amssymb}
\usepackage[dvips]{graphicx,psfrag}
\usepackage{latexsym}
\usepackage{cite}
\usepackage{wrapfig}
\usepackage{pstricks}
\usepackage{eepic}
\usepackage{epic}

\allowdisplaybreaks[1]

\newcommand{\eq}[1]{(\ref{#1})}
\newcommand{\nn}{\nonumber}
\newcommand{\ds}{\displaystyle}

\newcommand{\vev}[1]{\left\langle #1 \right\rangle}

\newcommand{\del}{\partial}

\newcommand{\asymeq}{\underset{asym}{\simeq}}

\newcommand{\bP}{\boldsymbol{P}}
\newcommand{\bQ}{\boldsymbol{Q}}

\newcommand{\bGamma}{\boldsymbol{\Gamma}}

\DeclareMathOperator{\tr}{tr}
\DeclareMathOperator{\str}{str}
\DeclareMathOperator{\diag}{diag}

\newtheorem{Theorem}{Theorem}
\newtheorem{Definition}{Definition}
\newtheorem{Proposition}{Proposition}
\newtheorem{Lemma}{Lemma}

\renewcommand{\thefootnote}{\fnsymbol{footnote}}

\makeatletter
\@addtoreset{equation}{section}

\begin{document}


\begin{titlepage}
\thispagestyle{empty} 
\begin{flushright}
arXiv:1011.5745 \\
\vspace{0.2cm} 
 November 2010 
\end{flushright}

\vspace{2.3cm}

\begin{center}
\noindent{\Large \textbf{
Stokes phenomena 
and non-perturbative completion \\
\vspace{0.3cm}
in the multi-cut two-matrix models
}}
\end{center}

\vspace{1cm}

\begin{center}
\noindent{Chuan-Tsung Chan\footnote{ctchan@thu.edu.tw}$^{,p}$, 
Hirotaka Irie\footnote{irie@phys.cts.nthu.edu.tw; irie@phys.ntu.edu.tw}$^{,q,r}$
and Chi-Hsien Yeh\footnote{d95222008@ntu.edu.tw}$^{,q}$}
\end{center}
\vspace{0.5cm}
\begin{center}
{\it 
$^{p}$Department of Physics, Tunghai University, Taiwan, 40704\\
\vspace{.3cm}
$^{q}$Department of Physics and Center for Theoretical Sciences, \\
National Taiwan University, Taipei 10617, Taiwan, R.O.C  \\
\vspace{.3cm}
$^{r}$National Center for Theoretical Sciences, \\
National Tsing-Hua University, Hsinchu 30013, Taiwan, R.O.C.%
\footnote{The affiliation of H. Irie from January 1st, 2011. }
}
\end{center}

\vspace{1.0cm}

\begin{abstract}
The Stokes multipliers in the matrix models are invariants in 
the string-theory moduli space 
and related to the D-instanton chemical potentials. 
They not only represent non-perturbative information but also play an important role 
in connecting various perturbative string theories in the moduli space. 
They are a key concept to the non-perturbative completion
of string theory and also expected to imply some remnant of strong coupling dynamics 
in M theory. 
In this paper, we investigate the non-perturbative completion problem consisting of two constraints on the Stokes multipliers. 
As the first constraint, 
Stokes phenomena which realize the multi-cut geometry 
are studied in the $\mathbb Z_k$ symmetric critical points of 
the multi-cut two-matrix models. 
Sequence of solutions to the constraints 
are obtained in general $k$-cut critical points. 
A discrete set of solutions and a continuum set of solutions are explicitly shown, 
and they can be classified 
by several constrained configurations of the Young diagram. 
As the second constraint, we discuss non-perturbative stability of backgrounds in terms of the Riemann-Hilbert 
problem. In particular, our procedure in the $2$-cut $(1,2)$ case (pure-supergravity case) 
completely fixes 
the D-instanton chemical potentials and results in the Hastings-McLeod solution 
to the Painlev\'e II equation. 
It is also stressed that the Riemann-Hilbert approach realizes an off-shell background 
independent formulation of non-critical string theory. 
\end{abstract}

\end{titlepage}

\newpage

\renewcommand{\thefootnote}{\arabic{footnote}}
\setcounter{footnote}{0}




\section{Introduction}

Non-critical string theory \cite{Polyakov} has provided interesting theoretical laboratories 
which uncover various intriguing features about string theory. 
This string theory is known as solvable system not only 
in the perturbative world-sheet formulation, Liouville theory 
\cite{KPZ,DDK,DOZZ,Teschner,FZZT,ZZ,sDOZZ,fuku-hoso,SeSh}, 
but also in the non-perturbative matrix-model formulation \cite{DSL,TwoMatString,GrossMigdal2,    BIPZ,KazakovSeries,Kostov1,Kostov2,BMPNonP,BDSS,Kostov3,MSS,DKK,TadaYamaguchiDouglas,DouglasGeneralizedKdV,Moore,GinspargZinnJustin,fkn,DVV,EynardZinnJustin,David,fy12,fy3}. 
Recently, among various kinds of matrix models, the multi-cut matrix models \cite{MultiCut} 
have turned out to be a fruitful system. The first discovery was on the two-cut 
matrix models \cite{GrossWitten,PeShe,DSS,Nappi,CDM,HMPN}, 
which were found to describe type 0 superstring theory \cite{TT,NewHat,UniCom}. 
Furthermore, the multi-cut two-matrix models were generally found to have a correspondence with 
the so-called fractional superstring theory \cite{irie2} and also with non-critical M theory 
as its strong-coupling dual theory \cite{CIY1}, which realizes 
the philosophy proposed in the Ho\v rava-Keeler non-critical M theory \cite{NonCriticalMTheory}. 

Quantitative analyses of critical points and perturbative amplitudes in 
the multi-cut two-matrix models have been carried out in \cite{CISY1,CIY1}. 
The main observables used there are macroscopic loop amplitudes (or resolvent) 
\cite{BIPZ,KazakovSeries,Kostov1,Kostov2,BDSS,Kostov3,MSS,DKK,Kris,
AnazawaIshikawaItoyama1,AnazawaIshikawaItoyama2,AnazawaItoyama,fim} 
which provide the information of spectral curves, 
the classical spacetime of this string theory \cite{fy3,MMSS,SeSh2}. 
A concrete expression for 
spectral curve is important because it provides relevant information for reproducing
all order perturbative amplitudes in the multi-cut two-matrix models by the method of 
topological recursions \cite{EynardLoop}. 

The main theme in this paper is, on the other hand, about 
{\em non-perturbative aspects of the multi-cut two-matrix models}. 
Non-perturbative aspects in matrix models have also been studied extensively \cite{GinspargZinnJustin,EynardZinnJustin,David,fy12,fy3,MultiCutUniversality,McGreevyVerlinde,Martinec,KMS,AKK,KazakovKostov,HHIKKMT,KOPSS,KKM,InstantonInCSFT,ChemicalPotentials1,ChemicalPotentials2,MMSS,SeSh2,fis,ChemicalPotentials3,fim,fi1,KurokiSugino,fi2,MSW1,EynardNPPartition,MarinoHMSolution,MSW2,EynardMarino,PasquettiSchiappa,GKM,KMR,MarinoLecture}.%
\footnote{See \cite{MarinoLecture} for a nice review of these recent progress. } 
The main concern is about non-perturbative contributions to the matrix-model 
free energy $\mathcal F(\mathcal C;g_{\rm str})$ on the large $N$ spectral curve $\mathcal C$:%
\footnote{We carefully put ``asym'' below the equation in order to emphasize 
that they are equal only in the asymptotic sense. }
\begin{align}
\mathcal F(\mathcal C;g_{\rm str}) \asymeq 
\sum_{n=0}^\infty \,g_{\rm str}^{2n-2}\,\mathcal F_n(\mathcal C)
+ \mathcal F_{\text{non-perturb.}}(\mathcal C;g_{\rm str}),
\qquad g_{\rm str} \to 0. \label{FormalExpansionMM1}
\end{align}
Here $\mathcal F_n(\mathcal C)$ is the genus-$n$ perturbative free energy 
on the spectral curve $\mathcal C$, and the information of the matrix-model potentials 
(so-called KP flows $\{t_n\}_{n\in \mathbb Z}$) is implicitly included in the spectral curve:
\begin{align}
\mathcal C= \mathcal C\bigl(\{t_n\}_{n\in \mathbb Z}\bigr),
\qquad\{t_n\}_{n\in \mathbb Z} \in \mathcal M^{(\rm non-norm.)}_{\rm string} \subset \mathbb C^{\infty}. 
\end{align}
Here $\mathcal M^{(\rm non-norm.)}_{\rm string}$ 
stands for the non-normalizable string-theory moduli space \cite{SeibergShenker}.%
\footnote{The normalizable string-theory moduli space $\mathcal M^{(\rm norm.)}_{\rm string}$ is known 
as the space of filling fraction \cite{EynardMarino} 
which parametrizes the on-shell string backgrounds. 
The off-shell backgrounds are defined in Section \ref{RHapproachSection}. }
The first quantitative implication was given in the early 90's and is about 
the strength of string non-perturbative corrections which are of order 
$\mathcal O(e^{-1/g_{str}})$ quantities \cite{Shenker}, i.e.~open-string (D-brane) 
degree of freedom \cite{Polchinski}: 
\begin{align}
\mathcal F_{\text{non-perturb.}}(\mathcal C;g_{\rm str})
= \sum_{I} \,\theta_{I} \, \exp\Bigl[{-\frac{1}{g_{\rm str}} 
\mathcal S_{\rm inst}^{(I)}(\mathcal C; g_{\rm str})}\Bigr]. 
\label{FormalExpansionMM2}
\end{align}
Here $I$ is a set of indices which labels multi-instanton sectors, $I=\{i_1,i_2,\cdots\}$, 
\begin{align}
\mathcal S_{\rm inst}^{(I)}(\mathcal C; g_{\rm str}) 
= \sum_{i\in I=\{i_1,i_2,\cdots\}} S_{\rm inst}^{(i)}(\mathcal C)+ \mathcal O(g_{\rm str}). 
\end{align}
Each primitive instanton action $S_{\rm inst}^{(i)}(\mathcal C)$ 
($i=1,2,\cdots,N_{\rm inst}$), is shown to correspond to a singular point 
of the spectral curve $\mathcal C$ \cite{McGreevyVerlinde,Martinec,KMS,AKK,UniCom,KazakovKostov,KOPSS,fi1}
and is identified with the ZZ-brane disk amplitudes in Liouville theory \cite{ZZ,fuku-hoso,SeSh}. 
It is worth mentioning that 
these instanton corrections including higher order $g_{\rm str}$ corrections 
$\mathcal S_{\rm inst}^{(I)}(\mathcal C; g_{\rm str})$ are generally expressed as theta functions 
on the spectral curve \cite{MultiCutUniversality,EynardNPPartition} and 
important in order to make the free energy $\mathcal F(\mathcal C;g_{\rm str})$ 
modular invariant under modular transformations of the spectral curve 
$\mathcal C$ and also to be background independent in the normalizable string-theory 
moduli space $\mathcal M^{(\rm norm.)}_{\rm string}$ 
(i.e.~the filling fractions) \cite{EynardNPPartition,EynardMarino}. 
The constant $\theta_I$ is called D-instanton chemical potential (or fugacity). 
These constants are understood as 
integration constants of corresponding string equations \cite{David}, that is, 
\begin{align}
\frac{\del \theta_I}{\del t_m}= 0,
\qquad m\in \mathbb Z,
\qquad \{t_n\}_{n\in \mathbb Z} \in \mathcal M^{(\rm non-norm.)}_{\rm string}, 
\end{align}
for the flows in the non-normalizable moduli space $\mathcal M^{(\rm non-norm.)}_{\rm string}$. 
It was shown \cite{fy3} 
that the only $N_{\rm inst}$ (i.e.~the number of 
primitive instantons) chemical potentials $\theta_i$ $(i=1,2,\cdots,N_{\rm inst})$ 
are independent among all the chemical potentials $\theta_I$. 

Although various aspects of matrix models have been understood well so far, 
there still remains an important issue regarding the D-instanton chemical potentials. 
This is also known as non-perturbative ambiguities of string theory. Therefore, 
{\em what is the physical requirement to determine the D-instanton chemical potentials?}
Although the actual matrix models should 
employ some particular universal values \cite{HHIKKMT}, 
they seem to be totally free parameters at least within continuum formulations based 
on string (or loop) equations. 
This point has been studied 
in the bosonic minimal/2D string theories 
\cite{HHIKKMT,ChemicalPotentials1,ChemicalPotentials2,ChemicalPotentials3}, 
in the type 0 $(1,2)$ superstring theory \cite{KKM}, 
in the collective string field theory \cite{InstantonInCSFT}, in the free-fermion formulation \cite{fis,fim} 
and in the topological string interpretations \cite{EynardMarino}. 
In this paper, we address this issue by solving {\em non-perturbative completion problem} 
within a continuous formulation for the critical points of the multi-cut two-matrix models. 
In practice, we pick up physically acceptable D-instanton chemical potentials which 
realize physically reasonable behaviors in the non-perturbative regime $g_{\rm str}\to \infty$. 
Our solutions are based on two physical requirements: 
One is {\em multi-cut boundary condition} (in Section \ref{MultiCutBCSection}) and 
the other is {\em non-perturbative stability of perturbative backgrounds} 
(in Section \ref{RHapproachSection}). 

The first requirement, the multi-cut boundary condition, is a non-perturbative constraint on 
the Baker-Akhiezer function system in these multi-cut critical points: 
\begin{align}
g_{\rm str}\frac{\del }{\del \zeta} \Psi(t;\zeta) = \mathcal Q(t;\zeta)\, \Psi(t;\zeta), \qquad
g_{\rm str}\frac{\del }{\del t} \Psi(t;\zeta) = \mathcal P(t;\zeta)\, \Psi(t;\zeta),\label{BAIntroIIP}
\end{align}
where the equation system here is expressed as an ordinary differential equation in $\zeta$ 
and its isomonodromy deformation system in $t$.%
\footnote{
The parameter $t$ is one of the parameters in the non-normalizable moduli space 
$\mathcal M^{(\rm non-norm.)}_{\rm string}$, which is usually a coupling of the most relevant operator or 
the world-sheet cosmological constant. } 
Note that the Lax operators 
in Eq.~\eq{BAIntroIIP} in the $k$-cut critical points are $k\times k$ matrix-valued operators \cite{fi1}.  
The idea of the first constraint is motivated by the non-perturbative relationship between the Baker-Akhiezer 
functions and cuts in the resolvent curves. This kind of relationship is discussed in terms of Airy function \cite{MMSS}. 
Specifically, the asymptotic expansion of the Airy function around the cut ($\zeta\to -\infty$) 
is expressed as%
\footnote{The asymptotic expansion of Airy function is reviewed in Appendix \ref{AppendixAiryFunction}. }
\begin{align}
{\rm Ai}(t;\zeta) &\asymeq
\Bigl(\frac{g_{\rm str} \pi }{(\zeta + t)^{1/2}}\Bigr)^{1/2}\,
\Bigl[
e^{-\frac{2}{3 g_{\rm str}} (\zeta + t)^{3/2}}
+i
e^{\frac{2}{3 g_{\rm str}} (\zeta + t)^{3/2}}
\Bigr]+\cdots,
\end{align}
where the relation to the resolvent (or macroscopic loop) operator $\mathcal R(\zeta)$ 
\cite{GrossMigdal2} is roughly expressed as 
\begin{align}
{\rm Ai}(t;\zeta) \sim  \exp\Bigl[N
\int^\zeta d\zeta'\, \mathcal R(\zeta')\Bigr],\qquad 
\mathcal R(\zeta) \equiv \frac{1}{N}\vev{\tr \frac{1}{\zeta-M}} -\frac{V'(\zeta)}{2}
\sim \sqrt{\zeta+t}, 
\end{align}
with the expectation value $\vev{\cdots }$ which 
is taken with respect to the Hermitian one-matrix model of a matrix $M$. 
From this expression, one observes that 
{\em the cut in the negative axes $(\zeta<-t)$ appears as a line where a 
competition between the exponents $e^{\pm\frac{2}{3 g_{\rm str}} (\zeta + t)^{3/2}}$ (i.e.~along the Stokes lines) happens}. 
Therefore, we interpret this as {\em a non-perturbative definition of the resolvent cuts}. 
This consideration turns out to be important in the fractional-superstring critical points of the 
multi-cut two-matrix models \cite{CIY1}, since most of the cuts in these critical points 
are created by this procedure and cannot be read from the algebraic equations 
of the resolvent spectral curve. 
However, as we will see in Section \ref{MultiCutBCSection}, this procedure 
do not necessarily create the necessary and sufficient $k$ cuts on the resolvent curve, even though 
the $k$-cut Baker-Akhiezer function Eq.~\eq{BAIntroIIP} is obtained from the assumption that 
the critical points have $k$ cuts around $\zeta \to \infty$. 
In view of this, we need to impose a physical constraint so that 
the resolvent curves in the $k$-cut critical points should have $k$ cuts around $\zeta\to \infty$. 
This constraint is expressed in terms of {\em Stokes multipliers} 
for the {\em possible Stokes phenomena} in this system.

\begin{figure}[tbp]
\begin{center}
\setlength{\unitlength}{1mm}
\begin{picture}(145,52)(-14,15)
\put(-15,40){\framebox(36,20){\shortstack{
String equations \\ \\
{
(Painlev\'e system)
}
}}}

\put(-7,25){\makebox(20,20){\rotatebox{90}{\Large $\in$}}}

\put(-12,14){\framebox(30,15){\shortstack{D-instanton \\ \\ fugacities}}}

\put(85,40){\framebox(50,20){\shortstack{
Orthonormal polynomials \\ \\
{
(Baker-Akhiezer system)
}
}}}

\put(100,25){\makebox(20,20){\rotatebox{90}{\Large $\in$}}}

\put(95,14){\framebox(30,15){\shortstack{Stokes data \\ \\ at $\zeta \to \infty$}}}

\thicklines
\put(40,48){\Large $=\!=\!=\!=\!=\!\Longrightarrow$}
\put(42,16){\Large $\Longleftarrow\!=\!=\!=\!=\!=$}
\put(23,53){\makebox(60,7.5){Inverse scattering method}}
\put(26,23){\makebox(60,7.5){\shortstack{Inverse monodromy method \\ \\ (Riemann-Hilbert problem)}}}
\end{picture}
  \end{center}
 \caption{\footnotesize The Riemann Hilbert approach and the D-instanton chemical potentials (or fugacities)}
 \label{RHapproachFig}
\end{figure}
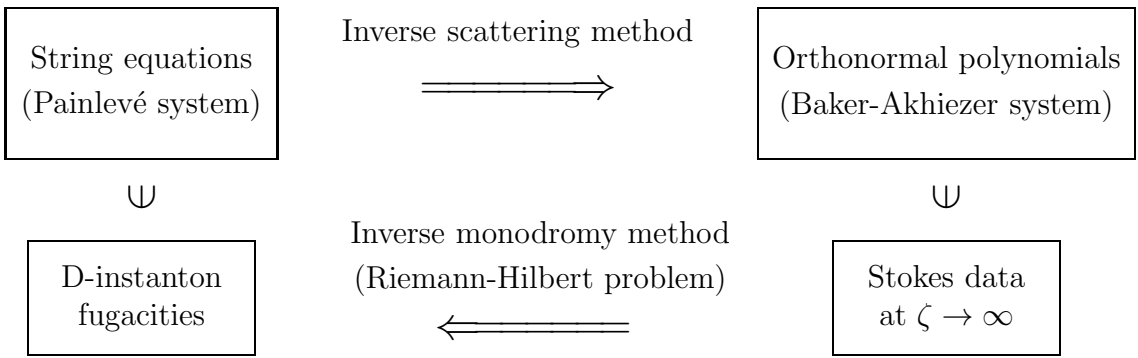

The second requirement, the non-perturbative stability of perturbative backgrounds, 
is imposed in the other formulation which is closely related to the Baker-Akhierzer function system: 
the so-called the Riemann-Hilbert (or inverse monodromy) approach \cite{RHcite, RHPIIcite,DeiftZhou} 
\cite{Moore}. 
A brief flowchart of this approach is shown in Fig.~\ref{RHapproachFig}. 
Details are given in Section \ref{RHapproachSection}, but in order to show how the Riemann-Hilbert 
approach works in resolving the issue, we here show the leading expression of the free energy 
(more precisely the two-point function of cosmological constant $t$) 
in the two-cut $(1,2)$ case: 
\begin{align}
\frac{\del^2 \mathcal F(t;g_{\rm str})}{\del t^2} = \bigl[f(t)\bigr]^2,
\qquad f(t)=\sum_{n} s_{n,2,1}   \int_{\mathcal K_n} \frac{d\lambda}{2\pi i} \,
 e^{g^{(2)}(t;\lambda)-g^{(1)}(t;\lambda)}+\cdots. \label{RHintegralIntroduction}
\end{align}
The parameter $s_{n,2,1}$ is {\em a Stokes multiplier} of the Baker-Akhierzer function system of 
the corresponding integrable system
and the contour $\mathcal K_n$ is {\em an anti-Stokes line} 
corresponding to the Stokes multiplier $s_{n,2,1}$. 
As one can suspect from the expression, 
the Riemann-Hilbert approach is directly related to the study of {\em Stokes phenomena} at 
$\zeta \to \infty$ in the ordinary differential equation of the Baker-Akhierzer system. 

In this expression, 
the function $g^{(j)}(t;\zeta)$ is {\em an arbitrary function} but should be 
properly chosen so that the integrals other than the ``leading'' expression 
shown in Eq.~\eq{RHintegralIntroduction} are negligible \cite{DeiftZhou}. From the matrix-model viewpoints 
(to be discussed in Section \ref{RHapproachSection}), this function can be interpreted as 
{\em an off-shell string background geometry of string theory}. 
Therefore, if one {\em chooses} $g^{(j)}(t;\zeta)$ 
as a {\em macroscopic loop amplitude} realized in the large $N$ limit of the matrix models, 
then the leading integral \eq{RHintegralIntroduction} 
becomes a similar expression to the mean field expression for a single eigenvalue of the 
matrix integral which appears in various studies in literature 
\cite{David, fy12,fy3,HHIKKMT,KKM,KazakovKostov}.%
\footnote{It is interesting that the Riemann-Hilbert expression gives a similar expression 
to the D-instanton operators obtained in the free-fermion formulation \cite{fy12,fy3}. }
Therefore, the Stokes multipliers $s_{n,2,1}$ in Eq.~\eq{RHintegralIntroduction} 
are directly identified as {\em the D-instanton chemical potentials} 
in the semi-classical saddle-point analysis. That is, {\em the first constraint 
is directly related to the constraint on the D-instanton chemical potentials.}
Furthermore, since the Riemann-Hilbert integral, Eq.~\eq{RHintegralIntroduction}, 
provides the complete integration representation 
based on the reference string background $g^{(j)}(t;\zeta)$, 
we can discuss {\em non-perturbative stability of the background $g^{(j)}(t;\zeta)$}, 
especially for the background which is obtained as large $N$ limit of the matrix models. 
This consideration for the stability is also expressed as a constraint on the Stokes multipliers and therefore 
the D-instanton chemical potentials. 
Originally, the mean field analyses include 
ambiguity of {\em choice of contour} and {\em weight of these contours} \cite{David} and this 
fact becomes a cause of the ambiguity about the D-instanton chemical potentials 
in continuum loop-equation systems. 
In the Riemann-Hilbert approach, however, these degrees of freedom are identified as anti-Stokes lines 
$\mathcal K_n$ and Stokes multipliers $s_{n,2,1}$, and they are tightly related to each other. 
As a consequence, the physical section of the D-instanton chemical potentials is obtained 
in the name of {\em non-perturbative completion}. 
This viewpoint is important in non-critical string theory 
because non-critical strings are sometimes defined as the large $N$ (i.e.~perturbative) expansion 
of unstable matrix-model critical points (e.g.~$(2,3)$ bosonic minimal string theory) 
and therefore the matrix-model description does not necessary guarantee 
non-perturbative completion of string theory.%
\footnote{Early investigations of non-perturbative complete string theories are found in \cite{BMPNonP,DSS,BMN,ComplexMatrixCite1}.}

As we will see in the coming sections, 
the above procedures completely determine the D-instanton chemical potentials in the two-cut $(1,2)$ critical 
points and results in the Hastings-McLeod solution \cite{HastingsMcLeod} to the Painlev\'e II equation 
(in Section \ref{TwoCutSectionSmallInstantonCond}). 
Actually it is known that this is the unique solution 
which realizes the two phases of the two-cut $(1,2)$ critical point of the two-cut matrix model,%
\footnote{It was shown by Hastings-McLeod \cite{HastingsMcLeod} that 
their solution is a unique solution to the Painlev\'e II equation, Eq.~\eq{PainleveIIEquation}, 
which realizes the following asymptotic behaviors of $f(t)$ 
on the two sides of infinity $t\to \pm \infty$: 
\begin{align}
\frac{1}{2}\ddot{f(t)}- f^3(t) + 2 t f(t)=0:\qquad 
f(t\to \infty)\sim 0,\qquad f(t\to -\infty) \sim \sqrt{-t}, 
\end{align}
which is the same behavior as the two-cut $(1,2)$ critical point of the two-cut matrix model 
discussed in \cite{UniCom}. For some mathematical derivation of this solution in the two-cut matrix models, 
see also \cite{BleherIts} which has been studied within the Riemann-Hilbert problem. }
and therefore the Hastings-McLeod solution is suitable for this critical point. 
An advantage of our work is the discovery of the actual physical 
requirements to obtain the correct solutions to the non-perturbative completion which are also 
applicable to the critical points with an arbitrary number of cuts. 

The next main developments shown in this paper is, therefore, 
an extension of our procedure to the general multi-cut cases 
(which even reaches to $\infty$-cut!). 
In particular, general structures of Stokes multipliers in the $k\times k$ 
isomonodromy systems are investigated in Section \ref{MultiCutStokesSection}, 
and a new way to identify non-trivial Stokes multipliers is proposed (Theorem \ref{TheoremIndexStokesMultipliers}) 
with terminology of {\em the profile of dominant exponents}. 
Furthermore, explicit solutions are obtained with help of the physical constraints, 
i.e.~the multi-cut boundary conditions (Theorem \ref{TheoremDiscreteSolutions} 
and \ref{TheoremContinuumSolutions} 
in Section \ref{AnsatzSolutionsSubSection}). 
In this sense, our solutions provide the multi-cut generalization 
of the Hastings-McLeod solution. 
Interestingly, we found that these solutions are labeled by 
constrained Young diagrams (Proposition \ref{AvalanchesProposition} in Section \ref{AnsatzSolutionsSubSection}). 
This result implies that there is a quite rich world beyond the non-perturbative horizon, 
and that the multi-cut matrix models provide fruitful fields for a quantitative study of these issues.

\

Organization of this paper is as follows:
In Section \ref{Section2StokesPheno}, 
after summarizing the asymptotic expansion of the ODE system in the multi-cut critical points, 
the general facts about Stokes phenomenon in ordinary differential equations are reviewed. 
As a warming up, the case of the two-cut $(1,2)$ critical point is also shown. 
In Section \ref{MultiCutStokesSection}, 
Stokes phenomena in the multi-cut critical points are studied. In particular, 
a systematic way of reading the Stokes multipliers in general cases is developed. 
In Section \ref{MultiCutBCSection}, 
the multi-cut boundary condition is proposed. 
In Section \ref{AnsatzSolutionsSubSection}, 
the discrete and continuum solutions are shown. 
In Section \ref{RHapproachSection}, the non-perturbative stability condition is studied in terms of the Riemann-Hilbert problem. 
Section \ref{SectionConclusionDiscussion} is devoted to conclusion and discussion. 

Context of Appendices is: 
Appendix \ref{AppendixAiryFunction} is about the Stokes phenomenon of Airy function (a review of \cite{MMSS}). 
Appendix \ref{AppendixLaxOP} is about calculation of Lax operators. 
Appendix \ref{ExamplesProfilesSections} is about supplements to Theorem \ref{ThmMCBCRec} 
and Theorem \ref{ThmConplementaryBCRec} 
with some examples of the multi-cut boundary-condition recursive equations. 
Appendix \ref{ProofOfSolutionSectionAppendix} is about derivation of continuum solutions. 
Appendix \ref{CalculationInConcreteSystems} is about calculation of the $3$-cut $(1,1)$ critical points 
and Appendix \ref{CalculationInConcreteSystems2} is about calculation of $4$-cut $(1,1)$ critical points. 

\section{Stokes phenomena in the ODE systems \label{Section2StokesPheno}}

Before we devote ourselves into the multi-cut systems, 
here we first review some general facts about Stokes phenomenon in ordinary differential equation systems, 
then we summarize the well-studied two-cut $(1,2)$ case. 
This two-cut system has been extensively studied not only in physical context \cite{GrossWitten,PeShe,DSS,Nappi,CDM,HMPN,UniCom,SeSh2,fi1} 
but also in mathematical context \cite{RHPIIcite,DeiftZhou,HastingsMcLeod,Kap12,Novok,IK,BI}, 
since it is related to the Hastings-McLeod solution \cite{HastingsMcLeod} of the Painlev\'e II system. 
For more comprehensive and rigorous reviews and references on the isomonodromy deformations, 
Stokes phenomenon and inverse monodromy problems, see \cite{ItsBook}. 
We also note that the idea of ismonodromy deformation was introduced in non-critical string theory by \cite{Moore}. 

\subsection{The ODE system and asymptotic expansions}
It was first proposed in \cite{fi1} that the multi-cut matrix models are 
controlled by multi-component KP hierarchy \cite{kcKP} and therefore by 
the following Baker-Akhiezer function system:
\begin{align}
\zeta \Psi(t;\zeta) 
&= \bP(t;\del)\, \Psi(t;\zeta),\label{BA11} \\ 
g_{\rm str}\frac{\del }{\del \zeta } \Psi(t;\zeta) 
&= \bQ(t;\del)\, \Psi(t;\zeta). \label{BA21}
\end{align}
Here the operator $\bP(t;\del)$ and $\bQ(t;\del)$ 
are $\hat p$-th and $\hat q$-th order differential operators in $\del\equiv g_{\rm str}\del_t$, respectively, 
which satisfy the Douglas (string) equation \cite{DouglasGeneralizedKdV}:
\begin{align}
\bigl[\bP(t;\del),\bQ(t;\del)\bigr]= g_{\rm str} I_k. 
\end{align}
Critical points in the multi-cut two-matrix models are characterized by 
these Lax operators and explicitly obtained in \cite{CISY1} with their critical potentials. 
There are two kinds of interesting critical points:~the $\mathbb Z_k$-symmetric 
critical points and fractional-superstring critical points. A brief summary of the corresponding 
Baker-Akhiezer function system is following:%
\footnote{For the derivation of these systems from the multi-cut two-matrix models, see \cite{CISY1}. }
\begin{itemize}
\item [1.]
The $\mathbb Z_k$-symmetric critical points 
are characterized by the following $k\times k$ Lax operators \cite{CISY1}: 
\begin{align}
\bP(t;\del) = \Gamma\, \del^{\hat p} 
+ \sum_{n=0}^{\hat p-1} U_n^{(Z_kP)}(t)\, \del^n, 
\qquad 
\bQ(t;\del) = \Gamma^{-1}\, \del^{\hat q} 
+ \sum_{n=0}^{\hat q-1} U_n^{(Z_kQ)}(t)\, \del^n,\label{PQOPZkSym}
\end{align}
with the shift matrix $\Gamma$,
\begin{align}
\Gamma = 
\begin{pmatrix}
0 & 1 &  \cr
   & 0 & 1 & \cr
   &    &  \ddots & \ddots & \cr
   &    &             &       0     &  1 \cr
1 &    &             &              &  0
\end{pmatrix}, 
\end{align}
and the $k\times k$ matrix-valued real coefficients $U_n^{(Z_kP)}(t)$ and $U_n^{(Z_kQ)}(t)$ 
which satisfy 
\begin{align}
U_n^{(Z_kP)}(t)=
\begin{pmatrix}
0 & * &  \cr
   & 0 & * & \cr
   &    &  \ddots & \ddots & \cr
   &    &             &       0     &  * \cr
* &    &             &              &  0
\end{pmatrix},
\qquad U_n^{(Z_kQ)}(t)=
\begin{pmatrix}
0 &  &  & & *\cr
 *  & 0 &  & \cr
   &  \ddots  &  \ddots &  & \cr
   &    &      *       &       0     &   \cr
 &    &             &        *      &  0
\end{pmatrix},\label{UZKQZKSYM}
\end{align}
as a result of the $\mathbb Z_k$ symmetry of the critical points. 
Macroscopic loop amplitudes (i.e.~off-critical resolvent amplitudes with $t\neq 0$) in this kind of critical points 
are also obtained in \cite{CISY1} with the Daul-Kazakov-Kostov prescription \cite{DKK} and 
expressed as the Jacobi polynomials or the third and fourth Chebyshev polynomials. 
In particular, the amplitudes in the the $k$-cut $(1,1)$ critical points are given as 
the eigenvalues of the Lax operators Eq.~\eq{PQOPZkSym} in the weak coupling limit $g_{\rm str}\to 0$:%
\footnote{In this paper, the equality $\simeq$ means that they are equal up to some similarity transformation. }
\begin{align}
\bP(t;\del) &\simeq \diag_{j=1}^k \Bigl(P^{(j)}_{\rm classical}(t;z)\Bigr) = 
\diag_{j=1}^k \Bigl(\omega^{j-1}\,x(z) \Bigr),\nn\\
\bQ(t;\del) &\simeq \diag_{j=1}^k \Bigl(Q^{(j)}_{\rm classical}(t;z)\Bigr) =
\diag_{j=1}^k \Bigl(\omega^{-(j-1)}\,y(z)\Bigr), 
\end{align}
with 
\begin{align}
x(z)
=  t \sqrt[k]{\bigl(z-c\bigr)^{l}\bigl(z-b\bigr)^{k-l}},\qquad 
y(z)
=  t \sqrt[k]{\bigl(z-c\bigr)^{k-l}\bigl(z-b\bigr)^{l}}
\end{align}
and $0=c\,l+b\,(k-l)$ and the dimensionless variable $z\equiv g_{\rm str} t^{-1}\del_t$. 
\item [2.]
The fractional-superstring critical points \cite{irie2}
are characterized by the following two kinds of Lax operators \cite{CISY1}: 
The first kind is given as 
\begin{align}
\bP(t;\del) = \Gamma\, \del^{\hat p} 
+ \sum_{n=0}^{\hat p-1} U_n^{(F_kP)}(t)\, \del^n, \qquad 
\bQ(t;\del) = \Gamma\, \del^{\hat q} 
+ \sum_{n=0}^{\hat q-1} U_n^{(F_kQ)}(t)\, \del^n. 
\end{align}
These Lax operators are derived 
from the $\omega^{1/2}$-rotated critical potentials. 
The second kind is given as 
\begin{align}
\bP(t;\del) = \Gamma^{(\rm real)}\, \del^{\hat p} 
+ \sum_{n=0}^{\hat p-1} U_n^{(R_kP)}(t)\, \del^n, \qquad 
\bQ(t;\del) = \Gamma^{(\rm real)}\, \del^{\hat q} 
+ \sum_{n=0}^{\hat q-1} U_n^{(R_kQ)}(t)\, \del^n, \label{RealPotentialLaxOP}
\end{align}
with the matrix $\Gamma^{(\rm real)}$,
\begin{align}
\Gamma^{(\rm real)} = 
\begin{pmatrix}
0 & 1 &  \cr
   & 0 & 1 & \cr
   &    &  \ddots & \ddots & \cr
   &    &             &       0     &  1 \cr
-1 &    &             &              &  0
\end{pmatrix}. 
\end{align}
These Lax operators are derived from the real critical potentials. 
In both cases, all the $k\times k$ matrix-valued coefficients $U_n^{(F_kP)}(t)$ and $U_n^{(F_kQ)}(t)$ 
(and $U_n^{(R_kP)}(t)$ and $U_n^{(R_kQ)}(t)$) are real functions. 
The macroscopic loop amplitudes in each case are obtained 
and given by the deformed Chebyshev functions \cite{CIY1}. 
\end{itemize}

In this paper, for the sake of simplicity, 
we concentrate on the $\hat p=1$ cases of the $\mathbb Z_k$-symmetric critical points. 
With this choice of critical points, the Lax operator $\bP(t;\del)$ becomes 
\begin{align}
\bP(t;\del)= \Gamma \del + H(t),  \label{DefOfOpPGeneralKCutHHHHH}
\end{align}
and the Baker-Akhiezer function for 
the eigenvalue problem of the operator $\bP(t;\del)$, Eq.~\eq{BA11}, 
is rewritten as 
\begin{align}
g_{\rm str} \frac{\del}{\del t} \Psi(t;\zeta) = \mathcal P(t;\zeta)\, \Psi(t;\zeta) 
\equiv \Gamma^{-1} \bigl[\zeta -H(t)\bigr] \, \Psi(t;\zeta), 
 \label{GeneralODEP}
\end{align}
and therefore Eq.~\eq{BA21} is also rewritten 
as a $k\times k$ matrix polynomial operator in $\zeta$:
\begin{align}
g_{\rm str} \frac{\del \Psi(t;\zeta)}{\del \zeta }= \mathcal Q(t;\zeta)\, \Psi(t;\zeta)\equiv \bQ(t;\del)\, \Psi(t;\zeta),
 \qquad \mathcal Q(t;\zeta)= \sum_{n=1}^{r} \mathcal Q_{-n}(t) \zeta^{n-1}.
 \label{GeneralODEQ}
\end{align}
Here we define $r$ as
\begin{align}
r \equiv \hat q+1>0,
\end{align}
which is referred to as the Poincar\'e index in literature. 
The advantage of this formulation is that the pair of Lax operators 
$(\bP(t;\del),\bQ(t;\del))$ becomes 
a pair of the polynomial operators $(\mathcal P(t;\zeta),\mathcal Q(t;\zeta))$, and 
the system can be expressed as an $k\times k$ first order ordinary differential equation (ODE) system. 
These systems are called 
the Zakharov-Shabat eigenvalue problem \cite{ZS} or AKNS hierarchy \cite{AKNS} in literature. 
Note that the Douglas equation becomes
\begin{align}
\bigl[\bP(t;\del),\bQ(t;\del)\bigr] = g_{\rm str} I_k \quad \Leftrightarrow \quad 
\bigl[g_{\str} \del_\zeta- \mathcal Q(t;\zeta), g_{\rm str}\del_t- \mathcal P(t;\zeta)\bigr] = 0,
\end{align}
in terms of these Lax operators. 

This ODE system Eq.~\eq{GeneralODEQ} has the $k$ independent order $k$ column vector solutions 
$\Psi^{(j)}(t;\zeta)$, ($j=1,2,\cdots,k$), and we here use the following matrix solution notation: 
\begin{align}
\Psi(t;\zeta) \equiv \Bigl(\Psi^{(1)}(t;\zeta),\cdots, \Psi^{(k)}(t;\zeta)\Bigr). 
\end{align}
As in the usual ODE, we consider formal expansion around $\zeta\to \infty$. 
However the point $\zeta \to \infty$ is an irregular singularity 
and the formal series expansion around this irregular point in general does not converge absolutely.  
Up to proper redefinition of the $k$ independent solutions, 
the formal series expansion of the solutions around $\zeta\to \infty$ is given as 
\begin{align}
\Psi_{asym}(t;\zeta) \equiv Y(t;\zeta) \, e^{\frac{1}{g_{\rm str}}\varphi(t;\zeta)} 
\equiv  \Bigl[I_k +\sum_{n=1} ^\infty \frac{Y_n(t)}{\zeta^n}\Bigr]\times 
\exp \Bigl[\frac{1}{g_{\rm str}}
\Bigl(\varphi_0 \ln \zeta - \sum_{m=-r,\neq 0}^\infty \frac{\varphi_m(t)}{m\,\zeta^m} \Bigr)\Bigr]. 
\label{FormalExpansionAsymptoticExpansionsMultiCut}
\end{align}
The coefficient matrices are obtained from the recursive equations, 
\begin{align}
0=-n g_{\rm str}Y_n(t) + \sum_{m=0}^{n+r} \Bigl[ Y_m (t) \,\varphi_{n-m}(t)- \mathcal Q_{n-m}(t) \,Y_m(t)\Bigr], 
\quad \bigl(n=-r,-r+1,\cdots\bigr). \label{RecAsym}
\end{align}
For convenience, we extend the indices of the coefficient matrices: 
\begin{align}
Y_0(t)=I_k,\quad Y_n(t)=0 \quad (n<0),\qquad \varphi_m(t) =\mathcal Q_m(t)=0 \quad (m<-r), 
\end{align}
and impose the following constraints on $Y_n(t)$ and $\varphi_n(t)$:
\begin{align}
\bigl[\Gamma^l,\varphi_n(t)\bigr]=0,\qquad \sum_{i=1}^k\bigl[Y_n(t)\bigr]_{i,i+l} = 0, \qquad (l =0,1,\cdots, k-1). 
\end{align}
This recursive relation then can be solved uniquely 
and all the expansion coefficient are written with the coefficient matrix-valued function $H(t)$ 
in Eq.~\eq{DefOfOpPGeneralKCutHHHHH}. 

On the other hand, it is also convenient to use a diagonal basis, 
$\widetilde \Psi_{asym}(t;\zeta)$,
which is defined by 
\begin{align}
\widetilde \Psi_{asym}(t;\zeta) &\equiv
\widetilde Y(t;\zeta) \, e^{\frac{1}{g_{\rm str}}\widetilde \varphi(t;\zeta)} 
\equiv  \Bigl[I_k +\sum_{n=1} ^\infty \frac{\widetilde Y_n(t)}{\zeta^n}\Bigr]\times 
\exp \Bigl[\frac{1}{g_{\rm str}}\Bigl(\widetilde \varphi_0 \ln \zeta - \sum_{m=-r,\neq 0}^\infty \frac{\widetilde \varphi_m(t)}{m\,\zeta^m} \Bigr)\Bigr] \nn\\
&\equiv U^\dagger \, \Psi_{asym}(t;\zeta) \, U,\label{AsymExpTilde}
\end{align}
where the matrix $U$ is given as 
\begin{align}
U_{jl} = \frac{1}{\sqrt{k}} \omega^{(j-1)(l-1)}, 
\qquad 
\Gamma\, U = U\, \Omega, \label{GammaOmega}
\end{align}
with $\Omega = \diag (1,\omega,\omega^2,\cdots,\omega^{k-1})$ and 
$\omega=e^{2\pi i/k}$. 
Since this is a similarity transformation, the coefficients also satisfy the same recursive 
relation \eq{RecAsym}. 
In this basis, the function $\widetilde \varphi(t;\zeta)$ is a diagonalized matrix 
and we write its eigenvalues as 
\begin{align}
\widetilde \varphi(t;\zeta) = \diag \bigl( \varphi^{(1)}(t;\zeta) ,\cdots, 
\varphi^{(k)}(t;\zeta)\bigr). \label{IEIKExponents}
\end{align}
The vector components of the formal series, 
$\widetilde \Psi_{asym} = \bigl(\widetilde \Psi_{asym}^{(1)}
,\cdots,\widetilde \Psi_{asym}^{(k)}\bigr)$, is given as 
\begin{align}
\widetilde \Psi_{asym}^{(j)}(t;\zeta) 
= \widetilde Y^{(j)}(t;\zeta)\, e^{\frac{1}{g_{\rm str}}\varphi^{(j)}(t;\zeta)},
\qquad (j=1,2,\cdots,k),
\label{FirstAppearanceOfAsymptoticExpansion}
\end{align}
with $\widetilde Y(t;\zeta) = \bigl(\widetilde Y^{(1)},\cdots, \widetilde Y^{(k)}\bigr)$.

Although the above formal solutions are formal series around the irregular singularity, 
they are related to the exact analytic solutions of the ODE system, $\widetilde \Psi(t;\zeta)$, 
in the sense of {\em asymptotic expansion}: 
\begin{align}
\widetilde \Psi(t;\zeta) \asymeq \widetilde \Psi_{asym}(t;\zeta) \, C, \label{AsympIEKIDKKEIH}
\end{align}
in some specific angular domain \cite{AsymptoticExpansionCite}: 
\begin{align}
\zeta \to \infty \in D(a,b) \equiv \bigl\{\zeta\in \mathbb C;\, 
a<\arg(\zeta)<b\bigr\}. 
\end{align}
An example of the angular domain is shown in Fig.~\ref{AngularDomainFigure}-a. 
Here $C$ is a proper coefficient matrix, and the meaning of asymptotic expansion is following:
\begin{Definition}
[asymptotic expansion]
For a holomorphic function $f(\zeta)$, 
{\em an asymptotic expansion of $f(\zeta)$ in a domain $D(a,b)$} is defined 
as a formal series $\sum_n f_n\zeta^{-n}$ such that there exists a constant $B_{R;a,b}^{(N)} \in \mathbb R$ which satisfies 
\begin{align}
\Bigl|f(\zeta) - \sum_{n=-r}^N \frac{f_n}{\zeta^n}\Bigr| 
< \frac{B_{R;a,b}^{(N)}}{ |\zeta|^{N}},
\qquad 
\zeta \in D(a,b) \cap \bigl\{\zeta\in \mathbb C;\, \bigl|\zeta\bigr|>R\bigr\}
\end{align}
for each integer $N = -r,-r+1,\cdots$ and sufficiently large $R \in \mathbb R$. 
This is written as 
\begin{align}
f(\zeta) \asymeq \sum_{n=-r}^\infty \frac{f_n}{\zeta^n},\qquad 
\zeta \to \infty \in D(a,b). 
\end{align}
\end{Definition}
The maximal angular domains are called {\em Stokes sectors}.

\subsection{General facts on Stokes phenomena in the ODE system \label{ReviewOfStokesODE}}
In this subsection, in order to understand the asymptotic expansion 
Eqs.~\eq{AsymExpTilde} and \eq{AsympIEKIDKKEIH}, we review some general theorem about 
the asympototic expansions and Stokes phenomena in the general $k\times k$ ODE systems,
\begin{align}
g_{\rm str} \frac{\del}{\del \zeta} \widetilde \Psi(t;\zeta) 
&= \Bigl[\widetilde {\mathcal Q}_{-r}\, \zeta^{r-1}
+ \widetilde {\mathcal Q}_{-r+1}(t)\, \zeta^{r-2}+ \cdots 
\widetilde {\mathcal Q}_{-1}(t) \Bigr] \widetilde \Psi(t;\zeta) \nn\\
&\equiv \widetilde {\mathcal Q}(t;\zeta)\,\widetilde \Psi(t;\zeta). 
\end{align}
Note that proof of the theorems appearing in this subsection can be found in \cite{ItsBook} 
and references therein. 
For sake of simplicity, we assume
\begin{align}
\widetilde {\mathcal Q}_{-r}= \diag \bigl(A_1,A_2,\cdots,A_k\bigr), \qquad 
A_i -A_j\neq 0,\qquad A_i \neq 0, \qquad (i,j=1,2,\cdots,k). \label{CoefOfLaxLeading}
\end{align}
Therefore, the exponents Eq.~\eq{IEIKExponents} are expressed as 
\begin{align}
\widetilde \varphi(t;\zeta) = \widetilde \varphi_0(t) \ln \zeta- \sum_{n=-r,n \neq 0}^\infty 
\frac{\widetilde \varphi_{n}(t)}{n \zeta^n}=\frac{1}{r}\widetilde{\mathcal Q}_{-r}\, \zeta^r + \cdots,\label{VerPhiDefinition}
\end{align}
and $\varphi_{-r}^{(i)} = A_i$ $(i=1,2,\cdots,k)$ also satisfies \eq{CoefOfLaxLeading}. 

\begin{figure}[htbp]
\begin{center}
\begin{picture}(180,190)(0,0)
\end{picture}
\begin{picture}(0,0)(220,-60)
\put(70,70){\line(0,1){10}\line(1,0){10}}
\put(73,75){$\zeta$}
\put(0,0){$0$}
\put(10,10){\line(0,-1){40}\line(-1,0){40}}
\put(10,10){\vector(0,1){70}\vector(1,0){70}}
\put(13,18){\line(3,-1){11}}
\put(15,23){\line(3,-1){17}}
\put(17,28){\line(3,-1){23}}
\put(19,34){\line(3,-1){30}}
\put(21,40){\line(3,-1){38}}
\put(23,46){\line(3,-1){46}}
\put(25,52){\line(3,-1){54}}
\put(27,58){\line(3,-1){52}}
\put(29,64){\line(3,-1){47}}
\put(31,70){\line(3,-1){40}}
\put(33,76){\line(3,-1){30}}
\put(35,82){\line(3,-1){20}}
\qbezier(39,10)(39,15)(37,19)
\qbezier(30,10)(31,20)(17,30)
\thicklines
\put(10,10){\line(1,3){25}}
\put(10,10){\line(3,1){70}}
\put(45,14){$a$}
\put(27,26){$b$}
\put(40,50){$D(a;b)$}
\put(16,-63){(a)}
\end{picture}
\begin{picture}(5,0)(0,-89)
\put(70,70){\line(0,1){10}\line(1,0){10}}
\put(73,75){$\zeta$}
\put(0,0){$0$}
{
\thicklines
\put(10,10){\rotatebox{-90}{\textcolor{red}{\line(1,0){70}}}}
\put(10,10){\rotatebox{-60}{\textcolor{green}{\line(1,0){70}}}}
\put(10,10){\rotatebox{-30}{\textcolor{blue}{\line(1,0){70}}}}
\put(10,10){\rotatebox{0}{\textcolor{red}{\line(1,0){70}}}}
\put(9.5,10){\rotatebox{30}{\textcolor{green}{\line(1,0){70}}}}
\put(9.5,10){\rotatebox{60}{\textcolor{blue}{\line(1,0){70}}}}
\put(9.5,10){\rotatebox{90}{\textcolor{red}{\line(1,0){70}}}}
\put(-25.5,10){\rotatebox{120}{\textcolor{green}{\line(1,0){70}}}}
\put(-51,10){\rotatebox{150}{\textcolor{blue}{\line(1,0){70}}}}
\put(-60,9.5){\rotatebox{180}{\textcolor{red}{\line(1,0){70}}}}
\put(-50,10){\rotatebox{-150}{\textcolor{green}{\line(1,0){70}}}}
\put(-25,10){\rotatebox{-120}{\textcolor{blue}{\line(1,0){70}}}}
}
\put(-8,0){
\rotatebox{0}{
\begin{picture}(0,0)(0,0)
\put(11,13){\line(3,-2){9}}
\put(12,18){\line(3,-2){24}}
\put(13,23){\line(3,-2){42}}
\put(14,28){\line(3,-2){56}}
\put(15,34){\line(3,-2){65}}
\put(16,40){\line(3,-2){63}}
\put(17,46){\line(3,-2){60}}
\put(18,52){\line(3,-2){58}}
\put(19,58){\line(3,-2){57}}
\put(20.5,64){\line(3,-2){55}}
\put(22,70){\line(3,-2){50}}
\put(23,76){\line(3,-2){40}}
\put(24,82){\line(3,-2){25}}
\thicklines
\put(10,10){\line(1,5){15}}
\put(10,10){\line(3,-1){70}}
\end{picture}}}
\put(58,47){\footnotesize (1,2)}
\put(70,14){\footnotesize (2,3)}
\put(33,74){\footnotesize (3,1)}
\put(-33,74){\footnotesize (1,2)}
\put(-70,14){\footnotesize (2,3)}
\put(-58,47){\footnotesize (3,1)}
\put(33,-58){\footnotesize (1,2)}
\put(0,-67){\footnotesize (2,3)}
\put(58,-33){\footnotesize (3,1)}
\put(-58,-33){\footnotesize (1,2)}
\put(0,82){\footnotesize (2,3)}
\put(-33,-58){\footnotesize (3,1)}
\put(3,-93){(b)}
\end{picture}
\end{center}
 \caption{\footnotesize a) An angular domain of $D(a,b)$. 
b) Stokes lines and Stokes sectors. 
This is the $3$-cut $(1,1)$ critical points. An example of Stokes sectors is also shown. 
In this critical point, there are three kinds of the Stokes lines ${\rm SL}_{i,j}$, $(i,j)=(1,2),(2,3),(3,1)$. 
Stokes sectors includes one and only one Stokes line of each kind. 
\label{AngularDomainFigure}}
\end{figure}
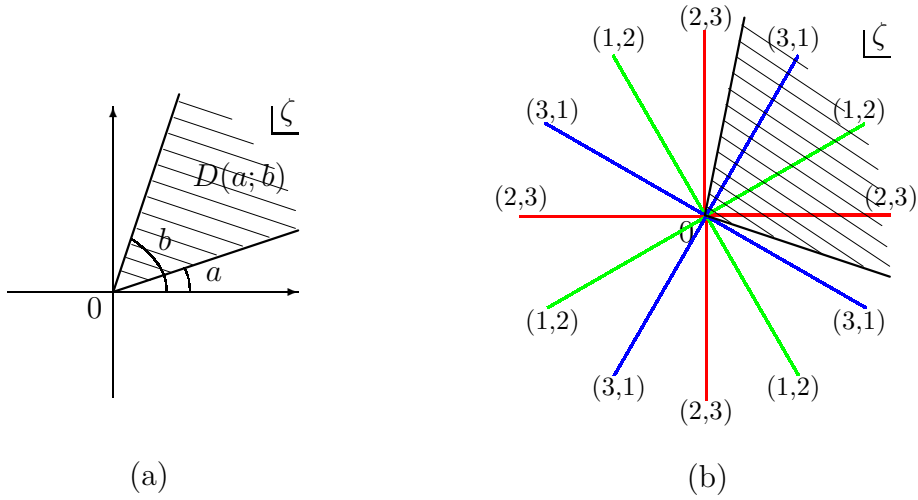

The meaning of the asymptotic expansion Eq.~\eq{AsymExpTilde} is that 
basically we ignore relatively small exponents. One takes 
some (small enough) anglular domain $D(a,e^{i\epsilon} a)$ then compares 
the relative magnitudes around $\zeta \to \infty$, for example,
\begin{align}
\bigl|e^{\varphi^{(j_1)}(t;\zeta)}\bigr|
<\bigl|e^{\varphi^{(j_2)}(t;\zeta)}\bigr|
<\cdots 
<\bigl|e^{\varphi^{(j_k)}(t;\zeta)}\bigr|,\qquad \zeta\to \infty \in D(a,e^{i\epsilon} a). 
\end{align}
Then one can obtain the following equality under the asymptotic expansion: 
\begin{align}
e^{\varphi^{(j_2)}(t;\zeta)} + \theta e^{\varphi^{(j_1)}(t;\zeta)} \asymeq e^{\varphi^{(j_2)}(t;\zeta)},\qquad\zeta \to \infty \in D(a,e^{i\epsilon} a). \label{NeglesibleExpAsym}
\end{align}
That is, the smaller exponents become practically invisible in view of the asymptotic expansion. 
Our interest is to identify angles of $\zeta$ 
where the the exponents, $\exp\bigl(\varphi^{(j)}(\zeta)\bigr)$ 
($i=1,2,\cdots,k$), change the relative magnitudes around $\zeta \to \infty$. 
This leads to the concept of {\em Stokes lines}: 
\begin{Definition}
[Stokes lines]
With the assumption \eq{CoefOfLaxLeading}, 
Stokes lines ${\rm SL}_{j,l}$ in this ODE system are defined 
for each pair of $(j,l)$ as 
\begin{align}
{\rm SL}_{j,l}&\equiv 
\Bigl\{\zeta \in \mathbb C;
{\rm Re}\bigl[\bigl( \varphi^{(j)}_{-r}-\varphi^{(l)}_{-r}\bigr)\zeta^r\bigr] = 0 \Bigr\}
= \bigcup_{n=0}^{2r-1} {\rm SL}_{j,l}^{(n)}, \label{StokesLinesDef}
\end{align}
which consists of $2r$ semi-infinite lines, ${\rm SL}_{j,l}^{(n)}$ $(n=0,1,\cdots,2r-1)$. 
The set of lines, ${\rm SL}$, denotes a set of whole Stokes lines, 
${\rm SL} \equiv \bigcup_{j,l} {\rm SL}_{j,l}$. 
\end{Definition}
An example of Stokes lines ${\rm SL}_{j,l}$ is shown in Fig.~\ref{AngularDomainFigure}-b. 
In particular, if the angular domain $D(a,b)$ of the asymptotic expansion includes a Stokes line, 
one cannot neglect the exponents as it happens in Eq.~\eq{NeglesibleExpAsym}. 
This leads to the following definition of {\em Stokes sectors}:

\begin{Definition}
[Stokes sectors]
A Stokes sector $D$ in the ODE system is an angular domain, $D= D(a,b)$, with angles $a$ and $b$ such that for each pair of $(j,l)$ there exist a unique 
Stokes line ${\rm SL}_{j,l}^{(n_{j,l})}$ which satisfies,
\begin{align}
{\rm SL}_{j,l}^{(n_{j,l})} \subset D=D(a,b), \label{DefStokesSector}
\end{align}
that is, except for this line ${\rm SL}_{j,l}^{(n_{j,l})}$ there is no other line ${\rm SL}_{j,l}^{(n_{j,l}')}$ 
($\neq {\rm SL}_{j,l}^{(n_{j,l})}$) which runs inside the domain, $D$. 
\end{Definition}
An example of the Stokes sectors (the $3$-cut $(1,1)$ critical point) 
is shown in Fig.~\ref{AngularDomainFigure}-b. 

Actually the definition of the Stokes sectors results in the following theorem \cite{AsymptoticExpansionCite}: 
\begin{Theorem}
For a given Stokes sector $D$, 
any solutions to the ODE system $\widetilde \Psi(t;\zeta)$ 
has the following asymptotic expansion: 
\begin{align}
\widetilde \Psi(t;\zeta) \asymeq \widetilde \Psi_{asym}(t;\zeta) \, C,\qquad 
\zeta \to \infty \in D, 
\end{align} 
with a matrix $C$. 
Furthermore, the coefficient matrix $C$ (i.e.~asymptotic expansion) is unique 
in the Stokes sector $D$. 
\end{Theorem}
This uniqueness enables us to define the following unique solution in a Stokes sector $D$:
\begin{Definition} 
[Canonical solution]
If the solution to the ODE system, $\widetilde \Psi_{can}(t;\zeta)$, 
has the asymptotic expansion with $C=I_k$ in a Stokes sector $D$, 
\begin{align}
\widetilde \Psi_{can}(t;\zeta) \asymeq \widetilde \Psi_{asym}(t;\zeta),
\qquad \zeta \to \infty \in D, 
\label{FirstAppearanceOfCanonicalSolutions}
\end{align}
this solution is called the canonical solution in the Stokes sector $D$. 
\end{Definition}

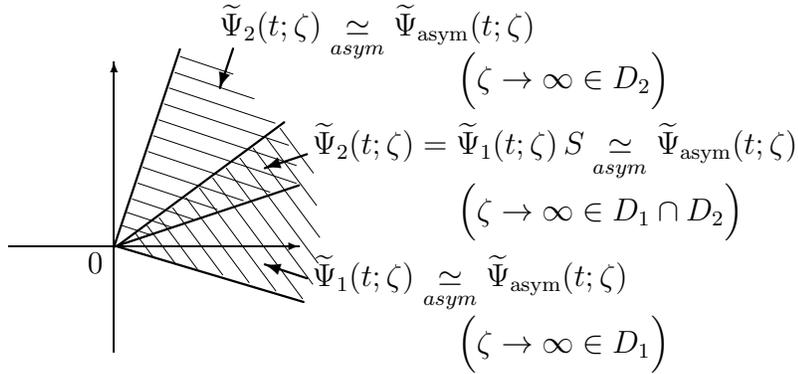
\begin{figure}[htbp]
\begin{center}
\begin{picture}(180,150)(0,0)
\end{picture}
\begin{picture}(0,0)(220,-30)
\put(0,0){$0$}
\put(10,10){\line(0,-1){40}\line(-1,0){40}}
\put(10,10){\vector(0,1){70}\vector(1,0){70}}
\put(13,18){\line(3,-1){11}}
\put(15,23){\line(3,-1){17}}
\put(17,28){\line(3,-1){23}}
\put(19,34){\line(3,-1){30}}
\put(21,40){\line(3,-1){38}}
\put(23,46){\line(3,-1){46}}
\put(25,52){\line(3,-1){54}}
\put(27,58){\line(3,-1){52}}
\put(29,64){\line(3,-1){47}}
\put(31,70){\line(3,-1){40}}
\put(33,76){\line(3,-1){30}}
\put(35,82){\line(3,-1){20}}
\put(-10,10){
\rotatebox{-35}{
\begin{picture}(0,0)(0,0)
\put(13,18){\line(3,-1){11}}
\put(15,23){\line(3,-1){17}}
\put(17,28){\line(3,-1){23}}
\put(19,34){\line(3,-1){30}}
\put(21,40){\line(3,-1){38}}
\put(23,46){\line(3,-1){46}}
\put(25,52){\line(3,-1){54}}
\put(27,58){\line(3,-1){52}}
\put(29,64){\line(3,-1){47}}
\put(31,70){\line(3,-1){40}}
\put(33,76){\line(3,-1){30}}
\put(35,82){\line(3,-1){20}}
\thicklines
\put(10,10){\line(1,3){25}}
\put(10,10){\line(3,1){70}}
\end{picture}}}
\thicklines
\put(10,10){\line(1,3){25}}
\put(10,10){\line(3,1){70}}
\put(85,-5){$\widetilde\Psi_1(t;\zeta)\asymeq \widetilde\Psi_{\rm asym}(t;\zeta)$}
\put(140,-30){$\Bigl(\zeta \to \infty \in D_1\Bigr)$}
\put(85,45){$\widetilde\Psi_2(t;\zeta)=\widetilde\Psi_1(t;\zeta)\, S \asymeq \widetilde\Psi_{\rm asym}(t;\zeta)$}
\put(140,20){$\Bigl(\zeta \to \infty \in D_1\cap D_2\Bigr)$}
\put(50,90){$\widetilde\Psi_2(t;\zeta) \asymeq \widetilde\Psi_{\rm asym}(t;\zeta)$}
\put(140,70){$\Bigl(\zeta \to \infty \in D_2\Bigr)$}
\put(55,85){\vector(-1,-3){5}}
\put(83,45){\vector(-3,-1){16}}
\put(83,-2){\vector(-3,1){16}}
\end{picture}
\end{center}
 \caption{\footnotesize  Explanation of Stokes phenomenon in ODE systems. For given two 
Stokes sectors, their canonical solutions are generally different by a Stokes matrix, $S$ in the intersection $D_1\cap D_2$. This behavior of analytic functions is called Stokes phenomenon. \label{StokesPhenomenonFigure}}
\end{figure}

This theorem on the other hand means that the asymptotic expansion is not unique 
if one chooses some angular domain $D'$ narrower than Stokes sectors. In particular, 
as is shown in Fig.~\ref{StokesPhenomenonFigure}, 
the intersection of two different Stokes sectors $D_1$ and $D_2$ is generally narrower 
than Stokes sectors, and therefore there appears 
difference between the canonical solutions $\widetilde\Psi_i(t;\zeta)$ of each sector $D_i (i=1,2)$: 
\begin{align}
\widetilde\Psi_2(t;\zeta) = \widetilde\Psi_1(t;\zeta) S, \qquad D_1 \cap D_2 \neq \emptyset. 
\label{FirstAppearanceOfStokesMatrix}
\end{align}
This $k\times k$ matrix $S$ which expresses the difference between $\widetilde\Psi_1(t;\zeta)$ and $\widetilde\Psi_2(t;\zeta)$ 
is called a {\em Stokes matrix} in the intersection $D_1\cap D_2$. 
This indicates that solutions in the ODE system generally have different asymptotic expansion 
in different Stokes sectors. 
This analytic behavior of the solutions is referred to 
as the {\em Stokes phenomenon} in the ODE system. 

A direct calculation shows that the Stokes matrices do not depend on $\zeta$, and 
furthermore, they do not depend on the deformation parameter $t$ either 
(as in \eq{GeneralODEP}): 
\begin{align}
\frac{dS}{d\zeta}=\frac{dS}{d t }= 0. 
\end{align}
This means that the Stokes matrices are understood 
as integration constants for the evolution system in the $t$ space. 
Therefore, these integrable deformations in the original multi-component KP hierarchy 
are also called {\em isomonodromy deformation system} \cite{RHcite}. 
This also leads us to the concept of {\em inverse monodromy approach} \cite{RHcite,RHPIIcite}, 
which is also briefly reviewed in Section \ref{RHapproachSection}. 

Components of Stokes matrices satisfy the following theorem (See \cite{ItsBook}, for example): 
\begin{Theorem}
[Stokes multipliers] 
For given two Stokes sectors $D_1$ and $D_2$ ($D_1 \cap D_2 \neq \emptyset$), 
components of their Stokes matrices, i.e.~Stokes multipliers, $S=(s_{i,j})$, satisfy 
\begin{align}
s_{j,j} = 1\qquad (j=1,2,\cdots,k),
\end{align}
and $s_{i,j}\, (i\neq j)$ can take non-zero values only when the exponents satisfy 
\begin{align}
{\rm Re}\bigl[\varphi_{-r}^{(i)}\zeta^r\bigr]< {\rm Re}\bigl[\varphi_{-r}^{(j)}\zeta^r\bigr]
\text{ for all angular range of } \zeta\to \infty  \in D_1\cap D_{2}\neq \emptyset. 
\end{align}
In particular, these Stokes multipliers, $s_{i,j}\, (i\neq j)$, 
are called ``non-trivial''. 
\label{TheoremStokesMultipliers}
\end{Theorem}
In this paper, we often refer to these facts about Stokes phenomena in ODE systems. 

\subsection{Stokes phenomena in the two-cut case \label{TwoCutStokesPheSubSection}}

In this subsection, we specialize the general consideration to the two-cut $(1,2)$ critical point. 

\subsubsection{The ODE system and asymptotic expansions in the two-cut case}
The string equation in this system is known as the Painlev\'e II equation \cite{PeShe,DSS}, 
\begin{align}
\frac{g_{\rm str}^2}{2}\ddot{f}- f^3 + 2 t f=0,\label{PainleveIIEquation}
\end{align}
which is equivalent to the following ODE system in $\zeta$ (Eq.~\eq{ODEandIsomono2cutQ}) with its 
isomonodromy deformations in $t$ (Eq.~\eq{ODEandIsomono2cutP}):%
\footnote{In the later discussion (from Section \ref{MultiCutStokesSection}), 
we also define a different basis: 
$\Psi(t;\zeta) \equiv U \widetilde \Psi(t;\zeta) U^\dagger$, 
with 
\begin{align}
U\sigma_3 U^\dagger = \sigma_1, \qquad U\sigma_1 U^\dagger = - \sigma_3, \qquad 
U \sigma_2 U^\dagger = \sigma_2. 
\end{align}
This basis naturally appears 
in the matrix-model calculations and is more suitable to read the Hermiticity 
of the multi-cut matrix models \cite{CISY1}. }
\begin{align}
g_{\rm str} \frac{\del}{\del \zeta} \widetilde \Psi(t;\zeta) 
&= \Bigl[\sigma_3 \zeta^2-\bigl(\sigma_1 f\bigr)\zeta  +\Bigl( -\frac{1}{2}f^2 + \mu \Bigr) \sigma_3- g_{\rm str}\frac{i}{2} \sigma_2 \dot{f} \Bigr] \widetilde \Psi(t;\zeta),  
\label{ODEandIsomono2cutQ}\\
g_{\rm str} \frac{\del}{\del t} \widetilde \Psi(t;\zeta) 
&= \Bigl[\sigma_3 \zeta - \sigma_1 \, f(t)\Bigr]\, \widetilde \Psi(t;\zeta). 
\label{ODEandIsomono2cutP}
\end{align}
Since this $2 \times 2$ first-order ODE system has two independent column vector solutions 
$\widetilde \Psi^{(1)}(t;\zeta) $ and $\widetilde \Psi^{(2)}(t;\zeta)$, we use the matrix notation 
for the solutions: 
\begin{align}
\widetilde \Psi(t;\zeta) = \Bigl(\widetilde \Psi^{(1)}(t;\zeta),
\widetilde \Psi^{(2)}(t;\zeta) \Bigr). 
\end{align}

At the point $\zeta \to \infty$, the ODE has an irregular singularity (of the Poincar\'e order $3$) 
and the formal expansion of the solutions \eq{AsymExpTilde} is given as 
\begin{align}
\widetilde \Psi_{asym}(\zeta;t) 
&= \Bigl[I_2 + \frac{i}{2\zeta }\sigma_2 f(t) + \mathcal O(1/\zeta^2)\Bigr] 
\exp\Bigl[\frac{1}{g_{\rm str}}
\Bigl(\frac{1}{3}\sigma_3 \zeta^3 + \mu \sigma_3 \zeta+ \mathcal O(1/\zeta)\Bigr) \Bigr] \nn\\
&\equiv \widetilde Y(t;\zeta) \, e^{\frac{1}{g_{\rm str}}\widetilde \varphi(t;\zeta)}. \label{AsymTwoCutIIIIIIIII}
\end{align}
This can be obtained with the recursion relation Eq.~\eq{RecAsym} (see also
in Appendix \ref{FSSTLaxOperatorAppendixxx}).
Note that the 
exponent $\widetilde \varphi(t; \zeta)$ is a diagonal matrix which satisfies 
$\widetilde \varphi(t;\zeta)\propto \sigma_3$, and then each vector solution 
$\widetilde \Psi_{asym}^{(i)}(t;\zeta)$ $(i=1,2)$ has different exponents:
\begin{align}
\widetilde \Psi_{asym}^{(i)}(t;\zeta) 
= \widetilde Y^{(i)}(t;\zeta)\, e^{\frac{1}{g_{\rm str}}\varphi^{(i)}(t;\zeta)},
\end{align}
with 
\begin{align}
\widetilde Y(t;\zeta) = \bigl(\widetilde Y^{(1)}(t;\zeta),\widetilde Y^{(2)}(t;\zeta) \bigr), 
\qquad \widetilde \varphi(t;\zeta) 
= \diag\bigl(\varphi^{(1)}(t;\zeta),\varphi^{(2)}(t;\zeta)\bigr). 
\end{align}

\subsubsection{Stokes sectors and Stokes matrices}
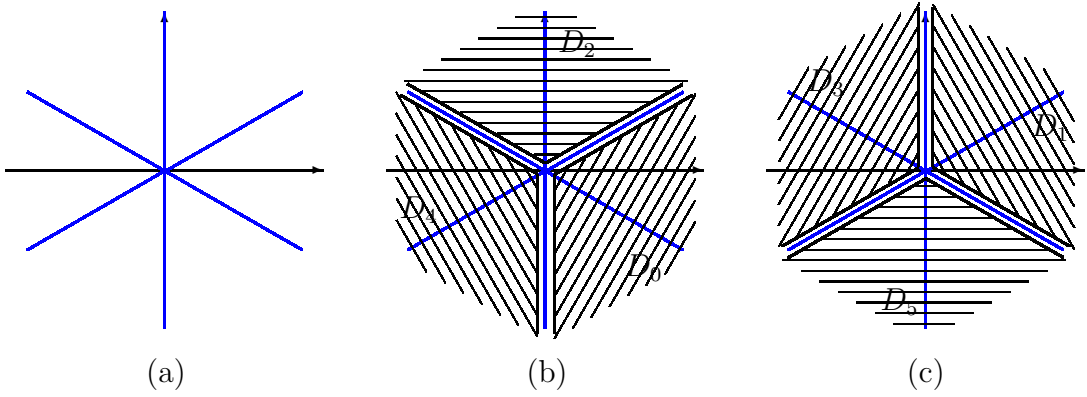
\begin{figure}[htbp]
\begin{center}
\begin{picture}(120,140)(0,0)
\begin{picture}(140,0)(90,-70)
\put(0,59){\vector(0,1){1}}
\put(0,0){\vector(1,0){60}}
\put(0,0){\line(-1,0){60}}
\thicklines
\put(0,0){\rotatebox{30}{\textcolor{blue}{\line(1,0){60}}}}
\put(0,0){\textcolor{blue}{\line(0,1){60}}}
\put(-52,0){\rotatebox{60}{\textcolor{blue}{\line(0,1){60}}}}
\put(0,0){\rotatebox{-30}{\textcolor{blue}{\line(1,0){60}}}}
\put(0,0){\textcolor{blue}{\line(0,-1){60}}}
\put(-52,-30){\rotatebox{-60}{\textcolor{blue}{\line(0,1){60}}}}
\put(-7,-80){(a)}
\end{picture}
\begin{picture}(140,0)(90,-70)
\put(0,59){\vector(0,1){1}}
\put(0,0){\vector(1,0){60}}
\put(0,0){\line(-1,0){60}}
\thicklines
\put(0,0){\rotatebox{30}{\textcolor{blue}{\line(1,0){60}}}}
\put(0,0){\textcolor{blue}{\line(0,1){60}}}
\put(-52,0){\rotatebox{60}{\textcolor{blue}{\line(0,1){60}}}}
\put(0,0){\rotatebox{-30}{\textcolor{blue}{\line(1,0){60}}}}
\put(0,0){\textcolor{blue}{\line(0,-1){60}}}
\put(-52,-30){\rotatebox{-60}{\textcolor{blue}{\line(0,1){60}}}}
\put(-7,-80){(b)}
\put(-55,-18){$D_4$}
\put(5,45){$D_2$}
\put(30,-40){$D_0$}
\end{picture}
\begin{picture}(0,0)(230,-70)
\put(-8,6){\line(1,0){9}}
\put(-15,10){\line(1,0){23}}
\put(-22,14){\line(1,0){37}}
\put(-29,18){\line(1,0){51}}
\put(-36,22){\line(1,0){65}}
\put(-43,26){\line(1,0){79}}
\put(-50,30){\line(1,0){93}}
\put(-57,34){\line(1,0){107}}
\put(-50,38){\line(1,0){93}}
\put(-43,42){\line(1,0){79}}
\put(-36,46){\line(1,0){65}}
\put(-29,50){\line(1,0){51}}
\put(-22,54){\line(1,0){37}}
\put(-15,58){\line(1,0){23}}
\thicklines
\put(-56,3){\rotatebox{60}{\line(0,1){60}}}
\put(-4,3){\rotatebox{30}{\line(1,0){60}}}
\end{picture}
\put(-236,70){\rotatebox{120}{
\begin{picture}(0,0)(0,0)
\put(-8,6){\line(1,0){9}}
\put(-15,10){\line(1,0){23}}
\put(-22,14){\line(1,0){37}}
\put(-29,18){\line(1,0){51}}
\put(-36,22){\line(1,0){65}}
\put(-43,26){\line(1,0){79}}
\put(-50,30){\line(1,0){93}}
\put(-57,34){\line(1,0){107}}
\put(-50,38){\line(1,0){93}}
\put(-43,42){\line(1,0){79}}
\put(-36,46){\line(1,0){65}}
\put(-29,50){\line(1,0){51}}
\put(-22,54){\line(1,0){37}}
\put(-15,58){\line(1,0){23}}
\thicklines
\put(-56,3){\rotatebox{60}{\line(0,1){60}}}
\put(-4,3){\rotatebox{30}{\line(1,0){60}}}
\end{picture}}}
\put(-235,70){\rotatebox{-120}{
\begin{picture}(0,0)(0,0)
\put(-8,6){\line(1,0){9}}
\put(-15,10){\line(1,0){23}}
\put(-22,14){\line(1,0){37}}
\put(-29,18){\line(1,0){51}}
\put(-36,22){\line(1,0){65}}
\put(-43,26){\line(1,0){79}}
\put(-50,30){\line(1,0){93}}
\put(-57,34){\line(1,0){107}}
\put(-50,38){\line(1,0){93}}
\put(-43,42){\line(1,0){79}}
\put(-36,46){\line(1,0){65}}
\put(-29,50){\line(1,0){51}}
\put(-22,54){\line(1,0){37}}
\put(-15,58){\line(1,0){23}}
\thicklines
\put(-56,3){\rotatebox{60}{\line(0,1){60}}}
\put(-4,3){\rotatebox{30}{\line(1,0){60}}}
\end{picture}}}
\begin{picture}(0,0)(90,-70)
\put(0,59){\vector(0,1){1}}
\put(0,0){\vector(1,0){60}}
\put(0,0){\line(-1,0){60}}
\thicklines
\put(0,0){\rotatebox{30}{\textcolor{blue}{\line(1,0){60}}}}
\put(0,0){\textcolor{blue}{\line(0,1){60}}}
\put(-52,0){\rotatebox{60}{\textcolor{blue}{\line(0,1){60}}}}
\put(0,0){\rotatebox{-30}{\textcolor{blue}{\line(1,0){60}}}}
\put(0,0){\textcolor{blue}{\line(0,-1){60}}}
\put(-52,-30){\rotatebox{-60}{\textcolor{blue}{\line(0,1){60}}}}
\put(-7,-80){(c)}
\put(40,13){$D_1$}
\put(-17,-53){$D_5$}
\put(-45,30){$D_3$}
\end{picture}
\put(-90,70){\rotatebox{60}{
\begin{picture}(0,0)(0,0)
\put(-8,6){\line(1,0){9}}
\put(-15,10){\line(1,0){23}}
\put(-22,14){\line(1,0){37}}
\put(-29,18){\line(1,0){51}}
\put(-36,22){\line(1,0){65}}
\put(-43,26){\line(1,0){79}}
\put(-50,30){\line(1,0){93}}
\put(-57,34){\line(1,0){107}}
\put(-50,38){\line(1,0){93}}
\put(-43,42){\line(1,0){79}}
\put(-36,46){\line(1,0){65}}
\put(-29,50){\line(1,0){51}}
\put(-22,54){\line(1,0){37}}
\put(-15,58){\line(1,0){23}}
\thicklines
\put(-56,3){\rotatebox{60}{\line(0,1){60}}}
\put(-4,3){\rotatebox{30}{\line(1,0){60}}}
\end{picture}}}
\put(-94,70){\rotatebox{180}{
\begin{picture}(0,0)(0,0)
\put(-8,6){\line(1,0){9}}
\put(-15,10){\line(1,0){23}}
\put(-22,14){\line(1,0){37}}
\put(-29,18){\line(1,0){51}}
\put(-36,22){\line(1,0){65}}
\put(-43,26){\line(1,0){79}}
\put(-50,30){\line(1,0){93}}
\put(-57,34){\line(1,0){107}}
\put(-50,38){\line(1,0){93}}
\put(-43,42){\line(1,0){79}}
\put(-36,46){\line(1,0){65}}
\put(-29,50){\line(1,0){51}}
\put(-22,54){\line(1,0){37}}
\put(-15,58){\line(1,0){23}}
\thicklines
\put(-56,3){\rotatebox{60}{\line(0,1){60}}}
\put(-4,3){\rotatebox{30}{\line(1,0){60}}}
\end{picture}}}
\put(-90,70){\rotatebox{-60}{
\begin{picture}(0,0)(0,0)
\put(-8,6){\line(1,0){9}}
\put(-15,10){\line(1,0){23}}
\put(-22,14){\line(1,0){37}}
\put(-29,18){\line(1,0){51}}
\put(-36,22){\line(1,0){65}}
\put(-43,26){\line(1,0){79}}
\put(-50,30){\line(1,0){93}}
\put(-57,34){\line(1,0){107}}
\put(-50,38){\line(1,0){93}}
\put(-43,42){\line(1,0){79}}
\put(-36,46){\line(1,0){65}}
\put(-29,50){\line(1,0){51}}
\put(-22,54){\line(1,0){37}}
\put(-15,58){\line(1,0){23}}
\thicklines
\put(-56,3){\rotatebox{60}{\line(0,1){60}}}
\put(-4,3){\rotatebox{30}{\line(1,0){60}}}
\end{picture}}}
\end{picture}
\end{center}
\caption{\footnotesize a) Stokes lines in the two-cut $(1,2)$ case. 
b) Stokes sectors of $D_0, D_2$ and $D_4$. 
c) Stokes sectors of $D_1, D_3$ and $D_5$. \label{TwoCutStokesLinesSectorsFigure}}
\end{figure}

In this case, there is only one kind of the Stokes lines ${\rm SL}_{1,2}$ 
which is given by \eq{StokesLinesDef} as 
\begin{align}
\zeta = |\zeta| e^{i \theta}:\qquad 
\theta = \frac{\pi}{6} + \frac{n\pi}{3}\qquad (n=0,1,\cdots,5). 
\end{align}
Therefore, Stokes sectors $D_n$ are given as 
\begin{align}
D_n = e^{n i \frac{\pi}{3}} D_0,\qquad (n=0,1,\cdots, 5),
\qquad D_0\equiv D\bigl(-\frac{\pi}{2}, \frac{\pi}{6}\bigr). 
\end{align}
This is shown in Fig.~\ref{TwoCutStokesLinesSectorsFigure}. The canonical solution on the Stokes sector $D_n$ 
is denoted by $\widetilde \Psi_n(t;\zeta)$. 
The Stokes matrices $S_n$ are now defined as 
\begin{align}
S_n \equiv \widetilde \Psi_n^{-1}(t;\zeta) \, \widetilde \Psi_{n+1}(t;\zeta), \qquad 
(n=0,1,\cdots,5),
\end{align}
and therefore components of the Stokes matrices are read as
\begin{align}
D_{2n} \cap D_{2n+1}:& 
\qquad S_{2n} = 
\begin{pmatrix}
1 & 0 \cr
s_{2n} & 1
\end{pmatrix};\qquad \Bigl(\bigl|e^{\varphi^{(1)}(t;\zeta)}\bigr|> \bigl|e^{\varphi^{(2)}(t;\zeta)}\bigr|,\quad \zeta \to \infty\Bigr), \nn\\
D_{2n+1} \cap D_{2n+2}:& \quad S_{2n+1} = 
\begin{pmatrix}
1 & s_{2n+1} \cr
0 & 1
\end{pmatrix};\qquad \Bigl(\bigl|e^{\varphi^{(1)}(t;\zeta)}\bigr|< \bigl|e^{\varphi^{(2)}(t;\zeta)}\bigr|,\quad \zeta \to \infty\Bigr). \label{StokesMatrix2Cut32}
\end{align}

\subsubsection{Three basic constraints on the Stokes multipliers}

The Stokes multipliers satisfy three constraints from the symmetry of the 
original ODE system. 

\paragraph{$\mathbb Z_2$ symmetry constraint}

This symmetry originates from the $\mathbb Z_2$ symmetry of the matrix model. 
That is, this is the reflection symmetry $M \to -M$ of the one-matrix models: 
\begin{align}
\mathcal Z=\int dM e^{-N \tr V(M)},\qquad V(-M) = V(M). 
\end{align}
In terms of the ODE system, this symmetry is expressed by the reflection of 
$\zeta \to -\zeta$: 
\begin{align}
g_{\rm str} \frac{\del \widetilde \Psi(t;-\zeta)}{\del\zeta} 
&= \bigl[- \widetilde {\mathcal Q}(t;-\zeta)\bigr]\, \widetilde \Psi(t;-\zeta) 
= \bigl[\sigma_1 \widetilde {\mathcal Q}(t;\zeta)\sigma_1\bigr] \, \widetilde \Psi(t;-\zeta), \nn\\
g_{\rm str} \frac{\del \widetilde \Psi(t;-\zeta)}{\del t} 
&= \bigl[ \widetilde {\mathcal P}(t;-\zeta)\bigr]\, \widetilde \Psi(t;-\zeta) 
= \bigl[\sigma_1 \widetilde {\mathcal P}(t;\zeta)\sigma_1\bigr] \, \widetilde \Psi(t;-\zeta).
\end{align}
Therefore, each canonical solution is mapped to another canonical solution as:
\begin{align}
\sigma_1 \widetilde \Psi_n(t;-\zeta) \sigma_1 = \widetilde \Psi_{n+3}(t;\zeta),\qquad 
(n=0,1,\cdots,5),
\end{align}
and the Stokes matrices are mapped as 
\begin{align}
S_{n+3} = \sigma_1 S_n \sigma_1,\qquad 
s_{n+3}=s_n, \qquad (n=0,1,\cdots, 5). 
\end{align}
Consequently there are only three independent Stokes multipliers,
\begin{align}
s_0 =s_3\equiv \alpha,
\qquad s_1=s_4 \equiv \beta, \qquad s_2=s_5=\gamma. \label{DefAlfBetGam}
\end{align}

\paragraph{Hermiticity constraint}
This originates from Hermiticity of the matrix models. In the two-cut cases, 
they are studied in \cite{HMPN,fi1}. This symmetry is expressed as%
\footnote{Note that we use the following convention of complex conjugation in this paper: 
$[f(\zeta)]^*=f^*(\zeta^*)=\sum_n f_n^* \zeta^*$, with a function 
$f(\zeta)\equiv \sum_n f_n \zeta^n$.  }
\begin{align}
\widetilde {\mathcal Q}^*(t;\zeta^*)
= \widetilde {\mathcal Q}(t;\zeta^*), \qquad 
\widetilde {\mathcal P}^*(t;\zeta^*) 
= \widetilde {\mathcal P}(t;\zeta^*). 
\end{align}
Therefore, each canonical solution is mapped to another canonical solution as:
\begin{align}
\widetilde \Psi_n^*(t;\zeta^*) = \widetilde \Psi_{7-n}(t;\zeta),\qquad 
(n=0,1,\cdots,5),
\end{align}
and the Stokes matrices are mapped as 
\begin{align}
S_{n}^* = S_{6-n}^{-1},\qquad 
s_n^*+s_{6-n}=0, \qquad (n=0,1,\cdots, 5). 
\end{align}
This reduces three independent Stokes multipliers $\alpha,\beta$ and $\gamma$ 
to be two real parameters: 
\begin{align}
\alpha^*+\alpha=0,\qquad \beta^*+\gamma=0. \label{HermitCond2Cut}
\end{align}

\paragraph{Monodromy free constraint}
The last constraint is the requirement that the solutions to the ODE system 
are single-valued functions. Note that the presence of non-trivial monodromy 
in the context of matrix models corresponds to introducing background RR flux 
and/or D0-branes in 0A string background. That is, 
the system becomes like the complex matrix models \cite{ComplexMatrixCite1,ComplexMatrixCite2,UniCom}. 
This constraint for the single-valued solutions is expressed as 
\begin{align}
\widetilde\Psi_{n}(t;\zeta) = \widetilde\Psi_n(t;e^{2\pi i }\zeta)=\widetilde\Psi_{n+6}(t;\zeta),\label{MonodromyFreeTwoCutInTermsOfPsi}
\end{align}
therefore 
\begin{align}
S_0\, S_1\, S_2\, S_3\, S_4 \,S_5=I_2, 
\end{align}
which results in 
\begin{align}
s_0+s_1+s_2 + s_0 s_1 s_3=\alpha(1-|\beta|^2)+\beta-\beta^*=0. \label{AlgCurveStokes2cut}
\end{align}
Taking all constraints Eqs.~\eq{DefAlfBetGam}, \eq{HermitCond2Cut}, \eq{AlgCurveStokes2cut} into consideration, 
we find that the Stokes multipliers have two real degrees of freedom, say $\beta$. 
Since the Painlev\'e equation II equation, Eq.~\eq{PainleveIIEquation}, is the second order ODE system, 
These two parameters are the non-perturbative ambiguity of the system. 

As is mentioned in Introduction, among these Stokes multipliers satisfying 
the algebraic relation \eq{AlgCurveStokes2cut}, there is a special value 
which realizes the perturbative behavior (in $t \to \pm \infty$) 
of the matrix models argued from the physical point of views \cite{UniCom}. 
This special value is given as 
\begin{align}
\alpha = 0,\qquad \beta = \pm 1, \label{HMLTwoCutSecAAAAAAIIIII}
\end{align} 
and corresponds to the Hastings-McLeod solution 
in the Painlev\'e II equation \cite{HastingsMcLeod}.
From the mathematical point of view, this solution also has a good analytic behavior 
along the real isomonodromy parameter (cosmological constant) $t$ 
\cite{HastingsMcLeod,CK}. 
From this two-cut example, we generally expect that there is a special class 
of solutions of the Stokes multipliers 
which corresponds to the physical D-instantion chemical potentials.
To generalize the solutions to the cases of arbitrary number of cuts, 
it is natural to ask the following question: 
{\em what is the physical requirements which specify the above multipliers?}
This is also related to the issue cited by \cite{HHIKKMT,KKM}:
{\em What is the boundary condition in continuum formulations which can fix 
the D-instanton chemical potentials in the matrix models?}
Our procedure (discussed in Section \ref{MultiCutBCSection} and Section \ref{RHapproachSection}) 
gives an answer to the question. 
In Section \ref{TwoCutCaseMultiCutBClaskeie} and then in Section \ref{TwoCutSectionSmallInstantonCond}, 
we will see that our physical requirements correctly choose
this particular parametrization Eq.~\eq{HMLTwoCutSecAAAAAAIIIII} of the Stokes multipliers. 

\section{Stokes phenomena in the multi-cut cases \label{MultiCutStokesSection}}

In this section, we develop general framework for Stokes phenomena 
in the general multi-cut critical points, and show explicitly {\em how the actual systems can be controlled}. 
Key information is provided by {\em profile of dominant exponents} 
(Theorem \ref{TheoremComponentsOfProfiles}), and with this terminology 
we propose a systematic way to read the non-trivial Stokes multipliers 
(Theorem \ref{TheoremIndexStokesMultipliers}). 
Since the following discussions are 
{\em valid in general $k\times k$ ODE systems} of the following type: 
\begin{align}
\frac{d \Psi(t;\zeta)}{d\zeta} = \bigl(\Gamma^{-\gamma} \zeta^{r-1} + \cdots\bigl)\Psi(t;\zeta),\qquad \text{g.c.d.}\,(k,\gamma)=1, 
\end{align}
we here develop the general framework without restricting to the $\mathbb Z_k$ 
symmetry ($\gamma=r$). 
The restriction to the $\mathbb Z_k$ symmetric cases only appear 
in Section \ref{ThreeConstraints11}. 

\subsection{Stokes lines and Stokes sectors}

First we focus on the Stokes lines,
\begin{align}
{\rm SL}_{j,l}:\qquad 
{\rm Re} 
\bigl[\bigl(\varphi_{-r}^{(j)}-\varphi_{-r}^{(l)}\bigr)\zeta^r\bigr]=0,
\end{align}
and the resulting Stokes sectors \eq{DefStokesSector}. 
The leading coefficient of the exponents, 
$\varphi_{-r}^{(j)}$, which we consider here is given as%
\footnote{Note that the cases of our interest in the later sections are 
the $\mathbb Z_k$-symmetric critical points, 
and as one can see in Appendix \ref{AppendixLaxOP}, 
the cases are given by $\gamma=r$. Also for future reference, 
we note that the fractional-superstring cases are given by $\gamma=r-2$. 
}
\begin{align}
\varphi_{-r}^{(j)} = \omega^{-\gamma (j-1)}. 
\end{align}
Consequently, the conditions on the Stokes lines 
(in terms of angle, $\zeta =|\zeta|e^{i\theta}$) are expressed as 
\begin{align}
{\rm Re} 
\bigl[\bigl(\varphi_{-r}^{(j)}-\varphi_{-r}^{(l)}\bigr)e^{ir\theta}\bigr]
=2\sin \bigl(r\theta - \pi \frac{\gamma (j+l-2)}{k}\bigr) 
\sin\bigl( \pi \frac{\gamma (j-l)}{k}\bigr). \label{PreStokesLine342}
\end{align}
First of all, if there is a pair of $(j,l)$ such that 
\begin{align}
\gamma (j-l)\in k \mathbb Z, \label{NonDegenerateExponents}
\end{align}
then the condition \eq{CoefOfLaxLeading} does not satisfy. 
This means that the highest exponents degenerate 
$(\varphi^{(j)}_{-r}-\varphi^{(l)}_{-r})\,\zeta^r=0$. 
In this case, we consider the next leading Stokes lines,
\begin{align}
{\rm Re}
\bigl[\bigl(\varphi_{-r+1}^{(j)} -\varphi_{-r+1}^{(l)}\bigr) \zeta^{r-1}\bigr] 
=0, 
\end{align}
or more generally we consider the following Stokes lines:%
\footnote{The physical interpretation 
of these general Stokes lines is the positions of eigenvalues in the matrix models. 
This viewpoint is also essential in this paper and discussed in Section 
\ref{MCBCSUBSection}. }
\begin{Definition}
[General Stokes lines]
The general Stokes lines ${\rm GSL}_{j,l}$ 
in this ODE system are defined for each pair of $(j,l)$ as 
\begin{align}
{\rm GSL}_{j,l}&\equiv 
\Bigl\{\zeta \in \mathbb C;
{\rm Re}\bigl[\varphi^{(j)}(t;\zeta)-\varphi^{(l)}(t;\zeta)\bigr] = 0 \Bigr\}
= \bigcup_{n=0}^{2r-1} {\rm GSL}_{j,l}^{(n)}, 
\end{align}
which consists of $2r$ semi-infinite lines, ${\rm GSL}_{j,l}^{(n)}$ $(n=0,1,\cdots,2r-1)$. 
The set of lines, ${\rm GSL}$, denotes a set of whole (general) Stokes lines, 
${\rm GSL} \equiv \bigcup_{j,l} {\rm GSL}_{j,l}$. 
\label{GeneralStokesLinesDef}
\end{Definition}
The situations \eq{NonDegenerateExponents} are also interesting critical points 
in the multi-cut matrix models, however here for sake of simplicity, we concentrate on the following cases, 
\begin{align}
\text{g.c.d.$\bigl(k,\gamma \bigr)=1$}, \label{NonDegenerateConditionCoprimeKR}
\end{align}
because Eq.~\eq{NonDegenerateExponents} becomes trivial in this case:
\begin{align}
\gamma (j-l)\in k \mathbb Z\qquad \Leftrightarrow \qquad j-l \in k \mathbb Z. 
\end{align}
Therefore Eq.~\eq{PreStokesLine342} gives 
the angle $\theta_{j,l}^{(n)}$ for the Stokes lines ${\rm SL}_{j,l}$ as
\begin{align}
\theta =\theta_{j,l}^{(n)} 
= \frac{kn+\gamma (j+l-2)}{rk} \pi,\qquad n\in \mathbb Z. \label{StokesLineTheta}
\end{align}
From this formula, one can read several basic information about the Stokes lines. 
An example of Stokes lines ($3$-cut $(1,1)$ case) is shown in Fig.~\ref{AngularDomainFigure}-b. 
For later convenience, we introduce the following terminology:
\begin{Definition}
[Segments] 
Angular domains in between two Stokes lines which do not include any Stokes lines 
are called segments. 
\end{Definition}
In our present cases with a coprime $(k,\gamma)$, there are $2rk$ distinct 
segments $\delta D_n$ ($n=0,1,\cdots,2rk-1$) given as 
\begin{align}
\delta D_n \equiv D\bigl(n \delta \theta -\delta\theta, n\delta \theta\bigr),
\qquad  \Bigl(n=0,1,\cdots,2rk-1;\,\delta \theta =\frac{\pi}{rk}\Bigr), \label{DefOfSegment}
\end{align}
which can fill the complex plane $\mathbb C$,
\begin{align}
\bigcup_{n=0}^{2rk-1} \, \overline{\delta D_n} = \mathbb C,\qquad \delta D_m \cap \delta D_{m'} = \emptyset\qquad \bigl(m\neq m'\bigr). 
\end{align}
According to the definition of Stokes sectors, Eq.~\eq{DefStokesSector}, 
we define the following most basic Stokes sectors, $D_n$: 
\begin{Definition} 
[Fine Stokes sectors/matrices]
The following angular domains $D_n$ 
\begin{align}
D_n = e^{n i \delta \theta} D_0,\qquad 
D_0 =D\bigl(-\delta \theta, k \delta \theta \bigr),\qquad (n=0,1,\cdots, 2rk-1),
\end{align}
are Stokes sectors of a coprime $(k,r)$ system with $k\geq 3$, 
which are referred to as fine Stokes sectors. 
The canonical solution of the fine Stokes sector $D_n$ is denoted as 
$\widetilde \Psi_n(t;\zeta)$ and 
the corresponding Stokes matrices $S_n$ are given as 
\begin{align}
\widetilde \Psi_{n+1}(t;\zeta) = \widetilde \Psi_n(t;\zeta)\, S_n, 
\label{StokesPhenoMultiCut111}
\end{align}
which is referred to as (fine) Stokes matrices. 
\end{Definition} 
Here we also define the other two kinds of Stokes sectors/matrices: 
First we define Stokes sectors/matrices which respect to the $\mathbb Z_k$ symmetry 
of the multi-cut matrix models:
\begin{Definition}
[Symmetric Stokes sectors/matrices]
The following subset of the fine Stokes sectors, 
\begin{align}
D_{2 nr}, \qquad (n=0,1,\cdots, k-1), \label{MultiCutDefinitionOfSymStokesSectors}
\end{align}
are referred to as symmetric Stokes sectors,%
\footnote{Note that this definition is not enough for the $k=3, r=2$ case. In these cases, 
we employ a modified version of the Stokes sectors, for example, $D_{n r}$. }
and the corresponding Stokes matrices $S_{2 r n}^{(\rm sym)}$ 
\begin{align}
S_{2 rn}^{(\rm sym)} \equiv \widetilde\Psi_{2rn}^{-1}(t;\zeta)\, \widetilde\Psi_{2r(n+1)}(t;\zeta) 
= S_{2rn}\cdot S_{2rn+1}\cdots S_{2r(n+1)-1}. \label{MultiSymStokes}
\end{align}
are referred to as symmetric Stokes matrices. \label{SymStokeSectorsDefinition}
\end{Definition}
Next we define the following economical Stokes sectors/matrices: 
\begin{Definition}
[Coarse Stokes sectors/matrices]
The following subset of the fine Stokes sectors, 
\begin{align}
D_{nk}, \qquad (n=0,1,\cdots, 2r-1), 
\end{align}
are referred to as coarse Stokes sectors, and 
the corresponding Stokes matrices $S_{nk}^{(\rm c)}$ are written as 
\begin{align}
S_{nk}^{(\rm coa)} \equiv \widetilde\Psi_{nk}^{-1}(t;\zeta)\, \widetilde\Psi_{(n+1)k}(t;\zeta) 
= S_{nk}\cdot S_{nk+1}\cdots S_{(n+1)k-1}. \label{MultiCoarseStokes}
\end{align}
are referred to as coarse Stokes matrices. 
\end{Definition}
Coarse Stokes sectors are most often used in the literature. 
However, in the following discussions, one will see that the fine Stokes matrices are 
more convenient for our calculations. 

\subsection{Stokes multipliers from the profile of dominant exponents\label{SectionProfileDom}}

In principle, 
one can use Theorem \ref{TheoremStokesMultipliers} to read 
the non-trivial (or non-zero) Stokes multipliers in each specific case. 
That is the problem of finding which components can take non-zero value 
in Stokes matrices. 
However, practically in general, it is tedious to use this standard way 
to read the non-trivial multipliers, especially in the higher 
$k\times k$ system with higher Poincar\'e index $r$. 
The purpose of this section is therefore 
to point out an interesting connection between 
the non-trivial Stokes multipliers and {\em profile of dominant exponents} 
which we develop in this subsection (Theorem \ref{TheoremComponentsOfProfiles} 
and \ref{TheoremIndexStokesMultipliers}). 
An important thing in this procedure is that these results 
make it easy to put data of the Stokes multipliers in computer, for example, 
in Mathematica program. 

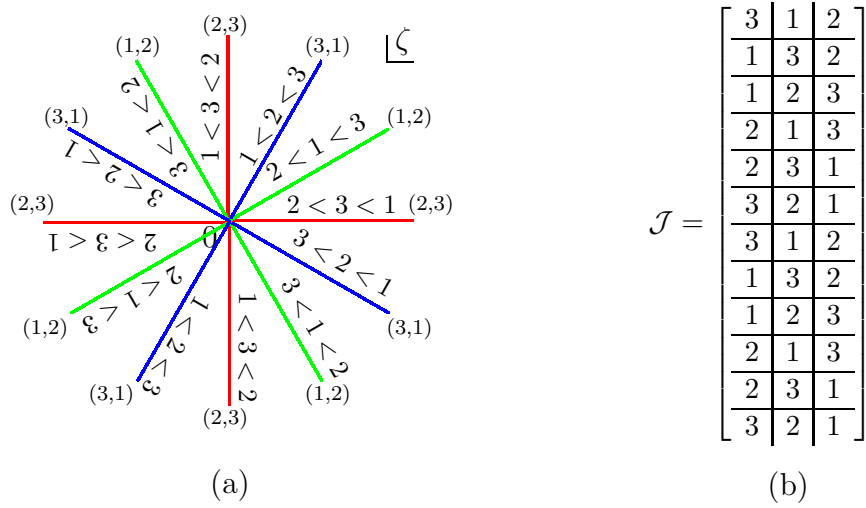
\begin{figure}[htbp]
\begin{center}
\begin{picture}(180,200)(0,0)
\end{picture}
\begin{picture}(5,0)(190,-89)
\put(70,70){\line(0,1){10}\line(1,0){10}}
\put(73,75){$\zeta$}
\put(0,0){$0$}
{
\thicklines
\put(10,10){\rotatebox{-90}{\textcolor{red}{\line(1,0){70}}}}
\put(10,10){\rotatebox{-60}{\textcolor{green}{\line(1,0){70}}}}
\put(10,10){\rotatebox{-30}{\textcolor{blue}{\line(1,0){70}}}}
\put(10,10){\rotatebox{0}{\textcolor{red}{\line(1,0){70}}}}
\put(9.5,10){\rotatebox{30}{\textcolor{green}{\line(1,0){70}}}}
\put(9.5,10){\rotatebox{60}{\textcolor{blue}{\line(1,0){70}}}}
\put(9.5,10){\rotatebox{90}{\textcolor{red}{\line(1,0){70}}}}
\put(-25.5,10){\rotatebox{120}{\textcolor{green}{\line(1,0){70}}}}
\put(-51,10){\rotatebox{150}{\textcolor{blue}{\line(1,0){70}}}}
\put(-60,9.5){\rotatebox{180}{\textcolor{red}{\line(1,0){70}}}}
\put(-50,10){\rotatebox{-150}{\textcolor{green}{\line(1,0){70}}}}
\put(-25,10){\rotatebox{-120}{\textcolor{blue}{\line(1,0){70}}}}
}
\put(58,47){\quad\scriptsize (1,2)}
\put(70,14){\,\, \scriptsize (2,3)}
\put(33,74){\,\,\,\scriptsize (3,1)}
\put(5,0){\rotatebox{0}{\footnotesize\put(27,13){$2<3<1$}}}
\put(0,0){\rotatebox{30}{\footnotesize\put(27,13){$2<1<3$}}}
\put(0,0){\rotatebox{60}{\footnotesize\put(27,13){$1<2<3$}}}
\put(0,5){\rotatebox{90}{\footnotesize\put(27,13){$1<3<2$}}}
\put(3,12){\rotatebox{120}{\footnotesize\put(27,13){$3<1<2$}}}
\put(6,18){\rotatebox{150}{\footnotesize\put(27,13){$3<2<1$}}}
\put(10,18){\rotatebox{180}{\footnotesize\put(27,13){$2<3<1$}}}
\put(7,18){\rotatebox{210}{\footnotesize\put(27,13){$2<1<3$}}}
\put(0,12){\rotatebox{240}{\footnotesize\put(27,13){$1<2<3$}}}
\put(1,10){\rotatebox{270}{\footnotesize\put(27,13){$1<3<2$}}}
\put(5,3){\rotatebox{300}{\footnotesize\put(27,13){$3<1<2$}}}
\put(4,4){\rotatebox{330}{\footnotesize\put(27,13){$3<2<1$}}}
\put(-33,74){\scriptsize (1,2)}
\put(-73,14){\scriptsize (2,3)}
\put(-60,47){\scriptsize (3,1)}
\put(33,-58){\,\,\,\scriptsize (1,2)}
\put(0,-67){\scriptsize (2,3)}
\put(58,-33){\quad \scriptsize (3,1)}
\put(-68,-33){\scriptsize (1,2)}
\put(0,82){\scriptsize (2,3)}
\put(-43,-58){\scriptsize (3,1)}
\put(3,-93){(a)}
\end{picture}
\begin{picture}(0,0)(30,-95)
\put(0,0){\small$\mathcal J = \left[
\begin{array}{c|c|c}
3 & 1 & 2 \cr
\hline
1 & 3 &2 \cr
\hline
1 & 2 & 3 \cr
\hline
2 & 1 & 3 \cr
\hline
2& 3 & 1 \cr
\hline
3 & 2 & 1 \cr
\hline
3 & 1 & 2 \cr
\hline
1& 3 & 2 \cr
\hline
1 & 2 & 3 \cr
\hline
 2 & 1 & 3 \cr
\hline
2& 3&1 \cr
\hline
3 & 2 & 1 \cr
\end{array}
\right]$}
\put(45,-100){(b)}
\end{picture}
\end{center}
 \caption{\footnotesize 
The two expressions for the profile of dominant exponents 
in the $3$-cut $(1,1)$ critical point. 
${\rm Re}[\varphi_{-2}^{(j_{l,1})}]
<{\rm Re}[\varphi_{-2}^{(j_{l,2})}]
<{\rm Re}[\varphi_{-2}^{(j_{l,3})}]$ is expressed as
$j_{l,1}<j_{l,2}<j_{l,3}$. 
a) The profile in 
the  $\zeta$ plane. b) The profile $\mathcal J$ in the table. 
In the same way, the dominance is expressed as 
$[j_{l,1}|j_{l,2}|j_{l,3}]$
\label{DominanceProfileAndZetaPlaneFigure}}
\end{figure}

Since there is no Stokes line in the segments defined in Eq.~\eq{DefOfSegment}, 
one can define the following ordered set $J_l$ of indices $j_{l,i}$: 
\begin{align}
J_l = 
\left[
\begin{array}{c|c|c|c}
j_{l,1} &j_{l,2} & \cdots & j_{l,k}
\end{array}
\right] \in \mathbb N^k,
\end{align}
which describes the profile of dominant exponents in the segment $D_l$
\begin{align}
{\rm Re}\bigl[
\varphi_{-r}^{(j_{l,1})} \zeta^r\bigr]<
{\rm Re}\bigl[\varphi_{-r}^{(j_{l,2})} \zeta^r \bigr]<\cdots<
{\rm Re}\bigl[
\varphi_{-r}^{(j_{l,k})} \zeta^r\bigr],
\qquad \zeta \in \delta D_l.
\end{align}
This sequence of numbers, $\mathcal J=\{J_l\}_{l=0}^{2rk}$, is referred to as 
{\em profile of dominant exponents}. Here we express the profile $\mathcal J$ as follows:
\begin{align}
\mathcal J= 
\left[
\begin{array}{c|c|c|c}
j_{2rk-1,1} & j_{2rk-1,2} & \cdots & j_{2rk-1,k} \cr
\hline
\vdots & \vdots & & \vdots \cr
\hline
j_{1,1} & j_{1,2} & \cdots & j_{1,k} \cr
\hline
j_{0,1} & j_{0,2} & \cdots & j_{0,k}
\end{array}
\right]
\end{align}
Note that the ordering of indices in the vertical direction is different from the usual matrix, 
and that elements are periodic in the index $l$, $J_l=J_{l+2rk}$. 
An example ($3$-cut $(1,1)$ critical point) and the relation to the $\zeta$ plane are shown in Fig.~\ref{DominanceProfileAndZetaPlaneFigure}. 

The non-trivial problem for the profile is then how to fill the numbers in the profiles. 
We found the following simple answer: 
\begin{Theorem}
[General components] 
The general components $j_{l,n}$ of the profile $\mathcal J$ 
with $\text{g.c.d.}\,(k,\gamma)=1$ are given as%
\footnote{In this paper, we use the floor-function notation for the gauss symbol, $\lfloor a\rfloor$, 
which means the largest integer less than or equal to $a$.}
\begin{align}
j_{l,n} \equiv 1 + \biggl(\Bigl\lfloor \frac{l}{2}\Bigr\rfloor 
+ (-1)^{k+l+n} \Bigl\lfloor \frac{k-n+1}{2}\Bigr\rfloor\biggr) m_1,\quad 
\mod k,
\end{align}
where $m_1$ is obtained by the Euclidean algorithm 
of $k n_1 + \gamma m_1 = 1$. 
\label{TheoremComponentsOfProfiles}
\end{Theorem}
Some comments on this formula are in order:
\begin{itemize}
\item In any segment profile $J_l =\left[
\begin{array}{c|c|c|c}
j_{l,1} &j_{l,2} & \cdots & j_{l,k}
\end{array}
\right]$, 
a pair of indices $(i,j)$ which 
change their relative dominance at angle $\theta=l\delta\theta$ 
are always next to each other, 
and they satisfy the following sum rule:%
\footnote{Note that the condition 
for $\theta_{i,j}^{(n)}=l\delta \theta$ is given as 
\begin{align}
\frac{kn + \gamma(i+j-2)}{rk} \pi = \frac{l}{rk}\pi \quad
\Leftrightarrow \quad
kn + \gamma(i+j-2) = l. 
\end{align}
This means that, for a given $l$, find a pair $(i,j)$ such that there exists an 
integer $n$. 
}
\begin{align}
i+j-2 \equiv  m_l \,(\equiv l m_1) \qquad \mod k,  \label{SumRuleStokesLines}
\end{align}
with an integer 
$m_l$ which is obtained by the Euclidean algorithm 
of $k n_l + \gamma m_l = l$. 
Therefore in particular, 
we represent these pairs as $(i|j)$ in the profile 
(See Eqs.~\eq{ExampleOfProfileFull11}). 
\item Theorem \ref{TheoremComponentsOfProfiles} can be recursively shown 
by using the sum rules Eq.~\eq{SumRuleStokesLines}, and its initial conditions $j_{0,k}=j_{1,k}=1=j_{k,1}=j_{k+1,1}$ which can be easily checked. 
\item The trajectories of the indices, for instance $a$ and $b$, are given as follows:
\begin{align}
&\underline{\text{$k$ is odd}} &&\underline{\text{$k$ is even}} \nn\\
&\mathcal J= 
\left[
\begin{array}{c|c|c|c|c|c|c}
\vdots & \vdots &\vdots & & \vdots & \vdots & \vdots  \cr
\hline
 & a & &  \cdots &b & &  \cr
\hline
a &  & & \cdots &&b & \cr
\hline
a &  & &\cdots && & b  \cr
\hline
 & a &  &   \cdots && & b   \cr
\hline
 &  & a &   \cdots &&b &    \cr
\hline
\vdots &\vdots & \vdots & & \vdots & \vdots & \vdots 
\end{array}
\right],
&&\mathcal J= 
\left[
\begin{array}{c|c|c|c|c|c|c}
\vdots & \vdots &\vdots & & \vdots & \vdots & \vdots  \cr
\hline
 & a & &  \cdots & & b&  \cr
\hline
a &  & & \cdots && &b \cr
\hline
a &  & &\cdots && & b  \cr
\hline
 & a &  &   \cdots && b&    \cr
\hline
 &  & a &   \cdots &b& &    \cr
\hline
\vdots &\vdots & \vdots & & \vdots & \vdots & \vdots 
\end{array}
\right] 
\end{align}
\item The above formula is not applicable in the case of $\text{g.c.d }(k,\gamma)\neq 1$, since some exponents degenerate, but one example 
of this kind is shown in Appendix \ref{CalculationInConcreteSystems2}. 
\end{itemize}
Here also two examples of the profiles $\mathcal J_{k,r}$ are shown 
for the case of $(k,r)=(3,2)$ and $(5,2)$ with $\gamma=r$ 
($\mathbb Z_k$ symmetry condition):
\begin{align}
\mathcal J_{3,2}= 
\left[
\begin{array}{c|c|c}
3 & (1 & 2) \cr
\hline
(1 & 3) &2 \cr
\hline
1 & (2 & 3) \cr
\hline
(2 & 1) & 3 \cr
\hline
2& (3 & 1) \cr
\hline
(3 & 2) & 1 \cr
\hline
3 & (1 & 2) \cr
\hline
(1& 3) & 2 \cr
\hline
1 & (2 & 3) \cr
\hline
 (2 & 1) & 3 \cr
\hline
2& (3&1) \cr
\hline
(3 & 2) & 1 \cr
\end{array}
\right],
\begin{picture}(0,0)(0,0)
\put(0,82){\mbox{$\leftarrow J_{11}$}}
\end{picture}
\begin{picture}(20,0)(0,0)
\put(0,-82){\mbox{$\leftarrow J_0$}}
\end{picture}\qquad 
\mathcal J_{5,2}= 
\left[
\begin{array}{c|c|c|c|c}
2 & (4 & 5) & (1 & 3) \cr
\hline
(4 & 2) & (1 & 5) & 3 \cr
\hline
4 & (1 & 2) & (3 & 5) \cr
\hline
(1 & 4) & (3 & 2) & 5 \cr
\hline
1 & (3 & 4) & (5 & 2) \cr
\hline
(3 & 1)& (5 & 4) & 2 \cr
\hline
3 & (5 & 1) & (2 & 4)\cr
\hline
(5 & 3)& (2 & 1) & 4 \cr
\hline
5 & (2 & 3) & (4 & 1) \cr
\hline
(2 & 5) & (4 & 3) & 1 \cr
\hline
2 & (4 & 5) & (1 & 3) \cr
\hline
(4 & 2) & (1 & 5) & 3 \cr
\hline
4 & (1 & 2) & (3 & 5) \cr
\hline
(1 & 4) & (3 & 2) & 5 \cr
\hline
1 & (3 & 4) & (5 & 2) \cr
\hline
(3 & 1)& (5 & 4) & 2 \cr
\hline
3 & (5 & 1) & (2 & 4)\cr
\hline
(5 & 3)& (2 & 1) & 4 \cr
\hline
5 & (2 & 3)& (4&1) \cr
\hline
(2 & 5)&(4 & 3) & 1 \cr
\end{array}
\right].
\begin{picture}(0,0)(0,0)
\put(0,140){\mbox{$\leftarrow J_{19}$}}
\end{picture}
\begin{picture}(20,0)(0,0)
\put(0,-140){\mbox{$\leftarrow J_0$}}
\end{picture}\label{ExampleOfProfileFull11}
\end{align}
One can observe that there is a $2k$ periodicity, $ j_{l+2k,n} = j_{l,n}$, or 
more precisely, 
a reflection by step $k$, $j_{l,n} = j_{l+k,k-n+1}$. 

Next we demonstrate how to read the non-trivial Stokes multipliers 
in some examples, and see the general rule. In the case of 
$(r,k;\gamma)=(2,5;2)$ 
and its symmetric Stokes matrix $S_0^{(\rm sym)}$, 
one first sees the dominance profile in the domain $D_0\cap D_{4}$, 
\begin{align}
D_0\cap D_{4} \supset
\left[
\begin{array}{c|c|c|c|c}
1 & 3 & 4 & 5 &  2 \cr
\hline
3 & 1 & 5 & 4 &  2
\end{array}
\right],
\begin{picture}(0,0)(0,0)
\put(0,8){\mbox{$\leftarrow J_{5}$}}
\end{picture}
\begin{picture}(20,0)(0,0)
\put(0,-8){\mbox{$\leftarrow J_4$}}
\end{picture}
\end{align}
and reads the ordering of magnitude: 
\begin{align}
(2) > (5), (4),(3),(1),\qquad (5)>(3),(1),\qquad (4)>(3),(1). 
\end{align}
This results in the following symmetric Stokes multipliers:
\begin{align}
S_0^{(\rm sym)} =
\begin{pmatrix}
1 & s_{0,1,2}^{(\rm sym)} & 0 & s_{0,1,4}^{(\rm sym)} & s_{0,1,5}^{(\rm sym)} \cr
0 & 1 & 0 & 0 & 0 \cr
0 & s_{0,3,2}^{(\rm sym)} & 1 & s_{0,3,4}^{(\rm sym)} & s_{0,3,5}^{(\rm sym)} \cr
0 & s_{0,4,2}^{(\rm sym)} & 0 & 1 & 0  \cr
0 & s_{0,5,2}^{(\rm sym)} & 0 & 0 & 1
\end{pmatrix}. 
\end{align}
In the same way, for the calculation of 
the fine Stokes matrix $S_0$, one first sees the dominance profile 
in the domain $D_0\cap D_1$, 
\begin{align}
D_0\cap D_1 \supset
\left[
\begin{array}{c|c|c|c|c}
1 & \bf 3 & \bf 4 & \underbar {\it 5} & \underbar {\it 2} \cr
\hline
\bf 3 & 1 & \underbar {\it 5} & \bf 4 & \underbar {\it 2} \cr
\hline
\bf 3 & \underbar {\it 5} & 1 & \underbar {\it 2} & \bf 4\cr
\hline
\underbar {\it 5} & \bf 3& \underbar {\it 2} & 1 & \bf 4 \cr
\hline
\underbar {\it 5} & \underbar  {\it 2} & \bf 3& \bf 4&1 \cr
\end{array}
\right],
\begin{picture}(0,0)(0,0)
\put(0,30){\mbox{$\leftarrow J_{5}$}}
\end{picture}
\begin{picture}(20,0)(0,0)
\put(0,-30){\mbox{$\leftarrow J_1$}}
\end{picture} \label{J25FineStokesProfile}
\end{align}
and reads the ordering of magnitude: 
\begin{align}
(4) > (3),\qquad (2)>(5). 
\end{align}
This results in the Stokes multipliers:
\begin{align}
S_0 =
\begin{pmatrix}
1 & 0 & 0 & 0 & 0 \cr
0 & 1 & 0 & 0 & 0 \cr
0 & 0 & 1 & s_{0,3,4} & 0 \cr
0 & 0 & 0 & 1 & 0 \cr
0 & s_{0,5,2} & 0 & 0 & 1
\end{pmatrix}. 
\end{align}
These are the standard way of reading the multipliers.%
\footnote{
From this procedure, one may notice that the simplest choice is the coarse Stokes sectors $S_{nk}^{(\rm coa)}$, 
because intersections have the definite order of magnitude: 
$D_0\cap D_5 \supset \left[
\begin{array}{c|c|c|c|c}
1 & 3& 4 & 5 &  2 \cr
\end{array}
\right]$ and the number of Stokes matrices is the smallest. 
This is the main reason why the coarse Stokes sectors 
are often used in literature. However, 
we will see that the coarse Stokes multipliers are not suitable for general 
formula of higher $k$ and $r$ at least 
in the $\mathbb Z_k$ symmetric critical points. }

However, one may notice that there is a relation between indices of 
non-zero Stokes multipliers $s_{0,i,j}$ in the Stokes matrix $S_0$ 
and the dominance-changing pairs $(j|i)$ in the profile $J_0$:
\begin{align}
s_{0,3,4},\quad s_{0,5,2} \qquad \leftrightarrow \qquad 
(2|5),\quad (4|3)\quad \in \quad
J_0 = 
\left[
\begin{array}{c|c|c|c|c}
(2 & 5) & (4 & 3) &  1 \cr
\end{array}
\right]. 
\end{align}
We claim that this observation is generally true: 
\begin{Theorem}
[Stokes multipliers from the profiles]
The non-zero Stokes multipliers in the fine Stokes matrix $S_l$ 
have a correspondence with dominance-changing pairs $(j|i)$ 
in the profile $J_l$ as follows:
\begin{align}
s_{l,i,j} \,\, (i\neq j) \text{ can take non-zero value}\quad \Leftrightarrow\quad  (j|i) \in J_l. 
\end{align}
Note that the orderings of indices $(i|j)$ and $s_{l,j,i}$ are opposite $i\leftrightarrow j$. 
\label{TheoremIndexStokesMultipliers}
\end{Theorem}
A proof is easy if one notices that intersections of fine Stokes sectors 
$D_n \cap D_{n+1}$ are always a half of the period of Stokes line formula 
Eq.~\eq{PreStokesLine342}. 
The other Stokes matrices, say $S_n^{(\rm sym)}$ and $S_n^{(\rm coa)}$, 
are written as a product of the fine Stokes matrices $S_n$ 
(as in \eq{MultiSymStokes} and \eq{MultiCoarseStokes}). For instance, 
\begin{align}
S_0^{(\rm sym)} &=
\begin{pmatrix}
1 & s_{0,1,2}^{(\rm sym)} & 0 & s_{0,1,4}^{(\rm sym)} & s_{0,1,5}^{(\rm sym)} \cr
0 & 1 & 0 & 0 & 0 \cr
0 & s_{0,3,2}^{(\rm sym)} & 1 & s_{0,3,4}^{(\rm sym)} & s_{0,3,5}^{(\rm sym)} \cr
0 & s_{0,4,2}^{(\rm sym)} & 0 & 1 & 0  \cr
0 & s_{0,5,2}^{(\rm sym)} & 0 & 0 & 1
\end{pmatrix}
= 
\begin{pmatrix}
1 & s_{2,1,2}+s_{1,1,4} s_{3,4,2} & 0 & s_{1,1,4} & s_{3,1,5} \cr
0 & 1 & 0 & 0 & 0 \cr
0 & s_{1,3,2}+s_{0,3,4}s_{3,4,2}& 1 & s_{0,3,4} & s_{2,3,5} \cr
0 & s_{3,4,2} & 0 & 1 & 0  \cr
0 & s_{0,5,2} & 0 & 0 & 1
\end{pmatrix},\nn\\
S_0^{(\rm coa)} &=
\begin{pmatrix}
1 & s_{0,1,2}^{(\rm coa)} & s_{0,1,3}^{(\rm coa)} & s_{0,1,4}^{(\rm coa)} & s_{0,1,5}^{(\rm coa)} \cr
0 & 1 & 0 & 0 & 0 \cr
0 & s_{0,3,2}^{(\rm coa)} & 1 & s_{0,3,4}^{(\rm coa)} & s_{0,3,5}^{(\rm coa)} \cr
0 & s_{0,4,2}^{(\rm coa)} & 0 & 1 & s_{0,4,5}^{(\rm coa)}  \cr
0 & s_{0,5,2}^{(\rm coa)} & 0 & 0 & 1
\end{pmatrix}= \nn\\
&= 
\begin{pmatrix}
1 & s_{2,1,2}+s_{1,1,4} s_{3,4,2} & s_{4,1,3} & s_{1,1,4} & s_{3,1,5}+s_{1,1,4}s_{4,4,5} \cr
0 & 1 & 0 & 0 & 0 \cr
0 & s_{1,3,2}+s_{0,3,4}s_{3,4,2}& 1 & s_{0,3,4} & s_{2,3,5}+s_{0,3,4}s_{4,4,5} \cr
0 & s_{3,4,2} & 0 & 1 & s_{4,4,5}  \cr
0 & s_{0,5,2} & 0 & 0 & 1
\end{pmatrix}
\end{align}
As one can see from these special examples, the Stokes multipliers are always related as
\begin{align}
s_{0,i,j}^{(\rm xxx)} = s_{*,i,j} + \cdots, 
\end{align}
and one can then show that the number of independent Stokes multipliers 
in each Stokes matrix $S_0^{(\rm sym)}$ and $S_0^{(\rm coa)}$ 
is the same and is supplied by 
the fine Stokes matrices. 
Consequently, the same statement also holds for these different kinds of Stokes multipliers: 
For example, 
\begin{align}
s_{2r(l-1),i,j}^{\rm (sym)}& \,\, (i\neq j) \text{ can take non-zero value}\nn\\
&\Leftrightarrow\quad  (j|i) \in J_n, \quad n=2r(l-1),2r(l-1)+1,\cdots, 2rl-1. \label{SymFineRelationCol}
\end{align}

\subsection{Three basic constraints on the Stokes matrices \label{ThreeConstraints11}}

Finally we show the three basic constraints on the Stokes multipliers, 
which result from the detail analysis of (the $\mathbb Z_k$-symmetric) 
critical points in the multi-cut 
two-matrix models \cite{CISY1} and also which provide 
a natural extension of the two-cut cases 
(See Section \ref{TwoCutStokesPheSubSection}). 
We should note that the conditions from the matrix models are given in  
the $\Gamma$-basis (or the matrix-model basis) $\Psi(t;\zeta)$ 
and the Stokes matrices are defined in the $\Omega$-basis (the diagonal basis) 
$\widetilde \Psi(t;\zeta)$, and they are related 
by a unitary transformation (See Eq.~\eq{AsymExpTilde} and Eq.~\eq{GammaOmega}). 

\paragraph{$\mathbb Z_k$ symmetry condition}
This condition is from the $Z_k$ symmetry 
in the multi-cut two-matrix models \cite{CISY1}
and generally expressed as%
\footnote{This is a direct consequence of Eqs.~\eq{UZKQZKSYM}. }
\begin{align}
\omega^{-1} \mathcal Q(t;\omega^{-1}\zeta) = \Omega^{-1}\, \mathcal Q(t;\zeta) \,\Omega, 
\qquad 
 \mathcal P(t;\omega^{-1}\zeta) = \Omega^{-1}\, \mathcal P(t;\zeta) \,\Omega, \label{ZkSymODEExpression}
\end{align}
with $\Omega^{-1} E_{i,i+1} \Omega = \omega E_{i,i+1}$. 
The constraint on the Stokes matrices are then obtained as 
\begin{align}
S_{n+ 2r}= \Gamma^{-1} S_n \Gamma, 
\qquad (n=0,1,\cdots, 2rk-1) \label{ZkSymCondition2323}
\end{align}
for the fine Stokes matrices $S_n$. 
A note for the derivation is following:
\begin{itemize}
\item [1.] Because of the condition Eq.~\eq{ZkSymODEExpression}, 
the canonical solution $\Psi_n(t;\zeta)$ for a Stokes sector $D_n$ satisfies 
\begin{align}
g_{\rm str} \frac{\del \bigl[\Omega\Psi_n(t;\omega^{-1}\zeta) \Omega^{-1}\bigr]}{\del \zeta} 
= \mathcal Q(t;\zeta) \, \bigl[\Omega\Psi_n(t;\omega^{-1}\zeta) \Omega^{-1}\bigr], 
\end{align}
and therefore one obtains 
\begin{align}
\Psi_{n+2r}(t;\zeta)
=\bigl[\Omega\Psi_n(t;\omega^{-1}\zeta) \Omega^{-1}\bigr]\asymeq \Psi_{\rm asym}(t;\zeta), \qquad \zeta \to \infty \in D_{n+2r}=D_n. 
\end{align}
\item [2.] By translating this relation into the $\Omega$-basis (diagonal basis),
\begin{align}
\Psi_{n+2r}(t;\zeta) 
= \Omega\Psi_n(t;\omega^{-1}\zeta) \Omega^{-1}
\quad \Leftrightarrow \quad 
\widetilde \Psi_{n+2r}(t;\zeta) 
= \Gamma^{-1} \widetilde \Psi_n(t;\omega^{-1}\zeta) \Gamma,
\end{align}
with $U^{-1}\Omega U=\Gamma^{-1}$, 
one obtains the relation of the Stokes matrices: 
\begin{align}
S_{n+ 2r} = \widetilde \Psi_{n+ 2r}^{-1}(t;\zeta)
\widetilde \Psi_{n+2r+1}(t;\zeta) 
=\Gamma^{-1} \widetilde \Psi_n^{-1}(t;\zeta) \widetilde \Psi_{n+1}(t;\zeta) \Gamma = \Gamma^{-1} S_n \Gamma. 
\end{align}
\end{itemize}
This condition means that 
only the first $2r$ Stokes matrices $S_n\, (n=0,1,\cdots,2r-1)$ 
are independent. Therefore, we use the first $2r$ dominance profiles
to identify the non-trivial Stokes multipliers: 
\begin{align}
\mathcal J^{(\rm sym)}_{k,r}\equiv
\left[ 
\begin{array}{c}
J_{2r-1} \cr 
\hline
\vdots \cr 
\hline 
J_1 \cr 
\hline
J_0
\end{array}
\right]
\qquad\Leftrightarrow\qquad S_{n}\qquad (n=0,1,\cdots, 2r-1). 
\end{align}
Here we show an examples of $k=5, r=\gamma =2$:
\begin{align}
&\mathcal J_{5,2}^{(\rm sym)}= 
\left[
\begin{array}{c|c|c|c|c}
3 & (5 & 1) & (2 & 4)\cr
\hline
(5 & 3)& (2 & 1) & 4 \cr
\hline
5 & (2 & 3)& (4&1) \cr
\hline
(2 & 5)&(4 & 3) & 1 \cr
\end{array}
\right]\!\!\!\!
\begin{array}{l}
:J_{3} \cr 
:J_2 \cr 
:J_1 \cr 
:J_0
\end{array}. 
\end{align}

\paragraph{Hermiticity condition}
This condition is from the hermiticity of the multi-cut matrix models \cite{CISY1} and generally expressed as 
\begin{align}
{\mathcal Q}^*(t;\zeta^*) = \mathcal Q(t;\zeta^*).  \label{HermiticityODEExpression}
\end{align}
The constraints on the Stokes matrices are then obtained as 
\begin{align}
S_n^*= \Delta\Gamma\, S_{(2r-1)k-n}^{-1}\, \Gamma^{-1}\Delta, 
\qquad \Delta_{i,j}=\delta_{i,k-j+1}, \qquad 
(n=0,1,\cdots, 2rk-1)\label{HermiticityCondStokes}
\end{align}
for the fine Stokes matrices $S_n$. 
A note for the derivation is following:
\begin{itemize}
\item [1.] Because of the condition Eq.~\eq{HermiticityODEExpression}, 
the canonical solution $\Psi_n(t;\zeta)$ for a Stokes sector $D_n$ satisfies 
\begin{align}
g_{\rm str} \frac{\del \Psi_n^*(t;\zeta)}{\del \zeta} 
= {\mathcal Q}(t;\zeta) \, \Psi_n^*(t;\zeta),
\end{align}
and therefore one obtains 
\begin{align}
\Psi_{(2r-1)k+1-n}(t;\zeta)= \Psi_n^*(t;\zeta) 
\asymeq \Psi_{\rm asym}(t;\zeta), \qquad 
\zeta \in D_{(2r-1)k+1-n}= D_n^*. 
\end{align}
\item [2.] By translating this relation into the $\Omega$-basis (diagonal basis),
\begin{align}
\Psi_{n}^*(t;\zeta) =\Psi_{(2r-1)k+1-n}(t;\zeta)
\quad \Leftrightarrow \quad
\widetilde \Psi_n^* 
&= U^2\, \widetilde \Psi_{(2r-1)k+1-n}(t;\zeta)\, U^{-2} \nn\\
&=\Delta \Gamma \, \widetilde \Psi_{(2r-1)k+1-n}(t;\zeta)\, \Gamma^{-1}\Delta,
\end{align}
with $U^* = U^{-1}$ and $U^2 = \Delta \Gamma$, 
one obtains the relation of the Stokes matrices
\begin{align}
S_n^* = \bigl[\widetilde \Psi_n^{-1}(t;\zeta) \widetilde \Psi_{n+1}(t;\zeta)\bigr]^*
&= \Delta \Gamma 
\bigl[\widetilde \Psi_{(2r-1)k+1-n}^{-1}(t;\zeta) 
\widetilde \Psi_{(2r-1)k-n}(t;\zeta)\bigr]
\Gamma^{-1} \Delta \nn\\
&= \Delta \Gamma 
\bigl[\widetilde \Psi_{(2r-1)k-n}^{-1}(t;\zeta) 
\widetilde \Psi_{(2r-1)k+1-n}(t;\zeta)\bigr]^{-1}
\Gamma^{-1} \Delta \nn\\
&= \Delta\Gamma\, S_{(2r-1)k-n}^{-1}\, \Gamma^{-1}\Delta. 
\end{align}
\end{itemize}

\paragraph{Monodromy free condition}
If the formal expansion satisfies $\varphi_0=0$ (discussed in Appendix \ref{AppendixLaxOP}), then 
the canonical solutions are the single-valued functions: 
\begin{align}
\widetilde\Psi_{n}(t;\zeta) = \widetilde\Psi_n(t;e^{2\pi i }\zeta)=\widetilde\Psi_{n+2kr}(t;\zeta),\label{MonodromyFreeMultiCutCutInTermsOfPsiGeneral}
\end{align}
therefore the Stokes matrices satisfy 
\begin{align}
S_0 \cdot S_1 \cdots S_{2rk-1} 
&= S_{0}^{(\rm coa)}\cdot S_{k}^{(\rm coa)}\cdots S_{k (2r-1)}^{(\rm coa)} \nn\\
&= S_{0}^{(\rm sym)}\cdot S_{2r}^{(\rm sym)}\cdots S_{2 r (k-1)}^{(\rm sym)}
= I_k. 
\end{align}
Note that, with the $\mathbb Z_k$-symmetry constraints, $2rk$ Stokes matrices are 
reduced to fundamental $2r$ Stokes matrices, $\{S_{n}\}_{n=0}^{2r-1}$, 
and also that the monodromy free condition is written as
\begin{align}
\bigl(S_0^{(\rm sym)}\,\Gamma^{-1}\bigr)^k = I_k.  \label{MFcondOriginal101}
\end{align}

\section{The multi-cut boundary condition and solutions \label{MultiCutBCSection}}
In the previous section, 
we developed the general framework of Stokes phenomena 
in the ODE systems 
which appear in the multi-cut matrix models. 
Mathematically, general solutions $\Psi(t;\zeta)$ for 
these isomonodromy systems 
(or equivalently for the corresponding Douglas (string) equations) 
are parametrized by the Stokes multipliers with three basic constraints 
discussed in Section \ref{ThreeConstraints11}. 
As is mentioned in Introduction, however, not all the solutions to these constraints can realize the critical points in the multi-cut matrix models. 
This consideration requires {\em additional physical constraints} on the 
Stokes multipliers. 
In this section, the first {\em physical constraint} is proposed, 
which we refer to as {\em multi-cut boundary conditions}. 
The second physical condition is proposed in Section \ref{RHapproachSection}. 

\subsection{Two different viewpoints about spectral curves}
Before we discuss the detail of the multi-cut boundary conditions, 
we first recall the set up of the multi-cut two-matrix models 
and the relationship between the Baker-Akhiezer function system (i.e.~the ODE system) 
and the resolvent operator which defines the spectral curves. 

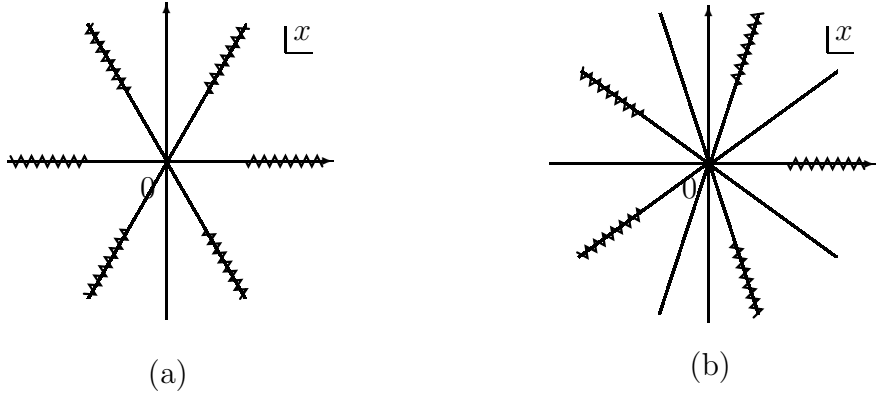
\begin{figure}[htbp]
\begin{center}
\begin{picture}(120,150)(0,0)
\begin{picture}(0,0)(48,-80)
\begin{picture}(0,0)(10,10)
\put(55,55){\line(0,1){10}\line(1,0){10}}
\put(58,60){$x$}
\put(0,0){$0$}
\end{picture}
\put(0,0){\rotatebox{90}{
\begin{picture}(0,0)(0,0)
\put(0,0){\vector(1,0){60}}
\put(0,0){\line(-1,0){60}}
\put(0,-59){\vector(0,-1){4}}
\thicklines
\put(0,0){\rotatebox{30}{\line(1,0){60}}}
\put(0,0){\line(0,1){60}}
\put(-52,0){\rotatebox{60}{\line(0,1){60}}}
\put(0,0){\rotatebox{-30}{\line(1,0){60}}}
\put(0,0){\line(0,-1){60}}
\put(-52,-30){\rotatebox{-60}{\line(0,1){60}}}
\put(0,30){\rotatebox{90}{
\put(0,0){\path(0,0)(1,-2)(3, 2)(5, -2)( 7, 2)( 9, -2)( 11, 2)( 13, -2)( 15, 2)( 17, -2)( 19, 2)(21, -2)( 23, 2)( 25, -2)( 27, 2)( 29, -2)}}}
\put(0,-30){\rotatebox{-90}{
\put(0,0){\path(0,0)(1,-2)(3, 2)(5, -2)( 7, 2)( 9, -2)( 11, 2)( 13, -2)( 15, 2)( 17, -2)( 19, 2)(21, -2)( 23, 2)( 25, -2)( 27, 2)( 29, -2)}}}
\put(26,15){\rotatebox{30}{
\put(0,0){\path(0,0)(1,-2)(3, 2)(5, -2)( 7, 2)( 9, -2)( 11, 2)( 13, -2)( 15, 2)( 17, -2)( 19, 2)(21, -2)( 23, 2)( 25, -2)( 27, 2)( 29, -2)}}}
\put(26,-15){\rotatebox{-30}{
\put(0,0){\path(0,0)(1,-2)(3, 2)(5, -2)( 7, 2)( 9, -2)( 11, 2)( 13, -2)( 15, 2)( 17, -2)( 19, 2)(21, -2)( 23, 2)( 25, -2)( 27, 2)( 29, -2)}}}
\put(-26,15){\rotatebox{150}{
\put(0,0){\path(0,0)(1,-2)(3, 2)(5, -2)( 7, 2)( 9, -2)( 11, 2)( 13, -2)( 15, 2)( 17, -2)( 19, 2)(21, -2)( 23, 2)( 25, -2)( 27, 2)( 29, -2)}}}
\put(-26,-15){\rotatebox{-150}{
\put(0,0){\path(0,0)(1,-2)(3, 2)(5, -2)( 7, 2)( 9, -2)( 11, 2)( 13, -2)( 15, 2)( 17, -2)( 19, 2)(21, -2)( 23, 2)( 25, -2)( 27, 2)( 29, -2)}}}
\end{picture}}}
\put(-7,-80){(a)}
\end{picture}
\begin{picture}(0,0)(-153,-83)
\begin{picture}(0,0)(10,13)
\put(55,55){\line(0,1){10}\line(1,0){10}}
\put(58,60){$x$}
\put(0,0){$0$}
\end{picture}
\put(0,0){\vector(0,1){60}}
\put(0,0){\line(0,-1){60}}
\put(0,0){\vector(1,0){63}}
\put(0,0){\line(-1,0){60}}
\thicklines
\put(0,0){\line(1,0){60}}
\put(0,0){\line(-1,0){60}}
\put(0,0){\rotatebox{36}{\line(1,0){60}}}
\put(0,0){\rotatebox{72}{\line(1,0){60}}}
\put(0,0){\rotatebox{-36}{\line(1,0){60}}}
\put(0,0){\rotatebox{-72}{\line(1,0){60}}}
\put(-18.77,0){\rotatebox{108}{\line(1,0){60}}}
\put(-48.5,0){\rotatebox{144}{\line(1,0){60}}}
\put(-19,0){\rotatebox{-108}{\line(1,0){60}}}
\put(-49.5,0){\rotatebox{-144}{\line(1,0){60}}}
\put(30,0){\rotatebox{0}{
\put(0,0){\path(0,0)(1,-2)(3, 2)(5, -2)( 7, 2)( 9, -2)( 11, 2)( 13, -2)( 15, 2)( 17, -2)( 19, 2)(21, -2)( 23, 2)( 25, -2)( 27, 2)( 29, -2)}}}
\put(10,30){\rotatebox{72}{
\put(0,0){\path(0,0)(1,-2)(3, 2)(5, -2)( 7, 2)( 9, -2)( 11, 2)( 13, -2)( 15, 2)( 17, -2)( 19, 2)(21, -2)( 23, 2)( 25, -2)( 27, 2)( 29, -2)}}}
\put(10,-30){\rotatebox{-72}{
\put(0,0){\path(0,0)(1,-2)(3, 2)(5, -2)( 7, 2)( 9, -2)( 11, 2)( 13, -2)( 15, 2)( 17, -2)( 19, 2)(21, -2)( 23, 2)( 25, -2)( 27, 2)( 29, -2)}}}
\put(-25,18){\rotatebox{144}{
\put(0,0){\path(0,0)(1,-2)(3, 2)(5, -2)( 7, 2)( 9, -2)( 11, 2)( 13, -2)( 15, 2)( 17, -2)( 19, 2)(21, -2)( 23, 2)( 25, -2)( 27, 2)( 29, -2)}}}
\put(-25,-18){\rotatebox{-144}{
\put(0,0){\path(0,0)(1,-2)(3, 2)(5, -2)( 7, 2)( 9, -2)( 11, 2)( 13, -2)( 15, 2)( 17, -2)( 19, 2)(21, -2)( 23, 2)( 25, -2)( 27, 2)( 29, -2)}}}
\put(-7,-80){(b)}
\end{picture}
\end{picture}
\end{center}
\caption{\footnotesize Examples of contours $\mathcal C^{(k)}$. (a) is 6-cut contour $\mathcal C^{(6)}$ and (b) is the 5-cut contour $\mathcal C^{(5)}$ 
which is equal to the 10-cut contour $\mathcal C^{(10)}$. 
For reference, the position of cuts (zig-zag lines) around $\zeta \to \infty$ is also 
denoted. \label{ContourExamplesFigure}}
\end{figure}

The definition of the multi-cut two-matrix models is given by the following matrix integral: 
\begin{align}
Z= \int_{\mathcal C_N^{(k)}\times \mathcal C_N^{(k)}} dX\,dY\, e^{-N{\rm tr}[V_1(X)+V_2(Y)-XY]}, \label{DefOfMultiCutMatrixModels}
\end{align}
with the matrix contour $\mathcal C_N^{(k)}$ of 
the following $N\times N$ $k$-cut normal matrix, 
\begin{align}
\mathcal C_N^{(k)} \equiv \Bigl\{ X = U \diag (x_1,x_2,\cdots,x_N) \,U^\dagger;\ 
U\in U(N),\quad x_j \in \bigcup_{n=0}^{k-1} e^{2\pi i\frac{n}{k}}\, \mathbb R\Bigr\}. 
\end{align}
The system of two-matrix models has the corresponding orthonormal polynomial system \cite{Mehta}:
\begin{align}
\alpha_n(x) = \frac{1}{\sqrt{h_n}} \Bigl(x^n + \cdots\Bigr),\qquad \beta_n(y) = 
\frac{1}{\sqrt{h_n}} \Bigl(y^n + \cdots\Bigr), 
\end{align}
with
\begin{align}
\delta_{n,m} 
= \int_{\mathcal C^{(k)}\times \mathcal C^{(k)}} dx\, dy\, e^{-N[V_1(x)+V_2(y)-xy]} 
\,\alpha_n(x)\,\beta_m(y). 
\end{align} 
Here the contour $\mathcal C^{(k)}$ is given as 
\begin{align}
\mathcal C^{(k)} 
= \Bigl\{x \in \bigcup_{n=0}^{k-1} e^{2\pi i\frac{n}{k}}\, \mathbb R\Bigr \},
\end{align}
an example of which is shown in Fig.~\ref{ContourExamplesFigure}. 

The Baker-Akhiezer systems (or the ODE systems) appear 
as the double scaling limit of the orthonormal polynomials $\alpha_n(x)$ 
(or their dual polynomials $\beta_n(y)$), which is given as follows: 
\begin{align}
\alpha_n(x) = a^{-\hat p/2}\,\Psi_{\rm orth}(\zeta;t),
\end{align}
with the following scaling relations of $a\to 0$: 
\begin{align}
x &= \omega^{-1/2} a^{\hat p/2} \zeta\to 0, 
\qquad \frac{n}{N} = \exp\bigl(-t a^{\frac{\hat p +\hat q-1}{2}}\bigr)\to 1,\nn\\
N^{-1} &= g_{\rm str}\, a^{\frac{\hat p + \hat q}{2}}\to 0,
\qquad \del_n=-a^{1/2} g_{\rm str} \del_t \equiv -a^{1/2} \del \to 0. \label{HowToTakeContinuumLimit}
\end{align}
The continuous function $\Psi_{\rm orth}(t;\zeta)$ is the scaling function of 
the orthonormal polynomials and 
satisfies the differential equations \eq{BA1} and \eq{BA2}:
\begin{align}
\zeta \Psi_{\rm orth}(t;\zeta) 
&= \bP(t;\del)\, \Psi_{\rm orth}(t;\zeta),\label{BA1} \\ 
g_{\rm str}\frac{\del }{\del \zeta } \Psi_{\rm orth}(t;\zeta) 
&= \bQ(t;\del)\, \Psi_{\rm orth}(t;\zeta). \label{BA2}
\end{align}
This means that the orthonormal polynomial system is {\em one of the solutions} 
to the differential equations \eq{BA1} and \eq{BA2}, and eventually the ODE systems 
\eq{GeneralODEP} and \eq{GeneralODEQ}. 
Consequently, the scaling function $\Psi_{\rm orth}(t;\zeta)$ is given 
by the canonical solutions $\Psi_n(t;\zeta)$ with some proper vector 
$X^{(n)} = {}^t\bigl(x_1^{(n)},x_2^{(n)},\cdots,x_k^{(n)}\bigr)$ as 
\begin{align}
\Psi_{\rm orth} (t;\zeta) = \widetilde \Psi_n(t;\zeta)\, X^{(n)}, \qquad  (n=0,1,\cdots). 
\label{DefOfX}
\end{align}
By taking into account the Stokes phenomena \eq{StokesPhenoMultiCut111}, these vectors $X^{(n)}$ 
of various Stokes sectors $D_n$ are related as follows:
\begin{align}
X^{(n)} = S_n\, X^{(n+1)},\qquad X^{(n+2rk)}=X^{(n)}.  \label{PreRecursion}
\end{align}
Note that the scaled orthonormal polynomials $\Psi_{\rm orth}(t;\zeta)$ 
are entire functions in $\zeta \in \mathbb C$ 
because the original orthonormal polynomials are also entire functions. 

On the other hand, another important approach to solving the multi-cut matrix models 
is the semi-classical approach with the resolvent operator $\mathcal R(x)$
of the matrix models,
\begin{align}
\mathcal R(x) = \vev{\frac{1}{N}\tr \frac{1}{x-X}} 
= \int_{\mathcal C^{(k)}} dz\, \frac{\rho(z)}{x-z}, \label{DefResol}
\end{align}
where $\rho(z)$ is the density function of eigenvalues of the matrix $X$. 
An important fact about the resolvent is that this operator is a single valued 
function in $x \in \mathbb C$ with a finite $N$ and 
the cuts appearing in the large $N$ limit are along the matrix-model contour $\mathcal C^{(k)}$ 
on the $x$ space (as shown in Fig.~\ref{ContourExamplesFigure}). 
These special cuts are called {\em physical cuts}. 
Interestingly, this resolvent operator is 
also related to the orthonormal polynomial solution 
$\Psi_{\rm orth}(t;\zeta)$ in the following way \cite{GrossMigdal2}:%
\footnote{For the precise relations, see Appendix A in \cite{CISY1}, for example. }
\begin{align}
\Psi_{\rm orth}(t;\zeta) \sim \vev{\det\bigl(x- X\bigr)} 
\sim \exp\Bigl[N \int^x dx' \mathcal R(x')\Bigr], \label{OrthResolCores}
\end{align}
with the scaling relation, $x = \omega^{-1/2} a^{\hat p/2} \zeta$, of Eq.~\eq{HowToTakeContinuumLimit}. 
This relation also indicates%
\footnote{
Although $\Psi_{\rm orth}(t;\zeta)$ is a vector valued function, 
the behaviors of exponents are the same among the vector components. Therefore, 
it is understood by taking one particular element of the function $\Psi_{\rm orth}(t;\zeta)$. 
}
\begin{align}
\mathcal R(x) \sim \lim_{g_{\rm str}\to 0}g_{\rm str} \frac{\del}{\del \zeta} 
\ln \Psi_{\rm orth}(t;\zeta).  \label{ResolventOrth}
\end{align}
Therefore, the two different observables, 
the semi-classical resolvent $\mathcal R(x)$ Eq.~\eq{DefResol} and 
the semi-classical orthonormal polynomials $\Psi_{\rm orth}(t;\zeta)$ 
Eq.~\eq{ResolventOrth}, provide two different viewpoints of spectral curves.  

\subsection{The multi-cut boundary conditions \label{MCBCSUBSection}}
As we carefully see the above two viewpoints of the spectral curve, 
one can notice that the realization of {\em the position of physical cuts} 
is not straightforward from the Baker-Akhierzer (or ODE) approach. 
The discontinuities of the scaling orthonormal polynomial of Eq.~\eq{DefOfX} 
around $\zeta\to \infty$ are {\em Stokes lines} and generally not distributed 
in the proper way expected in the semi-classical resolvent operator 
Eq.~\eq{DefResol}. 
This eventually means that not all the solutions to the ODE system 
Eq.~\eq{BA1} and Eq.~\eq{BA2} (therefore equivalently string equations) 
correspond to critical points of the multi-cut matrix models. 
The difference between these two viewpoints provides 
additional {\em physical constraints} not only on the vectors $X^{(n)}$ 
but also on the Stokes multipliers $s_{l,i,j}$ which are identified as 
integration constants of the string equations. 

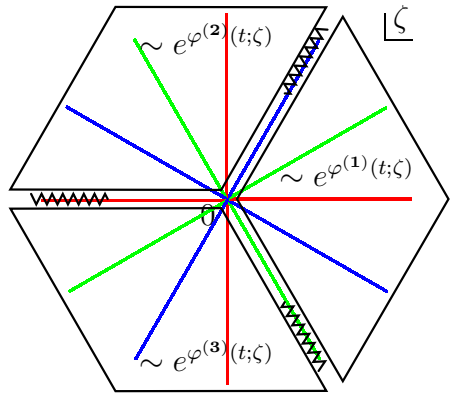
\begin{figure}[htbp]
\begin{center}
\begin{picture}(180,150)(0,0)
\end{picture}
\begin{picture}(5,0)(100,-55)
\put(70,70){\line(0,1){10}\line(1,0){10}}
\put(73,75){$\zeta$}
\put(0,0){$0$}
{
\thicklines
\put(10,10){\rotatebox{-90}{\textcolor{red}{\line(1,0){70}}}}
\put(10,10){\rotatebox{-60}{\textcolor{green}{\line(1,0){70}}}}
\put(10,10){\rotatebox{-30}{\textcolor{blue}{\line(1,0){70}}}}
\put(10,10){\rotatebox{0}{\textcolor{red}{\line(1,0){70}}}}
\put(9.5,10){\rotatebox{30}{\textcolor{green}{\line(1,0){70}}}}
\put(9.5,10){\rotatebox{60}{\textcolor{blue}{\line(1,0){70}}}}
\put(9.5,10){\rotatebox{90}{\textcolor{red}{\line(1,0){70}}}}
\put(-25.5,10){\rotatebox{120}{\textcolor{green}{\line(1,0){70}}}}
\put(-51,10){\rotatebox{150}{\textcolor{blue}{\line(1,0){70}}}}
\put(-60,9.5){\rotatebox{180}{\textcolor{red}{\line(1,0){70}}}}
\put(-50,10){\rotatebox{-150}{\textcolor{green}{\line(1,0){70}}}}
\put(-25,10){\rotatebox{-120}{\textcolor{blue}{\line(1,0){70}}}}
}
\put(32,50){\rotatebox{60}{\thicklines
\put(0,0){\path(0,0)(1,-2)(3, 2)(5, -2)( 7, 2)( 9, -2)( 11, 2)( 13, -2)( 15, 2)( 17, -2)( 19, 2)(21, -2)( 23, 2)( 25, -2)( 27, 2)( 29, -2)}}}
\put(32,-29){\rotatebox{-60}{\thicklines
\put(0,0){\path(0,0)(1,-2)(3, 2)(5, -2)( 7, 2)( 9, -2)( 11, 2)( 13, -2)( 15, 2)( 17, -2)( 19, 2)(21, -2)( 23, 2)( 25, -2)( 27, 2)( 29, -2)}}}
\put(-35,10){\rotatebox{180}{\thicklines
\put(0,0){\path(0,0)(1,-2)(3, 2)(5, -2)( 7, 2)( 9, -2)( 11, 2)( 13, -2)( 15, 2)( 17, -2)( 19, 2)(21, -2)( 23, 2)( 25, -2)( 27, 2)( 29, -2)}}}
\put(10,10){\rotatebox{0}{
\begin{picture}(0,0)(0,0)
\thicklines
\put(0,0){\path(0,0)(40,69)(80,0)(40,-69)(0,0)}
\end{picture}}}
\put(8,10){\rotatebox{120}{
\begin{picture}(0,0)(0,0)
\thicklines
\put(0,0){\path(0,0)(40,69)(80,0)(40,-69)(0,0)}
\end{picture}}}
\put(8,10){\rotatebox{-120}{
\begin{picture}(0,0)(0,0)
\thicklines
\put(0,0){\path(0,0)(40,69)(80,0)(40,-69)(0,0)}
\end{picture}}}
\put(30,15){$\sim e^{\varphi^{({\bf 1})}(t;\zeta)}$}
\put(-23,65){$\sim e^{\varphi^{({\bf 2})}(t;\zeta)}$}
\put(-23,-55){$\sim e^{\varphi^{({\bf 3})}(t;\zeta)}$}
\end{picture}
\end{center}
 \caption{\footnotesize 
The multi-cut boundary condition in the $3$-cut $(1,1)$ critical point. 
Although the general solutions to the Baker-Akhiezer function system 
can generally have ``twelve cuts'', 
there are only three cuts in the scaling orthonormal-polynomial solution 
$\Psi_{\rm orth}(t;\zeta)$. 
\label{MultiCutConditionFigure111333}}
\end{figure}

Next we formulate this physical constraint in the following way. 
Note that we here only care the leading behavior of $\zeta\to \infty$ 
for the Stokes lines. 
\begin{Definition} 
[Multi-cut boundary condition] 
The Baker-Akhiezer (or ODE) systems Eq.~\eq{BA1} and Eq.~\eq{BA2} 
are said to satisfy the multi-cut boundary condition, 
if there exists a special solution $\Psi_{\rm orth}(t;\zeta)$ 
which satisfies the following condition: 
\begin{itemize}
\item Stokes lines of the solution $\Psi_{\rm orth}(t;\zeta)$ 
around $\zeta \to \infty$ only exist 
along some special $k$ angles $\zeta \to \infty \times e^{i \chi_n}$: 
\begin{align}
\chi_n \equiv \chi_0 + \frac{2\pi n}{k}\qquad 
(n=0,1,2,\cdots,k-1), 
\end{align}
with a proper $\chi_0$ corresponding to each critical point. 
\item Therefore, there exist an ordered set of $k$ indices, 
$(a_1,a_2,\cdots,a_k)$, and a set of $k$ non-zero vectors, 
$(v_1,v_2,\cdots,v_k)$, such that the asymptotic expansions of the solution 
$\Psi_{\rm orth}(t;\zeta)$ in the angular domain 
$D(\chi_n,\chi_{n+1})$ are given as 
\begin{align}
\Psi_{\rm orth}(t;\zeta) \asymeq v_n\, e^{\varphi^{(a_n)}(t;\zeta)}+\cdots, \qquad 
\zeta \to \infty \in D(\chi_n,\chi_{n+1}),
\end{align}
and the expansions along the Stokes lines are given as the superposition: 
\begin{align}
\Psi_{\rm orth}(t;\zeta) \asymeq v_n\, e^{\varphi^{(a_n)}(t;\zeta)}+
v_{n+1}\, e^{\varphi^{(a_{n+1})}(t;\zeta)}\cdots, \qquad 
\zeta \to \infty \times e^{i\chi_{n+1}}. 
\end{align}
\end{itemize}\label{MultiCutBCDefinition}
\end{Definition}
Here appears a special angle $\chi_0$ which is determined by the 
critical points of the matrix models and is given as follows:
\begin{align}
\chi_0 =
\left\{
\begin{array}{ll}
\ds  \frac{\pi}{k}& \text{: $\mathbb Z_k$-symmetric cases, and $\omega^{1/2}$-rotated FSST cases} \cr
 0 & \text{: Real-potential FSST cases}
 \end{array}
 \right.. \label{MCBCchi0}
\end{align}
This angle $\chi_0$ comes from the scaling relation Eq.~\eq{HowToTakeContinuumLimit}, 
for detail discussion of which we should refer to \cite{CISY1}. 
An example of the boundary condition in the $\zeta$ plane is shown in Fig.~\ref{MultiCutConditionFigure111333}. 

Some comments are in order: 
\begin{itemize}
\item 
As in the definition of Stokes lines Eq.~\eq{StokesLinesDef}, 
we here only used the leading contributions of the exponents in $\zeta\to \infty$. 
If we also take $t\to \infty$, or equivalently if we just take $g_{\rm str} \to 0$, 
on the other hand, 
we naturally encounter the general Stokes lines of Definition \ref{GeneralStokesLinesDef}. 
Therefore, we interpret the general Stokes lines of the scaling orthonormal-polynomial solution as {\em the non-perturbative definition of physical cuts}. 
In particular, this definition guarantees {\em real eigenvalue-density function 
$\rho(\lambda) d\lambda$} along the physical cuts: 
\begin{align}
{\rm Re}\bigl[\pi i \rho(\lambda) d\lambda \bigr]\equiv 
{\rm Re}\bigl[d\varphi^{(j)}(t;\zeta) - d\varphi^{(l)}(t;\zeta)\bigr]
= 0,
\label{GeneralStokeslinesForCuts}
\end{align}
where $\zeta = \zeta(\lambda)$ is a local map from $\lambda \in \mathbb R$ to 
the generalized Stokes line ${\rm GSL}_{i,j}\subset \mathbb C$. 
Note that this definition naturally justifies the curved physical cuts 
observed in \cite{CIY1} which appear when the matrix-model potentials 
are perturbed with complex coefficients.%
\footnote{This consideration suggests that the position of 
physical cuts are not freely assigned and closely related to non-perturbative 
consistency with Stokes phenomena and therefore with D-instanton chemical potentials. }
This consideration is further extended to {\em off-shell backgrounds (or spectral curves)} 
in terms of the Riemann-Hilbert approach in Section \ref{RHapproachSection}. 
\item In the $\hat p>1$ cases, 
the exponents $\varphi^{(j)}(t;\zeta)$ have non-trivial cuts in the $\zeta$ plane, say
$\varphi^{(j)}(t;\zeta) \sim \zeta^{(\hat q+1)/\hat p}$. 
This $\hat p$-th root cut should be smeared by a proper supplement of exponents \cite{MMSS}. 
This is also reviewed in Appendix \ref{AppendixAiryFunction}. 
Since we concentrate on the $\hat p=1$ cases in this paper, 
this point in the general $k$-cut cases 
remains to be studied for future investigations. 
\item As it will be clear in Section \ref{MultiCutBCRecursionsSubSubSectioN}, 
the set of indices $(a_1,a_2,\cdots,a_k)$ in the multi-cut boundary condition (Definition \ref{MultiCutBCDefinition}) is generally given as 
\begin{align}
a_n = j_{2r(n-1),k} = n+(n-1) (r-\gamma)m_1,
\end{align}
with Theorem \ref{TheoremComponentsOfProfiles}. In particular, 
the $\mathbb Z_k$ symmetric cases ($\gamma=r$) is given as $a_n = n$. 
\end{itemize}
Next we apply this boundary condition to concrete systems. 
Before devoting ourselves into general cases, however, we first consider 
the multi-cut boundary condition in the two-cut case, as a warm-up exercise for the general systems. 

\subsubsection{The two-cut boundary condition \label{TwoCutCaseMultiCutBClaskeie}}

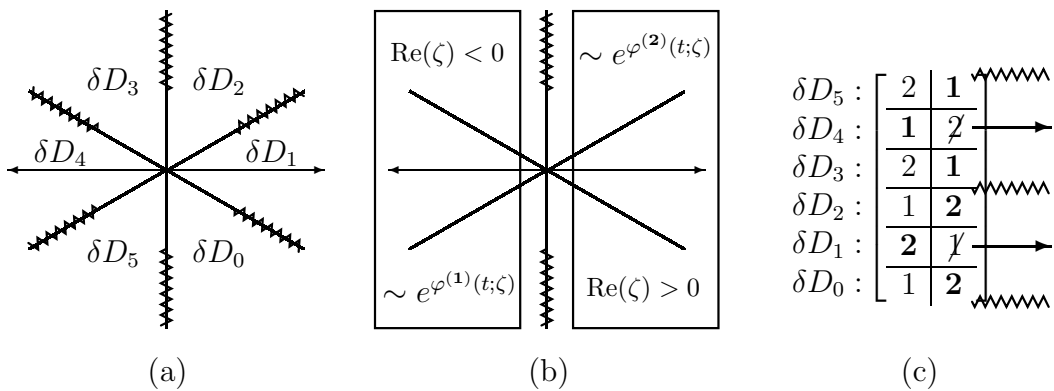
\begin{figure}[htbp]
\begin{center}
\begin{picture}(130,140)(0,0)
\begin{picture}(0,0)(70,-70)
\put(0,0){\vector(1,0){60}}
\put(0,0){\vector(-1,0){60}}
\thicklines
\put(0,0){\rotatebox{30}{\line(1,0){60}}}
\put(0,0){\line(0,1){60}}
\put(-52,0){\rotatebox{60}{\line(0,1){60}}}
\put(0,0){\rotatebox{-30}{\line(1,0){60}}}
\put(0,0){\line(0,-1){60}}
\put(-52,-30){\rotatebox{-60}{\line(0,1){60}}}
\put(0,30){\rotatebox{90}{
\put(0,0){\path(0,0)(1,-2)(3, 2)(5, -2)( 7, 2)( 9, -2)( 11, 2)( 13, -2)( 15, 2)( 17, -2)( 19, 2)(21, -2)( 23, 2)( 25, -2)( 27, 2)( 29, -2)}}}
\put(0,-30){\rotatebox{-90}{
\put(0,0){\path(0,0)(1,-2)(3, 2)(5, -2)( 7, 2)( 9, -2)( 11, 2)( 13, -2)( 15, 2)( 17, -2)( 19, 2)(21, -2)( 23, 2)( 25, -2)( 27, 2)( 29, -2)}}}
\put(26,15){\rotatebox{30}{
\put(0,0){\path(0,0)(1,-2)(3, 2)(5, -2)( 7, 2)( 9, -2)( 11, 2)( 13, -2)( 15, 2)( 17, -2)( 19, 2)(21, -2)( 23, 2)( 25, -2)( 27, 2)( 29, -2)}}}
\put(26,-15){\rotatebox{-30}{
\put(0,0){\path(0,0)(1,-2)(3, 2)(5, -2)( 7, 2)( 9, -2)( 11, 2)( 13, -2)( 15, 2)( 17, -2)( 19, 2)(21, -2)( 23, 2)( 25, -2)( 27, 2)( 29, -2)}}}
\put(-26,15){\rotatebox{150}{
\put(0,0){\path(0,0)(1,-2)(3, 2)(5, -2)( 7, 2)( 9, -2)( 11, 2)( 13, -2)( 15, 2)( 17, -2)( 19, 2)(21, -2)( 23, 2)( 25, -2)( 27, 2)( 29, -2)}}}
\put(-26,-15){\rotatebox{-150}{
\put(0,0){\path(0,0)(1,-2)(3, 2)(5, -2)( 7, 2)( 9, -2)( 11, 2)( 13, -2)( 15, 2)( 17, -2)( 19, 2)(21, -2)( 23, 2)( 25, -2)( 27, 2)( 29, -2)}}}
\put(-7,-80){(a)}
\put(30,3){$\delta D_1$}
\put(-50,3){$\delta D_4$}
\put(10,30){$\delta D_2$}
\put(-30,30){$\delta D_3$}
\put(10,-35){$\delta D_0$}
\put(-30,-35){$\delta D_5$}
\end{picture}
\begin{picture}(0,0)(-70,-70)
\put(0,0){\vector(1,0){60}}
\put(0,0){\vector(-1,0){60}}
\thicklines
\put(0,0){\rotatebox{30}{\line(1,0){60}}}
\put(0,0){\line(0,1){60}}
\put(-52,0){\rotatebox{60}{\line(0,1){60}}}
\put(0,0){\rotatebox{-30}{\line(1,0){60}}}
\put(0,0){\line(0,-1){60}}
\put(-52,-30){\rotatebox{-60}{\line(0,1){60}}}
\put(0,30){\rotatebox{90}{
\put(0,0){\path(0,0)(1,-2)(3, 2)(5, -2)( 7, 2)( 9, -2)( 11, 2)( 13, -2)( 15, 2)( 17, -2)( 19, 2)(21, -2)( 23, 2)( 25, -2)( 27, 2)( 29, -2)}}}
\put(0,-30){\rotatebox{-90}{
\put(0,0){\path(0,0)(1,-2)(3, 2)(5, -2)( 7, 2)( 9, -2)( 11, 2)( 13, -2)( 15, 2)( 17, -2)( 19, 2)(21, -2)( 23, 2)( 25, -2)( 27, 2)( 29, -2)}}}
\put(-7,-80){(b)}
\put(10,-60){\path(0,0)(0,120)(55,120)(55,0)(0,0)}
\put(-10,-60){\path(0,0)(0,120)(-55,120)(-55,0)(0,0)}
\put(12,40){$\sim e^{\varphi^{({\bf 2})}(t;\zeta)}$}
\put(15,-48){\footnotesize ${\rm Re}(\zeta)>0$}
\put(-62,-50){$\sim e^{\varphi^{({\bf 1})}(t;\zeta)}$}
\put(-59,41){\footnotesize ${\rm Re}(\zeta)<0$}
\end{picture}
\begin{picture}(0,0)(-150,-60)
\put(0,0){ $\begin{array}{c}
\delta D_5:  \cr
\delta D_4:  \cr
\delta D_3:  \cr
\delta D_2:  \cr
\delta D_1:  \cr
\delta D_0:  \cr
\end{array}\!\!\!\left[
\begin{array}{c|c}
2 & \bf 1 \cr
\hline
\bf 1 & \not\! 2 \cr
\hline
2 & \bf 1 \cr
\hline
1 & \bf 2 \cr
\hline
\bf 2 & \not\! 1 \cr
\hline
1 & \bf 2 \cr
\end{array}
\right]$
}
\thicklines
\put(77,3){\rotatebox{0}{
\put(0,0){\path(0,0)(1,-2)(3, 2)(5, -2)( 7, 2)( 9, -2)( 11, 2)( 13, -2)( 15, 2)( 17, -2)( 19, 2)(21, -2)( 23, 2)( 25, -2)( 27, 2)( 29, -2)}}}
\put(77,46){\rotatebox{0}{
\put(0,0){\path(0,0)(1,-2)(3, 2)(5, -2)( 7, 2)( 9, -2)( 11, 2)( 13, -2)( 15, 2)( 17, -2)( 19, 2)(21, -2)( 23, 2)( 25, -2)( 27, 2)( 29, -2)}}}
\put(77,-40){\rotatebox{0}{
\put(0,0){\path(0,0)(1,-2)(3, 2)(5, -2)( 7, 2)( 9, -2)( 11, 2)( 13, -2)( 15, 2)( 17, -2)( 19, 2)(21, -2)( 23, 2)( 25, -2)( 27, 2)( 29, -2)}}}
\put(77,-19){\vector(1,0){30}}
\put(77,26){\vector(1,0){30}}
\put(50,-70){(c)}
\end{picture}
\end{picture}
\end{center}
\caption{\footnotesize The positions of cuts in the two-cut $(1,2)$ ODE system. 
a) A general configuration of cuts for the general Stokes multipliers. There are 6 cuts. 
b) A configuration of cuts for the $(1,2)$ critical point in the two-cut matrix models. 
The boxes indicate the regions ${\rm Re}(\zeta)>0$ and ${\rm Re}(\zeta)<0$,
in which the asymptotic expansion is given by $\sim e^{\varphi^{(i)}(\zeta)}$ $(i=1,2)$. 
c) The profile of dominance depicted with the position of cuts and the weak coupling 
infinity $\zeta \to \pm \infty \in \mathbb R$. \label{TwoCutBCFigure}}
\end{figure}

Here we show how to solve the multi-cut boundary conditions 
in the two-cut $(1,2)$ case. The orthonormal polynomial $\Psi_{\rm orth}(t;\zeta)$ 
in a Stokes sector $D_n$ is generally given as a superposition of independent solutions, $\widetilde\Psi_n^{(j)}(t;\zeta)$: 
\begin{align}
\Psi_{\rm orth}(t;\zeta) = \widetilde\Psi_n(t;\zeta) X^{(n)} 
=  x_1^{(n)} \widetilde\Psi_n^{(1)}(t;\zeta)+x_2^{(n)} \widetilde\Psi_n^{(2)}(t;\zeta),\qquad 
\zeta\to \infty \in D_n. 
\end{align}
However this assumption results in the 6-cut geometry of resolvent 
as shown in Fig.~\ref{TwoCutBCFigure}-a, 
even though this system is called ``two-cut''. 
Therefore, one has to choose proper Stokes multipliers 
in order to satisfy the multi-cut boundary condition and therefore 
to obtain the geometry which only includes two cuts 
as shown in Fig.~\ref{TwoCutBCFigure}-b. 
 
The multi-cut boundary condition is then given as follows: 
Since we wish to erase the cuts of orthonormal polynomial 
\eq{DefOfX} along the Stokes lines of 
\begin{align}
\theta= \pm \frac{\pi }{3}, \quad \pm \frac{5\pi }{3}, 
\end{align}
we impose the following boundary condition on the vectors $X^{(n)}$:
\begin{align}
&X^{(0)} = 
\begin{pmatrix}
0\cr x_2^{(0)}
\end{pmatrix},
\qquad 
X^{(1)} = 
\begin{pmatrix}
0 \cr x_2^{(1)}
\end{pmatrix},
\qquad 
X^{(2)} = 
\begin{pmatrix}
x_1^{(2)} \cr x_2^{(2)}
\end{pmatrix}, \nn\\
&X^{(3)} = 
\begin{pmatrix}
x_1^{(3)} \cr 0
\end{pmatrix},
\qquad 
X^{(4)} = 
\begin{pmatrix}
x_1^{(4)} \cr 0
\end{pmatrix},
\qquad 
X^{(5)} = 
\begin{pmatrix}
x_1^{(5)} \cr x_2^{(5)}
\end{pmatrix},
\end{align}
where all the $x_i^{(n)}$ appearing here are non-zero. 
This can be also expressed in the dominance profile as in Fig.~\ref{TwoCutBCFigure}-c.
That is, if the Stokes sector $D_n$ includes the following profile, 
\begin{align}
\left[
\begin{array}{c|c|c|c|c|c|c|c}
m_1 &  \cdots & m_{I-1}& \bf m_I & \not\! m_{I+1} & \cdots & \not\! m_{k-1} 
& \not\!  m_k \cr
\end{array}
\right] \in D_n,
\end{align}
then the boundary condition can be read as 
\begin{align}
\Psi_{\rm orth}(t;\zeta)=\sum_{j=1}^I x_{m_j}^{(n)}\, \widetilde\Psi_n^{(m_j)}(t;\zeta),\qquad 
x_{m_I}^{(n)}\neq 0. 
\end{align}
Since these vectors are related 
with the Stokes matrix \eq{StokesMatrix2Cut32} 
(with the $\mathbb Z_2$ symmetry condition \eq{DefAlfBetGam}) as 
\begin{align}
X^{(n)} = S_n \, X^{(n+1)},\qquad X^{(n+6)}=X^{(n)}, 
\end{align}
one obtains the following conditions on the vectors $X^{(n)}$ 
and the Stokes multipliers: 
\begin{align}
&
\begin{pmatrix}
0\cr x_2^{(0)}
\end{pmatrix}
=
\begin{pmatrix}
0\cr x_2^{(1)}
\end{pmatrix},
\qquad 
\begin{pmatrix}
0\cr x_2^{(1)}
\end{pmatrix}
=
\begin{pmatrix}
x_1^{(2)}+\beta x_2^{(2)}\cr x_2^{(2)}
\end{pmatrix},
\qquad 
\begin{pmatrix}
x_1^{(2)}\cr x_2^{(2)}
\end{pmatrix}
=
\begin{pmatrix}
x_1^{(3)}\cr \gamma x_1^{(3)}
\end{pmatrix}, \nn\\
&
\begin{pmatrix}
x_1^{(3)} \cr 0
\end{pmatrix}
=
\begin{pmatrix}
x_1^{(4)} \cr 0
\end{pmatrix},
\qquad 
\begin{pmatrix}
x_1^{(4)} \cr 0
\end{pmatrix}
=
\begin{pmatrix}
x_1^{(5)} \cr \beta x_1^{(5)}+x_2^{(5)}
\end{pmatrix},
\qquad 
\begin{pmatrix}
x_1^{(5)} \cr x_2^{(5)}
\end{pmatrix}
=
\begin{pmatrix}
\gamma x_2^{(0)} \cr x_2^{(0)}
\end{pmatrix},
\end{align}
which results in 
\begin{align}
&\beta^2=1,\qquad \gamma^2=1,\qquad 1+\beta \gamma=0,\nn\\
&x_1^{(2)}=x_1^{(3)}=x_1^{(4)}=x_1^{(5)}=\gamma x_2^{(0)}\neq 0,\qquad
x_2^{(5)}=x_2^{(0)}=x_2^{(1)}=x_2^{(2)}=\gamma x_1^{(3)}\neq 0. 
\end{align}
Therefore, the solutions which are consistent with the Hermiticity condition \eq{HermitCond2Cut} and with the 
monodromy free condition \eq{AlgCurveStokes2cut} are given as
\begin{align}
\alpha \in i \mathbb R,\qquad \beta = -\gamma = \pm 1. 
\end{align}
Consequently, the solution to the multi-cut boundary condition in the two-cut case
has a real continuum parameter. However, as we will discuss in Section \ref{RHapproachSection} with the Riemann-Hilbert approach, 
this parameter $\alpha$ causes 
``exponentially growing non-perturbative corrections'' to the perturbative 
backgrounds (e.g.~one-cut/two-cut spectral curves), 
except when $\alpha=0$. Therefore, 
the multi-cut boundary condition (with ``the small instanton condition'') 
completely fix the D-instanton chemical potentials as we advertised 
at the end of Section \ref{Section2StokesPheno}. 

\subsubsection{The multi-cut boundary-condition recursions $(r=2)$ 
\label{MultiCutBCRecursionsSubSubSectioN}}
From here, we solve the multi-cut boundary condition for an arbitrary number of cuts, $k$. 
In order to solve the constraints, we use 
the symmetric Stokes sectors (See Definition \ref{SymStokeSectorsDefinition}),
\begin{align}
\Psi_{\rm orth}(t;\zeta) \asymeq \widetilde\Psi_{2r l}(t;\zeta)\, X^{(2r l)}, \qquad \zeta \to \infty 
\in D_{2rl},\quad 
(l=0,1,2,\cdots, k-1), 
\end{align}
and its Stokes matrices, 
$S^{(\rm sym)}_{2r l}= \Gamma^{-l}\, S^{(\rm sym)}_0 \,\Gamma^l$. 
For sake of simplicity, however, we here focus on the $r=2$ cases, 
and therefore $k=5,7,9,\cdots$.%
\footnote{Here $k=3$ is special because $k<2r=4$. This case is calculated separately 
in Appendix \ref{CalculationInConcreteSystems}. }
Some of the results can be easily generalized to the general $r$ cases. 

We first read the boundary condition 
in terms of the dominance profile: 
\begin{Proposition} [The multi-cut boundary condition on $X^{(n)}$] 
The multi-cut boundary condition in the general $k$-cut cases with $r=2$ is given as 
\begin{align}
&\underline{\text{\em The general $k$-cut cases}}\nn\\
&\underline{\text{$k=4k_0 + 1$}}\quad D_{2rn}:
&&\underline{\text{$k=4k_0 + 3$}}\quad D_{2rn}:\nn\\
&\left[
\begin{array}{c|c|c}
... & n+\frac{k+5}{2}+ \lfloor\frac{k-3}{4}\rfloor & \bf n+\lfloor\frac{k+3}{4}\rfloor \cr
\hline
... &n+\frac{k+3}{2}+ \lfloor\frac{k-3}{4}\rfloor & \bf n+\lfloor\frac{k+3}{4}\rfloor \cr
\hline
... &\bf n+\lfloor\frac{k+3}{4}\rfloor & \not\!{n}+{\not\!\!\frac{k+3}{2}+ \lfloor\not\!\!\frac{k-3}{4}\rfloor} \cr
\hline
... & \bf n+\lfloor\frac{k-1}{4}\rfloor & \not\! n+\not\!\!\frac{k+3}{2}+ \lfloor\not\!\!\frac{k-3}{4}\rfloor \cr
\hline
& \vdots & \vdots \cr
\hline
... & n+\frac{k+5}{2} & \bf n+2 \cr
\hline
... &n+\frac{k+3}{2} & \bf n+2 \cr
\hline
... &\bf n+2 & \not\! n+\not\!\!\frac{k+3}{2} \cr
\hline
... & \bf n+1 & \not\! n+\not\!\!\frac{k+3}{2} \cr
\hline
... &n+\frac{k+3}{2} & \bf n+1 \cr
\hline
... &n+\frac{k+1}{2} & \bf n+1 \cr
\end{array}
\right],
\begin{picture}(-10,0)(39,36.8)
\thicklines
\put(0,80.3){\path(0,0)(1,-2)(3, 2)(5, -2)( 7, 2)( 9, -2)( 11, 2)( 13, -2)( 15, 2)( 17, -2)( 19, 2)(21, -2)( 23, 2)( 25, -2)( 27, 2)( 29, -2)}
\put(0,110){\vector(1,0){30}}
\put(0,30){\vector(1,0){30}}
\put(0,0){\path(0,0)(1,-2)(3, 2)(5, -2)( 7, 2)( 9, -2)( 11, 2)( 13, -2)( 15, 2)( 17, -2)( 19, 2)(21, -2)( 23, 2)( 25, -2)( 27, 2)( 29, -2)}
\put(0,-30.9){\vector(1,0){30}}
\end{picture}
&&\left[
\begin{array}{c|c|c}
... & \bf n+\lfloor\frac{k+7}{4}\rfloor & \not\! n+\not\!\!\frac{k+3}{2}+ \lfloor\not\!\!\frac{k-3}{4}\rfloor \cr
\hline
... & \bf n+\lfloor\frac{k+3}{4}\rfloor & \not\! n+\not\!\!\frac{k+3}{2}+ \lfloor\not\!\!\frac{k-3}{4}\rfloor \cr
\hline
... & n+\frac{k+3}{2}+ \lfloor\frac{k-3}{4}\rfloor & \bf n+\lfloor\frac{k+3}{4}\rfloor \cr
\hline
...&n+\frac{k+1}{2}+ \lfloor\frac{k-3}{4}\rfloor & \bf n+\lfloor\frac{k+3}{4}\rfloor \cr
\hline
& \vdots & \vdots \cr
\hline
... & n+\frac{k+5}{2} & \bf n+2 \cr
\hline
... &n+\frac{k+3}{2} & \bf n+2 \cr
\hline
... & \bf n+2 & \not\! n+\not\!\!\frac{k+3}{2} \cr
\hline
... & \bf n+1 & \not\! n+\not\!\!\frac{k+3}{2} \cr
\hline
... &n+\frac{k+3}{2} & \bf n+1 \cr
\hline
... &n+\frac{k+1}{2} & \bf n+1 \cr
\end{array}
\right].
\begin{picture}(0,0)(39,36.8)
\thicklines
\put(0,110){\path(0,0)(1,-2)(3, 2)(5, -2)( 7, 2)( 9, -2)( 11, 2)( 13, -2)( 15, 2)( 17, -2)( 19, 2)(21, -2)( 23, 2)( 25, -2)( 27, 2)( 29, -2)}
\put(0,80.3){\vector(1,0){30}}
\put(0,30){\vector(1,0){30}}
\put(0,0){\path(0,0)(1,-2)(3, 2)(5, -2)( 7, 2)( 9, -2)( 11, 2)( 13, -2)( 15, 2)( 17, -2)( 19, 2)(21, -2)( 23, 2)( 25, -2)( 27, 2)( 29, -2)}
\put(0,-30.9){\vector(1,0){30}}
\end{picture}
\end{align}
Equivalently, the components of $X^{(4n)}$ ($r=2$ and $k\geq 5$) is given as 
\begin{align}
x_{n+i}^{(4n)} &\neq 0 \qquad (i=1,2,\cdots,\Bigl\lfloor\frac{k+3}{4}\Bigr\rfloor), \nn\\
x_{n+\frac{k+1}{2}+i}^{(4n)} &= 0 \qquad (i=1,2,\cdots, \Bigl\lfloor\frac{k+1}{4}\Bigr\rfloor), \label{BCcomp3233}
\end{align}
for $n=0,1,2,\cdots,k-1$. The constraints on the Stokes matrices are then imposed 
by Eq.~\eq{PreRecursion}. \label{MCBCProposition}
\end{Proposition}
It is then convenient to introduce a new vector 
$Y^{(4n)} =\bigl(y_{n,j}\bigr)_{j=1}^k\equiv \Gamma^n X^{(4n)}$, 
since the above boundary condition becomes simpler: 
\begin{align}
X^{(4n)} = 
\begin{pmatrix}
\vdots \cr
x_{n+1}^{(4n)} \neq 0 \cr
\vdots \cr
x_{n+\lfloor\frac{k+3}{4}\rfloor}^{(4n)} \neq 0 \cr
\vdots \cr
x_{n+\frac{k+3}{2}}^{(4n)} = 0\cr 
\vdots \cr
x_{n+ \frac{k+3}{2} + \lfloor\frac{k-3}{4}\rfloor}^{(4n)} = 0 \cr
\vdots
\end{pmatrix}, \qquad 
Y^{(4n)} = 
\begin{pmatrix}
y_{n,1} \neq 0 \cr
\vdots \cr
y_{n,\lfloor\frac{k+3}{4}\rfloor} \neq 0 \cr
\vdots \cr
y_{n,\frac{k+3}{2}} = 0\cr 
\vdots \cr
y_{n, \frac{k+3}{2} + \lfloor\frac{k-3}{4}\rfloor} = 0 \cr
\vdots
\end{pmatrix}. \label{BCforXYinMatrix}
\end{align}
Note that the periodicity of index $n$ follows:
\begin{align}
X^{(4n)} = X^{(4(n+k))},\qquad Y^{(4n)} =Y^{(4(n+k))},\qquad y_{n+k,j} = y_{n,j}. 
\end{align}
In terms of the vector $Y^{(4n)}$, 
the constraints on the Stokes multipliers Eq.~\eq{PreRecursion} are expressed as 
\begin{align}
X^{(4n)} = S_{4n}^{(\rm sym)} \, X^{(4(n+1))} \qquad \Leftrightarrow \qquad 
Y^{(4n)} = \bigl(S_{0}^{(\rm sym)} \Gamma^{-1}\bigr)\, Y^{(4(n+1))}. \label{BCRecurMatrix}
\end{align}
Therefore, in terms of components, we obtain the following recursive relations for $y_{n,i}$: 
\begin{align}
y_{n,i} = y_{n+1,i-1} + \sum_{j=1}^k s_{0,i,j}^{(\rm sym)}\, y_{n+1,j-1},\qquad y_{n+k,j} = y_{n,j}. \label{CompBCRecurMatrix}
\end{align}
This is the central equations for the multi-cut boundary condition. 
After some tedious calculations, the multi-cut boundary condition turns out to be
the following simple form:
\begin{Theorem} [The multi-cut BC recursions]
The recursion relation Eq.~\eq{CompBCRecurMatrix} with the multi-cut boundary condition 
Eq.~\eq{BCforXYinMatrix} in the $(k,r;\gamma)=(2m+1,2;2)$ case is equivalent to the following two recursion equations 
for $\{y_{n,1}\}_{n\in\mathbb Z}$: 
\begin{align}
\mathcal F_k[y_{n,1}] &= y_{n+m,1}+\sum_{j=1}^{\lfloor \frac{m}{2}\rfloor} s_{1,m+2-j,1+j}\times y_{n+2j-1,1}
+\sum_{j=1}^{\lfloor \frac{m+1}{2}\rfloor} s_{3,m+3-j,1+j}\times y_{n+2j-2,1}=0, \nn\\
\mathcal G_k[y_{n,1}] &= -y_{n,1}+\sum_{j=1}^{\lfloor \frac{m}{2}\rfloor} s_{0,k+1-j,1+j}\times y_{n+2j,1}+
\sum_{j=1}^{\lfloor \frac{m+1}{2}\rfloor} s_{2,k+2-j,1+j}\times y_{n+2j-1,1}=0, \label{MultiCutBC}
\end{align}
and linear expressions of the components $\{y_{n,i}\}_{
1\leq i \leq k}^{n\in\mathbb Z}$ 
in terms of $\{y_{n,1}\}_{n\in\mathbb Z}$: 
\begin{align}
y_{n,i} = y_{n,i}\bigl(\{y_{l,1}\}_{l\in \mathbb Z}\bigr).  \label{LinearEqForYBC}
\end{align}
Note that the coefficients in Eqs.~\eq{MultiCutBC} are understood as modulo $k$, say $s_{2,i,j}=s_{2,i+k,j}$. \label{ThmMCBCRec}
\end{Theorem}
The explicit expression for Eq.~\eq{LinearEqForYBC} is a bit long and therefore shown 
in Appendix \ref{ExamplesProfilesSections} with some examples. 
These recursive equations are the physical constraints which should be solved in addition to 
the basic constraints discussed in Section \ref{ThreeConstraints11}. 
In the general cases, the vectors $Y^{(n)}$ in terms of $\{y_{n,1}\}_{n\in\mathbb Z}$ are denoted as 
\begin{align}
Y^{(n)}\bigl(\{y_{l,1}\}_{l\in \mathbb Z}\bigr)\equiv 
\Bigl(y_{n,i}\bigl(\{y_{l,1}\}_{l\in \mathbb Z}\bigr)\Bigr)_{i=1}^k. 
\label{VectorYSolutions}
\end{align}
An important point is that all the Stokes multipliers $s_{l,i,j}$ in this expression are {\em fine Stokes multipliers}. 
Some detail derivation of this theorem can be found in \cite{CIY2}. 

Finally we also make a comment on the boundary condition 
for general $r\,(=2,3,\cdots)$.  In terms of the dominance profile, 
they are expressed as 
\begin{align}
D_{2r(n-1)} \supset
\left[
\begin{array}{c|c|c|c|c|c|c}
&& & & & \vdots & \vdots \cr
\hline
&& & & &A_3^{(n+1)} & \bf n+1 \cr
\hline
&& & & &A_{2r-1}^{(n)} & \bf n+1 \cr
\hline
&& & & & \bf n+1 & \not\!\! A_{2r-1}^{(n)} \cr
\hline
&& & &\bf n+1 & \not\!\! A_{2r-3}^{(n)}& \not\!\! A_{2r-1}^{(n)} \cr
\hline
&& &\bf n+1 & \not\!\! A_{2r-5}^{(n)} & \not\!\! A_{2r-1}^{(n)} & \not\!\! A_{2r-3}^{(n)} \cr
\hline
&& \rotatebox{70}{$\bf \ddots$} & \rotatebox{70}{$\bf \ddots$}&\rotatebox{70}{$\bf \ddots$} &  & \vdots \cr
\hline
&\bf n+1&\not\!\! A_{5}^{(n)}  & \not\!\! A_{2r-1}^{(n)}& &  & \vdots \cr
\hline
\bf n+1& \not\!\! A_{3}^{(n)}& \not\!\! A_{2r-1}^{(n)} & & &  & \vdots \cr
\hline
\bf n&\not\!\! A_{2r-1}^{(n)}& \not\!\! A_{3}^{(n)} & & &  & \vdots \cr
\hline
&\bf n& \not\!\! A_{2r-3}^{(n)}  & \not\!\! A_{3}^{(n)} & &  & \vdots \cr
\hline
&& \bf \ddots & \ddots &\ddots &  & \vdots \cr
\hline
&& &\bf n & \not\!\! A_7^{(n)} & \not\!\! A_3^{(n)} & \not\!\! A_5^{(n)} \cr
\hline
&& & &\bf n & \not\!\! A_5^{(n)} & \not\!\! A_3^{(n)} \cr
\hline
&& & & & \bf n & \not\!\! A_3^{(n)} \cr
\hline
&& & & &A_3^{(n)} & \bf n \cr
\hline
&& & & &A_{2r-1}^{(n-1)} & \bf n \cr
\end{array}
\right],
\begin{picture}(-10,0)(39,7)
\thicklines
\put(0,129){\vector(1,0){30}}
\put(0,0){\path(0,0)(1,-2)(3, 2)(5, -2)( 7, 2)( 9, -2)( 11, 2)( 13, -2)( 15, 2)( 17, -2)( 19, 2)(21, -2)( 23, 2)( 25, -2)( 27, 2)( 29, -2)}
\put(0,-128.5){\vector(1,0){30}}
\end{picture}
\label{GeneralrDiagram}
\end{align}
Here we define 
\begin{align}
A_i^{(n)}\equiv j_{i+2r(n-1),k}=n+\lfloor i/2\rfloor m_1
\end{align}
with Theorem \ref{TheoremComponentsOfProfiles}. 
Note that $A_{i+2r}^{(n)}=A_{i}^{(n+1)}$ and $A_0^{(n)}=A_1^{(n)}=n$. 
Therefore if the Stokes sector $D_{2r(n-1)}$ includes the indices $A_{i}^{(n)}$ 
of $i=3,5,\cdots,2r-1,2r+3,2r+5,\cdots,$ in the profiles, then we impose 
\begin{align}
x_{A_i^{(n)}}^{(2r(n-1))}=0\quad (i=3,5,\cdots)\qquad\text{and}
\qquad x_{n+i}^{(2r(n-1))}\neq 0\quad (i=0,1,\cdots). 
\end{align}
The ending points of these series (about $i$) depend on how many segments 
$D_{2r(n-1)}$ includes. 
This general classification could be tedious and we shall leave it for future study.

\subsubsection{The complementary boundary conditions \label{ComplementaryBCSubSection}}

It is suggestive to show which Stokes multipliers appear in the recursive equations Eqs.~\eq{MultiCutBC}. 
Here we show them by bold type in the profile of $\mathcal J_{k,2}^{(\rm sym)}$
(i.e.~Theorem \ref{TheoremIndexStokesMultipliers}):
\begin{align}
\left.
\begin{array}{c|c|c|c|c|c|c|c|c|c|c|c}
...  & \it k-1) & \bf ( 4 & \bf  \frac{k-1}{2}) & ( \it \frac{k+7}{2} & \it k) & (\bf 3 & \bf \frac{k+1}{2}) & (\it \frac{k+5}{2} &\it  1) & (\bf 2 & \bf \frac{k+3}{2}) \cr
\hline
...  & (\bf 4 & \bf k-1) & ( \it \frac{k+7}{2} & \it \frac{k-1}{2}) & (\bf 3 & \bf k) & (\it \frac{k+5}{2} &\it  \frac{k+1}{2}) & (\bf 2 & \bf 1) &  \frac{k+3}{2} \cr
\hline
...  & \bf \frac{k-3}{2}) & (\it \frac{k+7}{2} & \it k-1) & (\bf 3 & \bf \frac{k-1}{2}) & (\it \frac{k+5}{2} &\it k) & (\bf 2 & \bf \frac{k+1}{2}) &(\it \frac{k+3}{2} & \it 1) \cr
\hline
...   & (\it \frac{k+7}{2} & \it \frac{k-3}{2}) & (\bf 3 & \bf k-1) & (\it \frac{k+5}{2} & \it \frac{k-1}{2}) & (\bf 2 & \bf k) & (\it \frac{k+3}{2} &\it  \frac{k+1}{2}) & 1 \cr
\end{array}
\right]\!\!\!\!
\begin{array}{l}
:3 \cr 
:2 \cr 
:1 \cr 
:0
\end{array}. \label{ConstraintIndicesBCSolutionsProfile}
\end{align}
Note that exactly a half of multipliers 
$s_{l,i,j} \leftrightarrow (j|i)_l\in \mathcal J_{k,2}^{(\rm sym)}$ 
are picked up by the recursion. One may have the following question: {\em Are there similar equations 
which pick up exactly another half of the multipliers?} This can be positively answered. In general, we can expect 
that there are $(r-1)$ similar equations, each of which picks up a different set of Stokes multipliers.%
\footnote{This anticipation is further explicitly shown in the fractional-superstring critical points $(\gamma=r-2)$ with 
arbitrary Poincar\'e index $r$ \cite{CIY3}. }
These equations come from {\em complementary boundary conditions} 
which are given by the multi-cut boundary condition (Definition \ref{MultiCutBCDefinition}) 
with different initial angles $\chi_0$: 
\begin{align}
\chi_0 = \frac{\pi}{k}+ \frac{2\pi a}{rk} \qquad (a=1,2,\cdots,r-1).  \label{ComplabA}
\end{align}
The case of $a=0$ is the original multi-cut boundary condition Eq.~\eq{MCBCchi0}. 
In fact, in the $r=2$ case, the vectors $Y^{(4n)}$ in the recursion equation Eq.~\eq{BCRecurMatrix} are replaced by the following $\widetilde Y^{(4n)}$ 
\begin{align}
\widetilde Y^{(4n)} = 
\begin{pmatrix}
\widetilde y_{n,1} = 0 \cr
\vdots \cr
\widetilde y_{n,\lfloor\frac{k+3}{4}\rfloor} = 0 \cr
\vdots \cr
\widetilde y_{n,\frac{k+3}{2}} \neq 0\cr 
\vdots \cr
\widetilde y_{n, \frac{k+3}{2} + \lfloor\frac{k-3}{4}\rfloor} \neq 0 \cr
\vdots
\end{pmatrix},\label{BCforXYinMatrixComp}
\end{align}
satisfying the complementary boundary condition of Eq.~\eq{ComplabA} ($a=1$),
and consequently we found the following different recursion equations with $a=1$: 

\begin{Theorem} [The complementary BC recursion equations]
The recursion relation Eq.~\eq{CompBCRecurMatrix} 
with the complementary boundary condition 
($a=1$ of Eq.~\eq{ComplabA}) 
in the $(k,r;\gamma)=(2m+1,2;2)$ case is equivalent to the following two recursion equations 
for $\{\widetilde y_{n,m+2}\}_{n\in\mathbb Z}$: 
\begin{align}
\widetilde {\mathcal F}_k[\widetilde y_{n,m+2}] 
&=  \widetilde y_{m+n,m+2}
+\sum_{j=1}^{\lfloor \frac{m}{2}\rfloor}
s_{3,k+2-j,m+2+j}
\times \widetilde y_{n+2j-1,m+2} + \nn\\
&\qquad\qquad \qquad \quad+\sum_{j=1}^{\lfloor \frac{m+1}{2}\rfloor} 
s_{1,k+2-j,m+1+j}
\times \widetilde y_{n+2j-2,m+2}=0, \label{COMPRecEqFFFFFFFFF}\\
\widetilde{\mathcal G}_k[\widetilde y_{n,m+2}] 
&= -\widetilde y_{n,m+2}
+\sum_{j=1}^{\lfloor \frac{m}{2}\rfloor} 
s_{2,m+2-j,m+2+j}\times \widetilde y_{n+2j,m+2}
+ \nn\\
&\qquad\qquad \qquad \quad +\sum_{j=1}^{\lfloor \frac{m+1}{2}\rfloor} 
s_{0,m+2-j,m+1+j}\times \widetilde y_{n+2j-1,m+2}=0, \label{COMPRecEqGGGGGGGGG}
\end{align}
and linear expressions of the components $\{\widetilde y_{n,i}\}_{
1\leq i \leq k}^{n\in\mathbb Z}$ 
in terms of $\{\widetilde y_{n,m+2}\}_{n\in\mathbb Z}$: 
\begin{align}
\widetilde y_{n,i} = \widetilde y_{n,i}\bigl(\{\widetilde y_{l,m+2}\}_{l\in \mathbb Z}\bigr). 
\label{LinearEqForYBCComp}
\end{align}
Note that the coefficients in Eqs.~\eq{COMPRecEqGGGGGGGGG} are understood as modulo $k$, say $s_{2,i,j}=s_{2,i+k,j}$. \label{ThmConplementaryBCRec}
\end{Theorem}
The explicit expression for Eq.~\eq{LinearEqForYBCComp} is a bit long and therefore shown 
in Appendix \ref{ExamplesProfilesSections}. 
The vectors $\widetilde Y^{(n)}$ in terms of $\{\widetilde y_{n,m+2}\}_{n\in\mathbb Z}$ are also denoted by 
\begin{align}
\widetilde Y^{(n)} \bigl(\{\widetilde y_{l,m+2}\}_{l\in\mathbb Z}\bigr) 
=\Bigl(
\widetilde y_{n,i} \bigl(\{\widetilde y_{l,m+2}\}_{l\in\mathbb Z}\bigr)
\Bigr)_{i=1}^k, \label{YInTermsOfY1k3over2AllTheComponents}
\end{align}
In these recursion equations, exactly the complementary set of the multipliers 
$s_{l,i,j} \leftrightarrow (j|i)_l\in \mathcal J_{k,2}^{(\rm sym)}$ 
are picked up. 

\subsubsection{Useful reparametrization of the Stokes multipliers}

It is again suggestive to express these recursion equations as follows:%
\footnote{We point out the following interesting facts about these expressions. 
The algebraic equations defined by the recursion equations Eq.~\eq{FGRecInTermsOfTheta}, 
\begin{align}
\mathcal F_k(y) &\equiv y^{-n}\, \mathcal F_k\bigl[\{y_{j,1}\to y^j\}_{j\in\mathbb Z}\bigr]
= y^{m}+ \sum_{n=1}^{m} \theta_n \, y^{m-n}=0, \nn\\
\mathcal G_k(y) &\equiv y^{-n}\, \mathcal G_k\bigl[\{y_{j,1}\to y^j\}_{j\in\mathbb Z}\bigr]
=-\Bigl( 1 
+ \sum_{n=1}^{m} \theta_i^*\,  y^n \Bigr)=0
\end{align}
satisfy the following hermiticity relation: 
\begin{align}
\bigl[\mathcal F_k(y)\bigr]^{*} 
= - y^{-m}\mathcal G_k(y),\qquad \text{if $\qquad y^k=1$}. 
\end{align}
The same thing also happen for Eq.~\eq{FGRecInTermsOfThetaComp}. 
This therefore suggests that the solutions to the recursions $\{y_{n,1}\}_{n \in\mathbb Z}$ are given by 
$k$-th roots of unity, $y^k=1$. }
\begin{align}
\mathcal F_k \bigl[y_{n-m,1}\bigr] &= y_{n,1} 
+ \sum_{i=1}^{m} \theta_i\,  y_{n-i,1}=0, \nn\\
\mathcal G_k \bigl[y_{n,1}\bigr] &= -\Bigl( y_{n,1} 
+ \sum_{i=1}^{m} \theta_i^*\,  y_{n+i,1}\Bigr)=0; \label{FGRecInTermsOfTheta}\\
\widetilde{\mathcal F}_k \bigl[{\widetilde y}_{n-m,m+2}\bigr] &= {\widetilde y}_{n,m+2} 
+ \sum_{i=1}^{m} {\widetilde \theta}_i\,  {\widetilde y}_{n-i,m+2}=0, \nn\\
\widetilde{\mathcal G}_k \bigl[{\widetilde y}_{n,m+2}\bigr] &= -\Bigl( {\widetilde y}_{n,m+2} 
+ \sum_{i=1}^{m} {\widetilde \theta}_i^*\,  {\widetilde y}_{n+i,m+2}\Bigr)=0. 
\label{FGRecInTermsOfThetaComp}
\end{align}
The complex conjugation $\theta_n^*$ (and ${\widetilde\theta}_n^*$) comes from the hermiticity condition 
of Stokes multipliers Eq.~\eq{HermiticityCondStokes}. 
It is also interesting to see the index $n$ of the parameters $\theta_n$ (and ${\widetilde \theta}_n$)
in terms of the dominance profile: 
\begin{align}
\left.
\begin{array}{ccccccccccccc}
\cdots  & \it m-5 & )( & \bf m-4 & )(  & \it m-3 &  )( &\bf  m-2 &  )( & \it m-1 &  )( &\bf m  &)\cr
\hline
\cdots  & )( &  \bf 5 & )(  & \it 4 & )( & \bf 3  & )( &\it 2 & )( & \bf 1   &  ) & \cr
\hline
\cdots  & \bf m-5 & )( & \it m-4 & )( & \bf m-3 &  )( &\it  m-2 &  )( & \bf  m-1 &  )(&\it m  &)\cr
\hline
\cdots   & )( &\it 5  & )( & \bf 4 & )( &\it 3  & )( & \bf 2 &  )(&\it 1  &  ) &
\end{array}
\right]\!\!\!\!
\begin{array}{l}
:3 \cr 
:2 \cr 
:1 \cr 
:0
\end{array}. \label{ConstraintIndicesBCSolutionsProfileLLL}
\end{align}
Here bold type is again the coefficients of the multi-cut BC recursions Eqs.~\eq{MultiCutBC}. 
An important thing here is that these complementary boundary conditions are used to 
obtain explicit solutions of the Stokes multipliers (although they are not related to the physical 
boundary conditions). 

Finally, in order to write the explicit relation between the Stokes multipliers and 
the parameters $\theta_n$, 
we introduce integers $L_{l,i,j}\ (l,i,j\in \mathbb Z)$ which are defined as 
\begin{align}
0\leq L_{l,i,j} <k,\qquad L_{l,i,j} \equiv (-1)^{l-1}(i-j) \mod k. 
\end{align}
In particular, we pick up the following set of indices $(l;i,j)$:
\begin{align}
\underline{\text{$k=4k_0+1:$}}\quad L_{l,i,j}+ \Bigl\lfloor\frac{l-1}{2} \Bigr\rfloor \in 2\mathbb Z+1;
\qquad 
\underline{\text{$k=4k_0+3:$}}\quad L_{l,i,j}+ \Bigl\lfloor\frac{l}{2} \Bigr\rfloor \in 2\mathbb Z, 
\label{ConstraintIndicesBCSolutions}
\end{align}
and the relation is given as follows:
\begin{Proposition} 
[The $\theta_n$ parametrization]
The fine Stokes multipliers $s_{l,i,j}$ are parametrized by $k-1$ complex parameters 
$\{\theta_n,\widetilde \theta_n\}_{n=1}^{\lfloor \frac{k}{2}\rfloor}$ as 
\begin{align}
\text{$(j|i) \in \mathcal J_{k,2}^{(\rm sym)}$ satisfying Eq.~\eq{ConstraintIndicesBCSolutions}}: \qquad 
s_{l,i,j} = 
\left\{
\begin{array}{ll}
\theta_{L_{l,i,j}}& (l=1,3)\cr
-\theta^*_{L_{l,i,j}}&(l=0,2) 
\end{array}
\right.,\label{ThetaNotation1}
\end{align}
and 
\begin{align}
\text{$(j|i) \in \mathcal J_{k,2}^{(\rm sym)}$ \underline{not} satisfying Eq.~\eq{ConstraintIndicesBCSolutions}}: \qquad 
s_{l,i,j} = 
\left\{
\begin{array}{ll}
\widetilde \theta_{L_{l,i,j}}& (l=1,3)\cr
-\widetilde \theta^*_{L_{l,i,j}}&(l=0,2) 
\end{array}
\right.. \label{ThetaNotation2}
\end{align}
Therefore, this is a one to one correspondence up to the hermiticity condition Eq.~\eq{HermiCondSec437}. 
\label{ThetaParamProposition}
\end{Proposition}

\subsection{Solutions in the general $k$-cut cases \label{AnsatzSolutionsSubSection}}
Before solving the boundary conditions, here we summarize the equations to be solved: 
After imposing the $\mathbb Z_k$ symmetry condition \eq{ZkSymCondition2323}, 
\begin{align}
\underline{\text{$\mathbb Z_k$ symmetry:}}&&S^{(\rm sym)}_{2r l}= \Gamma^{-l}\, S^{(\rm sym)}_0 \,\Gamma^l,\qquad (l=1,2,\cdots,k-1) \label{ZkSymCondSectionAnsatzSolutions}
\end{align}
the system becomes 
\begin{align}
\underline{\text{Multi-cut BC recursion:}}&& Y^{(4n)} = \bigl(S_{0}^{(\rm sym)} \Gamma^{-1}\bigr)\, Y^{(4(n+1))} \label{MultiCutBCRec3322211}\\
\underline{\text{Monodromy free condition:}}&&
\bigl(S_0^{(\rm sym)}\,\Gamma^{-1}\bigr)^k = I_k, \qquad 
 \label{MFcond2} \\
 \underline{\text{Hermiticity condition:}}&&
 S_n^*= \Delta\Gamma\, S_{(2r-1) k-n}^{-1}\, \Gamma^{-1}\Delta, \label{HermiCondSec437}
\end{align}
for $n=0,1,2,\cdots,k-1$. 
In general, the patterns of solutions become complicated if we increase the number of cuts. 
However, we here show two kinds of special solutions which 
can be generalized to the cases with an arbitrary number of cuts. 

Before showing explicit solutions, we mention a key point of solving the above equations. 
The main difficulty is from the monodromy free condition \eq{MFcond2}. 
We here note that following fact:%
\footnote{The opposite is non-trivial but can be shown by using Jordan normal form. }
\begin{Lemma} 
If the matrix $S_{0}^{(\rm sym)} \Gamma^{-1}$ 
is diagonalizable and its eigenvalues $\lambda_j$ are $k$-th roots of unity $\lambda_j^k=1$, 
then the monodromy free condition \eq{MFcond2} is satisfied. The opposite is also true. 
\label{MFcond3}
\end{Lemma}
Our strategy of finding solutions is now to show 
that $S_{0}^{(\rm sym)} \Gamma^{-1}$ is diagonlizable. 
Below we list two types of explicit solutions: 

\subsubsection{Discrete solutions and configurations of avalanches \label{AvalanchesSubSubSection}}

\begin{Theorem} [Discrete Solution]
The following Stokes multipliers $s_{l,i,j}$ 
(written with $\theta_n,\widetilde \theta_n$ of Proposition \ref{ThetaParamProposition}) 
are solutions to the multi-cut boundary condition 
in the $\mathbb Z_k$ symmetric $(\hat p,\hat q)=(1,1)$ $k$-cut critical points 
($k=2m+1, \gamma=r=2$):
\begin{align}
\theta_n = {\sigma}_{n}
(\{-\omega^{n_j}\}_{j=1}^{m}),\qquad 
\widetilde \theta_n
= {\sigma}_{n}
(\{-\omega^{\widetilde n_j}\}_{j=1}^{m}),\qquad 
(n=1,2,\cdots,m)\label{DiscreteSolTheta}
\end{align}
with the symmetric polynomials $\sigma_n$ among $\{x_i\}_{i=1}^N$ of degree $n$:%
\footnote{Here $\sigma_n$ stands for the symmetric polynomials. Do not be confused with the Pauli matrices. }
\begin{align}
\sigma_n(\{x_i\}_{i=1}^N) \equiv \sum_{1\leq i_1<i_2<\cdots<i_n\leq N} x_{i_1}x_{i_2}\cdots x_{i_n}, 
\label{SymPolynomialsIKEKT}
\end{align}
if and only if the integers $\bigl(n_1,n_2,\cdots,n_{m}; \widetilde n_1,\widetilde n_2,\cdots, \widetilde n_{m}\bigr)$ of Eq.~\eq{DiscreteSolTheta} satisfy 
\begin{align}
\left\{
\begin{array}{ll}
\ds n_i\not\equiv n_j, \quad \widetilde n_i \not\equiv \widetilde n_j \mod k &\ds (i\neq j), \cr
\ds -\sum_{j=1}^{m} 
\bigl(n_j+\widetilde n_j\bigr) \not\equiv n_i, \,\,
\widetilde n_i \mod k &\ds 
(i = 1,2,\cdots, m), 
\end{array}
\right..
\label{DiscreteSolIndices}
\end{align}
\label{TheoremDiscreteSolutions}
\end{Theorem}
Comments on the conditions Eqs.~\eq{DiscreteSolIndices} are following:
\begin{itemize}
\item In this solution, one can find the $(k-1)$ explicit eigenvectors 
of the matrix $S_{0}^{(\rm sym)} \Gamma^{-1}$:
\begin{align}
\omega^{-n_i} {\mathcal Y}_i
= \bigl(S_{0}^{(\rm sym)} \Gamma^{-1}\bigr)\, {\mathcal Y}_i,
\qquad \omega^{-{\widetilde n}_j} \widetilde{\mathcal Y}_j 
= \bigl(S_{0}^{(\rm sym)} \Gamma^{-1}\bigr)\, \widetilde {\mathcal Y}_j,\qquad (i,j=1,2,\cdots, m),
\end{align}
with the vectors of the BC recursion equations, Eq.~\eq{VectorYSolutions} 
and Eq.~\eq{YInTermsOfY1k3over2AllTheComponents}:
\begin{align}
\mathcal Y_i \equiv Y^{(0)}\bigl[\{ y_{n,1}\to \omega^{n\times n_i}\}_{n\in\mathbb Z}\bigr],\qquad 
\widetilde{\mathcal Y}_j \equiv {\widetilde Y}^{(0)}\bigl[\{ {\widetilde y}_{n,m+2}\to \omega^{n \times \widetilde n_j}\}_{n\in\mathbb Z}\bigr]. 
\end{align}
They are distinct only when 
$n_i\not\equiv n_j \ (i\neq j)$ and $\widetilde n_i \not\equiv \widetilde n_j \ (i\neq j)$ with modulo $k$. 
\item Noting that $\det S_{0}^{(\rm sym)} \Gamma^{-1}=1$, one concludes 
that the eigenvalue of the remaining eigenvector is given by $\omega^{-n_0}$ 
with $n_0 \equiv -\sum_{j=1}^m (n_j + \widetilde n_j)$. This eigenvector becomes distinct only when 
$n_i \not\equiv n_0 \not\equiv \widetilde n_j \ (i,j=1,2,\cdots,m)$.%
\footnote{As a side remark, here we show 
the eigenvector $\mathcal Y_0$ of the eigenvalue $\eta^{-1} \equiv \omega^{-n_0}$ 
when $n_j=\widetilde n_j\ (j=1,2,\cdots,m$: 
\begin{align}
\mathcal Y_0 &\equiv Y^{(0)}
\bigl(\{ y_{j,1}\to \eta^{j}\}_{j\in\mathbb Z}\bigr) + \widetilde Y^{(0)}
\bigl(\{{\widetilde y}_{j,m+2 }\to \eta^{j}(-1)^{m}\eta^{1/2}\}_{j\in\mathbb Z} \bigr) \nn\\
&= 
\Bigl(
1,  \eta, \cdots, \eta^{\lfloor \frac{m}{2}\rfloor}, \underbrace{0,  \cdots, 0}_{\rm II}, (-1)^{m}\eta^{1/2},
\cdots, (-1)^{m}\eta^{1/2+\lfloor \frac{m-1}{2}\rfloor},
\underbrace{0,\cdots, 0}_{\rm IV}
\Bigr)^{t}. 
\end{align}
Note that all the components of $\mathcal V_0$ in the region II and region IV vanish (See Eqs.~\eq{RegionABCDIV} for definition of the regions).}
\end{itemize}
With these considerations, one can prove the above theorem. 
Below we show a graphical expression of the conditions 
Eq.~\eq{DiscreteSolIndices} in terms of Young diagram. 
\begin{itemize}
\item [1.] The following transformation is an automorphism among the solutions to the conditions \eq{DiscreteSolIndices}: 
\begin{align}
&\bigl(n_1,n_2,\cdots,n_{m};
\widetilde n_1,\widetilde n_2,\cdots,\widetilde n_{m}\bigr)\nn\\
&\qquad \to \qquad \bigl(n_1+1,n_2+1,\cdots,n_{m}+1;
\widetilde n_1+1,\widetilde n_2+1,\cdots,\widetilde n_{m}+1\bigr),
\label{ShiftOfYoungDiagram231}
\end{align}
which also maps $n_0 \,\bigl(\equiv -\sum_{j=1}^{m} 
(n_j+\widetilde n_j)\bigr)$ as $n_0 \to n_0+1$. 
\item[2.] By choosing the following representative of the solutions as $n_0 \equiv 0$, 
and by properly choosing the ordering of the indices, one can rewrite 
the conditions \eq{DiscreteSolIndices} as 
\begin{align}
\sum_{j=1}^{\lfloor \frac{k}{2}\rfloor} \bigl(n_j+\widetilde n_j\bigr) 
\equiv 0, \qquad 
\left\{
\begin{array}{l}
1 \leq n_1 < n_2<\cdots < n_{\lfloor \frac{k}{2} \rfloor}\leq k-1, \cr
1 \leq \widetilde n_1 < \widetilde n_2<\cdots < \widetilde n_{\lfloor \frac{k}{2} \rfloor}\leq k-1
\end{array}
\right..
\label{YoungDiagramRule222}
\end{align}
\item [3.] Therefore, these indices can be expressed in terms of Young diagram. 
Here is an example (a solution in the $11$-cut case). 
\begin{align}
(n_1,n_2,n_3,n_4,n_5)=(1,2,4,6,9)\qquad \Leftrightarrow\qquad 
\begin{picture}(120,50)(-10,0)
\matrixput(0,0)(10,0){11}(0,10){6}{\line(1,0){10}}
\matrixput(0,0)(10,0){12}(0,10){5}{\line(0,1){10}}
\thicklines
\matrixput(0,0)(10,0){1}(0,10){6}{\textcolor{blue}{\line(1,0){10}}}
\matrixput(0,0)(10,0){2}(0,10){5}{\textcolor{blue}{\line(0,1){10}}}
\matrixput(0,0)(10,0){1}(0,10){5}{\textcolor{blue}{\line(1,1){10}}}
\matrixput(0,10)(10,0){1}(0,10){5}{\textcolor{blue}{\line(1,-1){10}}}
\matrixput(10,10)(10,0){1}(0,10){5}{\textcolor{blue}{\line(1,0){10}}}
\matrixput(10,10)(10,0){2}(0,10){4}{\textcolor{blue}{\line(0,1){10}}}
\matrixput(10,10)(10,0){1}(0,10){4}{\textcolor{blue}{\line(1,1){10}}}
\matrixput(10,20)(10,0){1}(0,10){4}{\textcolor{blue}{\line(1,-1){10}}}
\matrixput(20,20)(10,0){2}(0,10){4}{\textcolor{blue}{\line(1,0){10}}}
\matrixput(20,20)(10,0){3}(0,10){3}{\textcolor{blue}{\line(0,1){10}}}
\matrixput(20,20)(10,0){2}(0,10){3}{\textcolor{blue}{\line(1,1){10}}}
\matrixput(20,30)(10,0){2}(0,10){3}{\textcolor{blue}{\line(1,-1){10}}}
\matrixput(40,30)(10,0){2}(0,10){3}{\textcolor{blue}{\line(1,0){10}}}
\matrixput(40,30)(10,0){3}(0,10){2}{\textcolor{blue}{\line(0,1){10}}}
\matrixput(40,30)(10,0){2}(0,10){2}{\textcolor{blue}{\line(1,1){10}}}
\matrixput(40,40)(10,0){2}(0,10){2}{\textcolor{blue}{\line(1,-1){10}}}
\matrixput(60,40)(10,0){3}(0,10){2}{\textcolor{blue}{\line(1,0){10}}}
\matrixput(60,40)(10,0){4}(0,10){1}{\textcolor{blue}{\line(0,1){10}}}
\matrixput(60,40)(10,0){3}(0,10){1}{\textcolor{blue}{\line(1,1){10}}}
\matrixput(60,50)(10,0){3}(0,10){1}{\textcolor{blue}{\line(1,-1){10}}}
\put(-10,1){\footnotesize 1}
\put(-10,11.7){\footnotesize 2}
\put(-10,22){\footnotesize 4}
\put(-10,32){\footnotesize 6}
\put(-10,42.5){\footnotesize 9}
\end{picture}\,\, . 
\end{align}
That is, the $i$-th row from the bottom has $n_i$ sky boxes in the diagram. 
\end{itemize}
We also draw $m\times k$ total boxes for later convenience. 
In particular, the upper-left Young diagram (written with
\begin{picture}(11,11)(0,1)
\thicklines
\matrixput(0,0)(10,0){1}(0,10){2}{\textcolor{blue}{\line(1,0){10}}}
\matrixput(0,0)(10,0){2}(0,10){1}{\textcolor{blue}{\line(0,1){10}}}
\matrixput(0,0)(10,0){1}(0,10){1}{\textcolor{blue}{\line(1,1){10}}}
\matrixput(0,10)(10,0){1}(0,10){1}{\textcolor{blue}{\line(1,-1){10}}}
\end{picture})
is referred to as {\em sky} and 
the lower-right Young diagram (written with
\begin{picture}(11,11)(0,1)
\matrixput(0,0)(10,0){1}(0,10){2}{\line(1,0){10}}
\matrixput(0,0)(10,0){2}(0,10){1}{\line(0,1){10}}
\end{picture})
is as {\em snow}. 
The pair $(n_j;\widetilde {n}_j)_j$ is denoted as 
\begin{align}
&
(n_1,n_2,n_3,n_4,n_5;\widetilde n_1,\widetilde n_2,\widetilde n_3,\widetilde n_4,\widetilde n_5)=(1,2,4,6,9;3,5,7,8,10) \nn\\
&\qquad \Leftrightarrow\qquad 
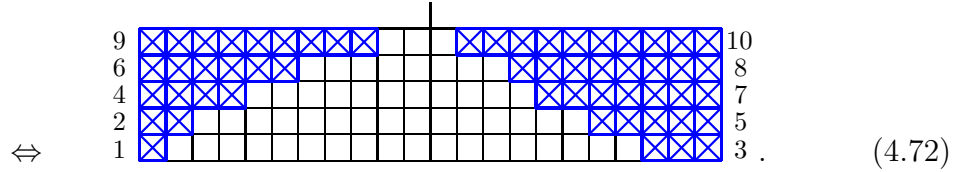
\begin{picture}(240,60)(-10,0)
\matrixput(0,0)(10,0){22}(0,10){6}{\line(1,0){10}}
\matrixput(0,0)(10,0){23}(0,10){5}{\line(0,1){10}}
\thicklines
\matrixput(110,0)(0,0){1}(0,10){6}{\line(0,1){10}}
\matrixput(0,0)(10,0){1}(0,10){6}{\textcolor{blue}{\line(1,0){10}}}
\matrixput(0,0)(10,0){2}(0,10){5}{\textcolor{blue}{\line(0,1){10}}}
\matrixput(0,0)(10,0){1}(0,10){5}{\textcolor{blue}{\line(1,1){10}}}
\matrixput(0,10)(10,0){1}(0,10){5}{\textcolor{blue}{\line(1,-1){10}}}
\matrixput(10,10)(10,0){1}(0,10){5}{\textcolor{blue}{\line(1,0){10}}}
\matrixput(10,10)(10,0){2}(0,10){4}{\textcolor{blue}{\line(0,1){10}}}
\matrixput(10,10)(10,0){1}(0,10){4}{\textcolor{blue}{\line(1,1){10}}}
\matrixput(10,20)(10,0){1}(0,10){4}{\textcolor{blue}{\line(1,-1){10}}}
\matrixput(20,20)(10,0){2}(0,10){4}{\textcolor{blue}{\line(1,0){10}}}
\matrixput(20,20)(10,0){3}(0,10){3}{\textcolor{blue}{\line(0,1){10}}}
\matrixput(20,20)(10,0){2}(0,10){3}{\textcolor{blue}{\line(1,1){10}}}
\matrixput(20,30)(10,0){2}(0,10){3}{\textcolor{blue}{\line(1,-1){10}}}
\matrixput(40,30)(10,0){2}(0,10){3}{\textcolor{blue}{\line(1,0){10}}}
\matrixput(40,30)(10,0){3}(0,10){2}{\textcolor{blue}{\line(0,1){10}}}
\matrixput(40,30)(10,0){2}(0,10){2}{\textcolor{blue}{\line(1,1){10}}}
\matrixput(40,40)(10,0){2}(0,10){2}{\textcolor{blue}{\line(1,-1){10}}}
\matrixput(60,40)(10,0){3}(0,10){2}{\textcolor{blue}{\line(1,0){10}}}
\matrixput(60,40)(10,0){4}(0,10){1}{\textcolor{blue}{\line(0,1){10}}}
\matrixput(60,40)(10,0){3}(0,10){1}{\textcolor{blue}{\line(1,1){10}}}
\matrixput(60,50)(10,0){3}(0,10){1}{\textcolor{blue}{\line(1,-1){10}}}
\matrixput(190,0)(10,0){3}(0,10){6}{\textcolor{blue}{\line(1,0){10}}}
\matrixput(190,0)(10,0){4}(0,10){5}{\textcolor{blue}{\line(0,1){10}}}
\matrixput(190,0)(10,0){3}(0,10){5}{\textcolor{blue}{\line(1,1){10}}}
\matrixput(190,10)(10,0){3}(0,10){5}{\textcolor{blue}{\line(1,-1){10}}}
\matrixput(170,10)(10,0){2}(0,10){5}{\textcolor{blue}{\line(1,0){10}}}
\matrixput(170,10)(10,0){3}(0,10){4}{\textcolor{blue}{\line(0,1){10}}}
\matrixput(170,10)(10,0){2}(0,10){4}{\textcolor{blue}{\line(1,1){10}}}
\matrixput(170,20)(10,0){2}(0,10){4}{\textcolor{blue}{\line(1,-1){10}}}
\matrixput(150,20)(10,0){2}(0,10){4}{\textcolor{blue}{\line(1,0){10}}}
\matrixput(150,20)(10,0){3}(0,10){3}{\textcolor{blue}{\line(0,1){10}}}
\matrixput(150,20)(10,0){2}(0,10){3}{\textcolor{blue}{\line(1,1){10}}}
\matrixput(150,30)(10,0){2}(0,10){3}{\textcolor{blue}{\line(1,-1){10}}}
\matrixput(140,30)(10,0){1}(0,10){3}{\textcolor{blue}{\line(1,0){10}}}
\matrixput(140,30)(10,0){2}(0,10){2}{\textcolor{blue}{\line(0,1){10}}}
\matrixput(140,30)(10,0){1}(0,10){2}{\textcolor{blue}{\line(1,1){10}}}
\matrixput(140,40)(10,0){1}(0,10){2}{\textcolor{blue}{\line(1,-1){10}}}
\matrixput(120,40)(10,0){2}(0,10){2}{\textcolor{blue}{\line(1,0){10}}}
\matrixput(120,40)(10,0){3}(0,10){1}{\textcolor{blue}{\line(0,1){10}}}
\matrixput(120,40)(10,0){2}(0,10){1}{\textcolor{blue}{\line(1,1){10}}}
\matrixput(120,50)(10,0){2}(0,10){1}{\textcolor{blue}{\line(1,-1){10}}}
\put(-10,1){\footnotesize 1}
\put(-10,11.7){\footnotesize 2}
\put(-10,22){\footnotesize 4}
\put(-10,32){\footnotesize 6}
\put(-10,42.5){\footnotesize 9}
\put(225,1){\footnotesize 3}
\put(225,11.7){\footnotesize 5}
\put(225,22){\footnotesize 7}
\put(225,32){\footnotesize 8}
\put(222,42.5){\footnotesize 10}
\end{picture}\,\, . 
\end{align}
Therefore, the graphical meaning of the conditions Eqs.~\eq{DiscreteSolIndices} is following:
\begin{itemize}
\item The number of the boxes \begin{picture}(11,11)(0,1)
\matrixput(0,0)(10,0){1}(0,10){2}{\line(1,0){10}}
\matrixput(0,0)(10,0){2}(0,10){1}{\line(0,1){10}}
\end{picture} (amount of snow) is always a multiple of $k$, and 
the following configurations are allowed solutions in the $7$-cut case:
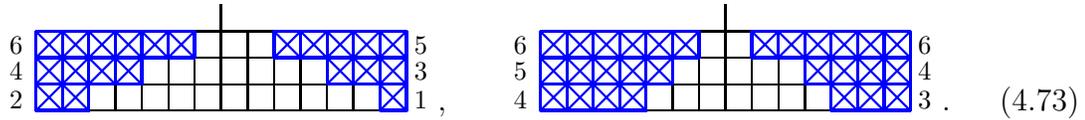
\begin{align}
\begin{picture}(160,35)(-10,0)
\matrixput(0,0)(10,0){14}(0,10){4}{\line(1,0){10}}
\matrixput(0,0)(10,0){15}(0,10){3}{\line(0,1){10}}
\thicklines
\matrixput(70,0)(0,0){1}(0,10){4}{\line(0,1){10}}
\matrixput(0,0)(10,0){2}(0,10){4}{\textcolor{blue}{\line(1,0){10}}}
\matrixput(0,0)(10,0){3}(0,10){3}{\textcolor{blue}{\line(0,1){10}}}
\matrixput(0,0)(10,0){2}(0,10){3}{\textcolor{blue}{\line(1,1){10}}}
\matrixput(0,10)(10,0){2}(0,10){3}{\textcolor{blue}{\line(1,-1){10}}}
\matrixput(20,10)(10,0){2}(0,10){3}{\textcolor{blue}{\line(1,0){10}}}
\matrixput(20,10)(10,0){3}(0,10){2}{\textcolor{blue}{\line(0,1){10}}}
\matrixput(20,10)(10,0){2}(0,10){2}{\textcolor{blue}{\line(1,1){10}}}
\matrixput(20,20)(10,0){2}(0,10){2}{\textcolor{blue}{\line(1,-1){10}}}
\matrixput(40,20)(10,0){2}(0,10){2}{\textcolor{blue}{\line(1,0){10}}}
\matrixput(40,20)(10,0){3}(0,10){1}{\textcolor{blue}{\line(0,1){10}}}
\matrixput(40,20)(10,0){2}(0,10){1}{\textcolor{blue}{\line(1,1){10}}}
\matrixput(40,30)(10,0){2}(0,10){1}{\textcolor{blue}{\line(1,-1){10}}}
\matrixput(130,0)(10,0){1}(0,10){4}{\textcolor{blue}{\line(1,0){10}}}
\matrixput(130,0)(10,0){2}(0,10){3}{\textcolor{blue}{\line(0,1){10}}}
\matrixput(130,0)(10,0){1}(0,10){3}{\textcolor{blue}{\line(1,1){10}}}
\matrixput(130,10)(10,0){1}(0,10){3}{\textcolor{blue}{\line(1,-1){10}}}
\matrixput(110,10)(10,0){2}(0,10){3}{\textcolor{blue}{\line(1,0){10}}}
\matrixput(110,10)(10,0){3}(0,10){2}{\textcolor{blue}{\line(0,1){10}}}
\matrixput(110,10)(10,0){2}(0,10){2}{\textcolor{blue}{\line(1,1){10}}}
\matrixput(110,20)(10,0){2}(0,10){2}{\textcolor{blue}{\line(1,-1){10}}}
\matrixput(90,20)(10,0){2}(0,10){2}{\textcolor{blue}{\line(1,0){10}}}
\matrixput(90,20)(10,0){3}(0,10){1}{\textcolor{blue}{\line(0,1){10}}}
\matrixput(90,20)(10,0){2}(0,10){1}{\textcolor{blue}{\line(1,1){10}}}
\matrixput(90,30)(10,0){2}(0,10){1}{\textcolor{blue}{\line(1,-1){10}}}
\put(-10,1){\footnotesize 2}
\put(-10,11.7){\footnotesize 4}
\put(-10,21.7){\footnotesize 6}
\put(143,1){\footnotesize 1}
\put(143,11.7){\footnotesize 3}
\put(143,21.7){\footnotesize 5}
\end{picture}\,,
\qquad
\begin{picture}(160,35)(-10,0)
\matrixput(0,0)(10,0){14}(0,10){4}{\line(1,0){10}}
\matrixput(0,0)(10,0){15}(0,10){3}{\line(0,1){10}}
\thicklines
\matrixput(70,0)(0,0){1}(0,10){4}{\line(0,1){10}}
\matrixput(0,0)(10,0){4}(0,10){4}{\textcolor{blue}{\line(1,0){10}}}
\matrixput(0,0)(10,0){5}(0,10){3}{\textcolor{blue}{\line(0,1){10}}}
\matrixput(0,0)(10,0){4}(0,10){3}{\textcolor{blue}{\line(1,1){10}}}
\matrixput(0,10)(10,0){4}(0,10){3}{\textcolor{blue}{\line(1,-1){10}}}
\matrixput(40,10)(10,0){1}(0,10){3}{\textcolor{blue}{\line(1,0){10}}}
\matrixput(40,10)(10,0){2}(0,10){2}{\textcolor{blue}{\line(0,1){10}}}
\matrixput(40,10)(10,0){1}(0,10){2}{\textcolor{blue}{\line(1,1){10}}}
\matrixput(40,20)(10,0){1}(0,10){2}{\textcolor{blue}{\line(1,-1){10}}}
\matrixput(50,20)(10,0){1}(0,10){2}{\textcolor{blue}{\line(1,0){10}}}
\matrixput(50,20)(10,0){2}(0,10){1}{\textcolor{blue}{\line(0,1){10}}}
\matrixput(50,20)(10,0){1}(0,10){1}{\textcolor{blue}{\line(1,1){10}}}
\matrixput(50,30)(10,0){1}(0,10){1}{\textcolor{blue}{\line(1,-1){10}}}
\matrixput(110,0)(10,0){3}(0,10){4}{\textcolor{blue}{\line(1,0){10}}}
\matrixput(110,0)(10,0){4}(0,10){3}{\textcolor{blue}{\line(0,1){10}}}
\matrixput(110,0)(10,0){3}(0,10){3}{\textcolor{blue}{\line(1,1){10}}}
\matrixput(110,10)(10,0){3}(0,10){3}{\textcolor{blue}{\line(1,-1){10}}}
\matrixput(100,10)(10,0){1}(0,10){3}{\textcolor{blue}{\line(1,0){10}}}
\matrixput(100,10)(10,0){2}(0,10){2}{\textcolor{blue}{\line(0,1){10}}}
\matrixput(100,10)(10,0){1}(0,10){2}{\textcolor{blue}{\line(1,1){10}}}
\matrixput(100,20)(10,0){1}(0,10){2}{\textcolor{blue}{\line(1,-1){10}}}
\matrixput(80,20)(10,0){2}(0,10){2}{\textcolor{blue}{\line(1,0){10}}}
\matrixput(80,20)(10,0){3}(0,10){1}{\textcolor{blue}{\line(0,1){10}}}
\matrixput(80,20)(10,0){2}(0,10){1}{\textcolor{blue}{\line(1,1){10}}}
\matrixput(80,30)(10,0){2}(0,10){1}{\textcolor{blue}{\line(1,-1){10}}}
\put(-10,1){\footnotesize 4}
\put(-10,11.7){\footnotesize 5}
\put(-10,21.7){\footnotesize 6}
\put(143,1){\footnotesize 3}
\put(143,11.7){\footnotesize 4}
\put(143,21.7){\footnotesize 6}
\end{picture}\,.
\end{align}
\item Neither $n_i$ and $\widetilde n_i$ ($i=1,2,\cdots, m$) 
can be $0$ or $k$, therefore the following configurations are not allowed:
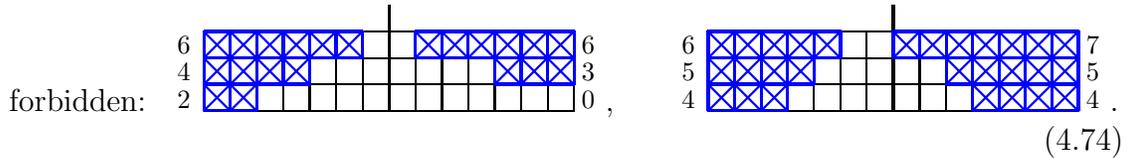
\begin{align}
\text{forbidden:}\quad
\begin{picture}(160,35)(-10,0)
\matrixput(0,0)(10,0){14}(0,10){4}{\line(1,0){10}}
\matrixput(0,0)(10,0){15}(0,10){3}{\line(0,1){10}}
\thicklines
\matrixput(70,0)(0,0){1}(0,10){4}{\line(0,1){10}}
\matrixput(0,0)(10,0){2}(0,10){4}{\textcolor{blue}{\line(1,0){10}}}
\matrixput(0,0)(10,0){3}(0,10){3}{\textcolor{blue}{\line(0,1){10}}}
\matrixput(0,0)(10,0){2}(0,10){3}{\textcolor{blue}{\line(1,1){10}}}
\matrixput(0,10)(10,0){2}(0,10){3}{\textcolor{blue}{\line(1,-1){10}}}
\matrixput(20,10)(10,0){2}(0,10){3}{\textcolor{blue}{\line(1,0){10}}}
\matrixput(20,10)(10,0){3}(0,10){2}{\textcolor{blue}{\line(0,1){10}}}
\matrixput(20,10)(10,0){2}(0,10){2}{\textcolor{blue}{\line(1,1){10}}}
\matrixput(20,20)(10,0){2}(0,10){2}{\textcolor{blue}{\line(1,-1){10}}}
\matrixput(40,20)(10,0){2}(0,10){2}{\textcolor{blue}{\line(1,0){10}}}
\matrixput(40,20)(10,0){3}(0,10){1}{\textcolor{blue}{\line(0,1){10}}}
\matrixput(40,20)(10,0){2}(0,10){1}{\textcolor{blue}{\line(1,1){10}}}
\matrixput(40,30)(10,0){2}(0,10){1}{\textcolor{blue}{\line(1,-1){10}}}
\matrixput(110,10)(10,0){3}(0,10){3}{\textcolor{blue}{\line(1,0){10}}}
\matrixput(110,10)(10,0){4}(0,10){2}{\textcolor{blue}{\line(0,1){10}}}
\matrixput(110,10)(10,0){3}(0,10){2}{\textcolor{blue}{\line(1,1){10}}}
\matrixput(110,20)(10,0){3}(0,10){2}{\textcolor{blue}{\line(1,-1){10}}}
\matrixput(80,20)(10,0){3}(0,10){2}{\textcolor{blue}{\line(1,0){10}}}
\matrixput(80,20)(10,0){4}(0,10){1}{\textcolor{blue}{\line(0,1){10}}}
\matrixput(80,20)(10,0){3}(0,10){1}{\textcolor{blue}{\line(1,1){10}}}
\matrixput(80,30)(10,0){3}(0,10){1}{\textcolor{blue}{\line(1,-1){10}}}
\put(-10,1){\footnotesize 2}
\put(-10,11.7){\footnotesize 4}
\put(-10,21.7){\footnotesize 6}
\put(143,1){\footnotesize 0}
\put(143,11.7){\footnotesize 3}
\put(143,21.7){\footnotesize 6}
\end{picture}\,,
\qquad
\begin{picture}(160,35)(-10,0)
\matrixput(0,0)(10,0){14}(0,10){4}{\line(1,0){10}}
\matrixput(0,0)(10,0){15}(0,10){3}{\line(0,1){10}}
\thicklines
\matrixput(70,0)(0,0){1}(0,10){4}{\line(0,1){10}}
\matrixput(0,0)(10,0){3}(0,10){4}{\textcolor{blue}{\line(1,0){10}}}
\matrixput(0,0)(10,0){4}(0,10){3}{\textcolor{blue}{\line(0,1){10}}}
\matrixput(0,0)(10,0){3}(0,10){3}{\textcolor{blue}{\line(1,1){10}}}
\matrixput(0,10)(10,0){3}(0,10){3}{\textcolor{blue}{\line(1,-1){10}}}
\matrixput(30,10)(10,0){1}(0,10){3}{\textcolor{blue}{\line(1,0){10}}}
\matrixput(30,10)(10,0){2}(0,10){2}{\textcolor{blue}{\line(0,1){10}}}
\matrixput(30,10)(10,0){1}(0,10){2}{\textcolor{blue}{\line(1,1){10}}}
\matrixput(30,20)(10,0){1}(0,10){2}{\textcolor{blue}{\line(1,-1){10}}}
\matrixput(40,20)(10,0){1}(0,10){2}{\textcolor{blue}{\line(1,0){10}}}
\matrixput(40,20)(10,0){2}(0,10){1}{\textcolor{blue}{\line(0,1){10}}}
\matrixput(40,20)(10,0){1}(0,10){1}{\textcolor{blue}{\line(1,1){10}}}
\matrixput(40,30)(10,0){1}(0,10){1}{\textcolor{blue}{\line(1,-1){10}}}
\matrixput(100,0)(10,0){4}(0,10){4}{\textcolor{blue}{\line(1,0){10}}}
\matrixput(100,0)(10,0){5}(0,10){3}{\textcolor{blue}{\line(0,1){10}}}
\matrixput(100,0)(10,0){4}(0,10){3}{\textcolor{blue}{\line(1,1){10}}}
\matrixput(100,10)(10,0){4}(0,10){3}{\textcolor{blue}{\line(1,-1){10}}}
\matrixput(90,10)(10,0){1}(0,10){3}{\textcolor{blue}{\line(1,0){10}}}
\matrixput(90,10)(10,0){2}(0,10){2}{\textcolor{blue}{\line(0,1){10}}}
\matrixput(90,10)(10,0){1}(0,10){2}{\textcolor{blue}{\line(1,1){10}}}
\matrixput(90,20)(10,0){1}(0,10){2}{\textcolor{blue}{\line(1,-1){10}}}
\matrixput(70,20)(10,0){2}(0,10){2}{\textcolor{blue}{\line(1,0){10}}}
\matrixput(70,20)(10,0){3}(0,10){1}{\textcolor{blue}{\line(0,1){10}}}
\matrixput(70,20)(10,0){2}(0,10){1}{\textcolor{blue}{\line(1,1){10}}}
\matrixput(70,30)(10,0){2}(0,10){1}{\textcolor{blue}{\line(1,-1){10}}}
\put(-10,1){\footnotesize 4}
\put(-10,11.7){\footnotesize 5}
\put(-10,21.7){\footnotesize 6}
\put(143,1){\footnotesize 4}
\put(143,11.7){\footnotesize 5}
\put(143,21.7){\footnotesize 7}
\end{picture}\,.
\end{align}
\item The solutions cannot have vertical cliffs, therefore the following configurations 
are not allowed: 
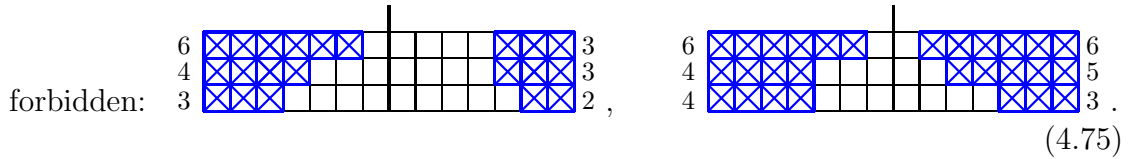
\begin{align}
\text{forbidden:}\quad
\begin{picture}(160,35)(-10,0)
\matrixput(0,0)(10,0){14}(0,10){4}{\line(1,0){10}}
\matrixput(0,0)(10,0){15}(0,10){3}{\line(0,1){10}}
\thicklines
\matrixput(70,0)(0,0){1}(0,10){4}{\line(0,1){10}}
\matrixput(0,0)(10,0){3}(0,10){4}{\textcolor{blue}{\line(1,0){10}}}
\matrixput(0,0)(10,0){4}(0,10){3}{\textcolor{blue}{\line(0,1){10}}}
\matrixput(0,0)(10,0){3}(0,10){3}{\textcolor{blue}{\line(1,1){10}}}
\matrixput(0,10)(10,0){3}(0,10){3}{\textcolor{blue}{\line(1,-1){10}}}
\matrixput(20,10)(10,0){2}(0,10){3}{\textcolor{blue}{\line(1,0){10}}}
\matrixput(20,10)(10,0){3}(0,10){2}{\textcolor{blue}{\line(0,1){10}}}
\matrixput(20,10)(10,0){2}(0,10){2}{\textcolor{blue}{\line(1,1){10}}}
\matrixput(20,20)(10,0){2}(0,10){2}{\textcolor{blue}{\line(1,-1){10}}}
\matrixput(40,20)(10,0){2}(0,10){2}{\textcolor{blue}{\line(1,0){10}}}
\matrixput(40,20)(10,0){3}(0,10){1}{\textcolor{blue}{\line(0,1){10}}}
\matrixput(40,20)(10,0){2}(0,10){1}{\textcolor{blue}{\line(1,1){10}}}
\matrixput(40,30)(10,0){2}(0,10){1}{\textcolor{blue}{\line(1,-1){10}}}
\matrixput(120,0)(10,0){2}(0,10){4}{\textcolor{blue}{\line(1,0){10}}}
\matrixput(120,0)(10,0){3}(0,10){3}{\textcolor{blue}{\line(0,1){10}}}
\matrixput(120,0)(10,0){2}(0,10){3}{\textcolor{blue}{\line(1,1){10}}}
\matrixput(120,10)(10,0){2}(0,10){3}{\textcolor{blue}{\line(1,-1){10}}}
\matrixput(110,10)(10,0){2}(0,10){3}{\textcolor{blue}{\line(1,0){10}}}
\matrixput(110,10)(10,0){3}(0,10){2}{\textcolor{blue}{\line(0,1){10}}}
\matrixput(110,10)(10,0){2}(0,10){2}{\textcolor{blue}{\line(1,1){10}}}
\matrixput(110,20)(10,0){2}(0,10){2}{\textcolor{blue}{\line(1,-1){10}}}
\put(-10,1){\footnotesize 3}
\put(-10,11.7){\footnotesize 4}
\put(-10,21.7){\footnotesize 6}
\put(143,1){\footnotesize 2}
\put(143,11.7){\footnotesize 3}
\put(143,21.7){\footnotesize 3}
\end{picture}\,,
\qquad
\begin{picture}(160,35)(-10,0)
\matrixput(0,0)(10,0){14}(0,10){4}{\line(1,0){10}}
\matrixput(0,0)(10,0){15}(0,10){3}{\line(0,1){10}}
\thicklines
\matrixput(70,0)(0,0){1}(0,10){4}{\line(0,1){10}}
\matrixput(0,0)(10,0){4}(0,10){4}{\textcolor{blue}{\line(1,0){10}}}
\matrixput(0,0)(10,0){5}(0,10){3}{\textcolor{blue}{\line(0,1){10}}}
\matrixput(0,0)(10,0){4}(0,10){3}{\textcolor{blue}{\line(1,1){10}}}
\matrixput(0,10)(10,0){4}(0,10){3}{\textcolor{blue}{\line(1,-1){10}}}
\matrixput(40,20)(10,0){1}(0,10){2}{\textcolor{blue}{\line(1,0){10}}}
\matrixput(40,20)(10,0){2}(0,10){1}{\textcolor{blue}{\line(0,1){10}}}
\matrixput(40,20)(10,0){1}(0,10){1}{\textcolor{blue}{\line(1,1){10}}}
\matrixput(40,30)(10,0){1}(0,10){1}{\textcolor{blue}{\line(1,-1){10}}}
\matrixput(50,20)(10,0){1}(0,10){2}{\textcolor{blue}{\line(1,0){10}}}
\matrixput(50,20)(10,0){2}(0,10){1}{\textcolor{blue}{\line(0,1){10}}}
\matrixput(50,20)(10,0){1}(0,10){1}{\textcolor{blue}{\line(1,1){10}}}
\matrixput(50,30)(10,0){1}(0,10){1}{\textcolor{blue}{\line(1,-1){10}}}
\matrixput(110,0)(10,0){3}(0,10){4}{\textcolor{blue}{\line(1,0){10}}}
\matrixput(110,0)(10,0){4}(0,10){3}{\textcolor{blue}{\line(0,1){10}}}
\matrixput(110,0)(10,0){3}(0,10){3}{\textcolor{blue}{\line(1,1){10}}}
\matrixput(110,10)(10,0){3}(0,10){3}{\textcolor{blue}{\line(1,-1){10}}}
\matrixput(90,10)(10,0){2}(0,10){3}{\textcolor{blue}{\line(1,0){10}}}
\matrixput(90,10)(10,0){3}(0,10){2}{\textcolor{blue}{\line(0,1){10}}}
\matrixput(90,10)(10,0){2}(0,10){2}{\textcolor{blue}{\line(1,1){10}}}
\matrixput(90,20)(10,0){2}(0,10){2}{\textcolor{blue}{\line(1,-1){10}}}
\matrixput(80,20)(10,0){2}(0,10){2}{\textcolor{blue}{\line(1,0){10}}}
\matrixput(80,20)(10,0){3}(0,10){1}{\textcolor{blue}{\line(0,1){10}}}
\matrixput(80,20)(10,0){2}(0,10){1}{\textcolor{blue}{\line(1,1){10}}}
\matrixput(80,30)(10,0){2}(0,10){1}{\textcolor{blue}{\line(1,-1){10}}}
\put(-10,1){\footnotesize 4}
\put(-10,11.7){\footnotesize 4}
\put(-10,21.7){\footnotesize 6}
\put(143,1){\footnotesize 3}
\put(143,11.7){\footnotesize 5}
\put(143,21.7){\footnotesize 6}
\end{picture}\,.
\end{align}
\end{itemize}
One of the ways to exhaust 
the solutions is first to take the most steepest configurations, 
and then to consider possible ways for snow to slide on the surface 
with satisfying the condition \eq{YoungDiagramRule222}, for example:
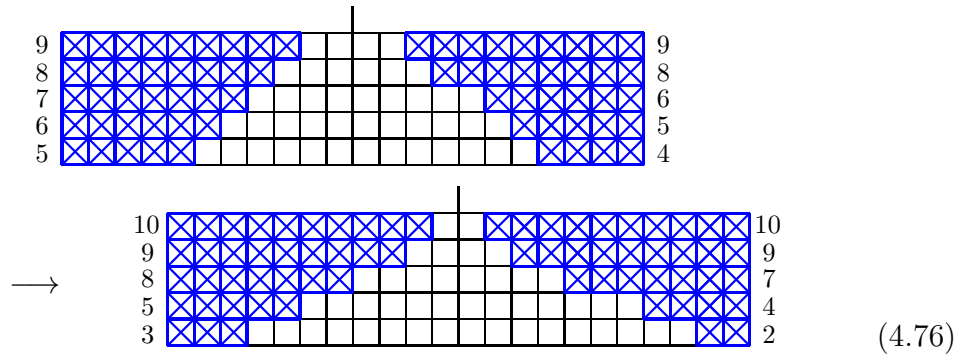
\begin{align}
\begin{picture}(240,60)(0,0)
\matrixput(0,0)(10,0){22}(0,10){6}{\line(1,0){10}}
\matrixput(0,0)(10,0){23}(0,10){5}{\line(0,1){10}}
\thicklines
\matrixput(110,0)(0,0){1}(0,10){6}{\line(0,1){10}}
\matrixput(0,0)(10,0){5}(0,10){6}{\textcolor{blue}{\line(1,0){10}}}
\matrixput(0,0)(10,0){6}(0,10){5}{\textcolor{blue}{\line(0,1){10}}}
\matrixput(0,0)(10,0){5}(0,10){5}{\textcolor{blue}{\line(1,1){10}}}
\matrixput(0,10)(10,0){5}(0,10){5}{\textcolor{blue}{\line(1,-1){10}}}
\matrixput(50,10)(10,0){1}(0,10){5}{\textcolor{blue}{\line(1,0){10}}}
\matrixput(50,10)(10,0){2}(0,10){4}{\textcolor{blue}{\line(0,1){10}}}
\matrixput(50,10)(10,0){1}(0,10){4}{\textcolor{blue}{\line(1,1){10}}}
\matrixput(50,20)(10,0){1}(0,10){4}{\textcolor{blue}{\line(1,-1){10}}}
\matrixput(60,20)(10,0){1}(0,10){4}{\textcolor{blue}{\line(1,0){10}}}
\matrixput(60,20)(10,0){2}(0,10){3}{\textcolor{blue}{\line(0,1){10}}}
\matrixput(60,20)(10,0){1}(0,10){3}{\textcolor{blue}{\line(1,1){10}}}
\matrixput(60,30)(10,0){1}(0,10){3}{\textcolor{blue}{\line(1,-1){10}}}
\matrixput(70,30)(10,0){1}(0,10){3}{\textcolor{blue}{\line(1,0){10}}}
\matrixput(70,30)(10,0){2}(0,10){2}{\textcolor{blue}{\line(0,1){10}}}
\matrixput(70,30)(10,0){1}(0,10){2}{\textcolor{blue}{\line(1,1){10}}}
\matrixput(70,40)(10,0){1}(0,10){2}{\textcolor{blue}{\line(1,-1){10}}}
\matrixput(80,40)(10,0){1}(0,10){2}{\textcolor{blue}{\line(1,0){10}}}
\matrixput(80,40)(10,0){2}(0,10){1}{\textcolor{blue}{\line(0,1){10}}}
\matrixput(80,40)(10,0){1}(0,10){1}{\textcolor{blue}{\line(1,1){10}}}
\matrixput(80,50)(10,0){1}(0,10){1}{\textcolor{blue}{\line(1,-1){10}}}
\matrixput(180,0)(10,0){4}(0,10){6}{\textcolor{blue}{\line(1,0){10}}}
\matrixput(180,0)(10,0){5}(0,10){5}{\textcolor{blue}{\line(0,1){10}}}
\matrixput(180,0)(10,0){4}(0,10){5}{\textcolor{blue}{\line(1,1){10}}}
\matrixput(180,10)(10,0){4}(0,10){5}{\textcolor{blue}{\line(1,-1){10}}}
\matrixput(170,10)(10,0){3}(0,10){5}{\textcolor{blue}{\line(1,0){10}}}
\matrixput(170,10)(10,0){4}(0,10){4}{\textcolor{blue}{\line(0,1){10}}}
\matrixput(170,10)(10,0){3}(0,10){4}{\textcolor{blue}{\line(1,1){10}}}
\matrixput(170,20)(10,0){3}(0,10){4}{\textcolor{blue}{\line(1,-1){10}}}
\matrixput(160,20)(10,0){1}(0,10){4}{\textcolor{blue}{\line(1,0){10}}}
\matrixput(160,20)(10,0){2}(0,10){3}{\textcolor{blue}{\line(0,1){10}}}
\matrixput(160,20)(10,0){1}(0,10){3}{\textcolor{blue}{\line(1,1){10}}}
\matrixput(160,30)(10,0){1}(0,10){3}{\textcolor{blue}{\line(1,-1){10}}}
\matrixput(140,30)(10,0){2}(0,10){3}{\textcolor{blue}{\line(1,0){10}}}
\matrixput(140,30)(10,0){3}(0,10){2}{\textcolor{blue}{\line(0,1){10}}}
\matrixput(140,30)(10,0){2}(0,10){2}{\textcolor{blue}{\line(1,1){10}}}
\matrixput(140,40)(10,0){2}(0,10){2}{\textcolor{blue}{\line(1,-1){10}}}
\matrixput(130,40)(10,0){1}(0,10){2}{\textcolor{blue}{\line(1,0){10}}}
\matrixput(130,40)(10,0){2}(0,10){1}{\textcolor{blue}{\line(0,1){10}}}
\matrixput(130,40)(10,0){1}(0,10){1}{\textcolor{blue}{\line(1,1){10}}}
\matrixput(130,50)(10,0){1}(0,10){1}{\textcolor{blue}{\line(1,-1){10}}}
\put(-10,1){\footnotesize 5}
\put(-10,11.7){\footnotesize 6}
\put(-10,22){\footnotesize 7}
\put(-10,32){\footnotesize 8}
\put(-10,42.5){\footnotesize 9}
\put(225,1){\footnotesize 4}
\put(225,11.7){\footnotesize 5}
\put(225,22){\footnotesize 6}
\put(225,32){\footnotesize 8}
\put(225,42.5){\footnotesize 9}
\end{picture}\,\, \nn\\
%
\begin{picture}(60,0)(0,0)
\put(0,20){$\longrightarrow$}
\end{picture}
\begin{picture}(200,60)(0,0)
\matrixput(0,0)(10,0){22}(0,10){6}{\line(1,0){10}}
\matrixput(0,0)(10,0){23}(0,10){5}{\line(0,1){10}}
\thicklines
\matrixput(110,0)(0,0){1}(0,10){6}{\line(0,1){10}}
\matrixput(0,0)(10,0){3}(0,10){6}{\textcolor{blue}{\line(1,0){10}}}
\matrixput(0,0)(10,0){4}(0,10){5}{\textcolor{blue}{\line(0,1){10}}}
\matrixput(0,0)(10,0){3}(0,10){5}{\textcolor{blue}{\line(1,1){10}}}
\matrixput(0,10)(10,0){3}(0,10){5}{\textcolor{blue}{\line(1,-1){10}}}
\matrixput(30,10)(10,0){2}(0,10){5}{\textcolor{blue}{\line(1,0){10}}}
\matrixput(30,10)(10,0){3}(0,10){4}{\textcolor{blue}{\line(0,1){10}}}
\matrixput(30,10)(10,0){2}(0,10){4}{\textcolor{blue}{\line(1,1){10}}}
\matrixput(30,20)(10,0){2}(0,10){4}{\textcolor{blue}{\line(1,-1){10}}}
\matrixput(50,20)(10,0){2}(0,10){4}{\textcolor{blue}{\line(1,0){10}}}
\matrixput(50,20)(10,0){3}(0,10){3}{\textcolor{blue}{\line(0,1){10}}}
\matrixput(50,20)(10,0){2}(0,10){3}{\textcolor{blue}{\line(1,1){10}}}
\matrixput(50,30)(10,0){2}(0,10){3}{\textcolor{blue}{\line(1,-1){10}}}
\matrixput(70,30)(10,0){2}(0,10){3}{\textcolor{blue}{\line(1,0){10}}}
\matrixput(70,30)(10,0){3}(0,10){2}{\textcolor{blue}{\line(0,1){10}}}
\matrixput(70,30)(10,0){2}(0,10){2}{\textcolor{blue}{\line(1,1){10}}}
\matrixput(70,40)(10,0){2}(0,10){2}{\textcolor{blue}{\line(1,-1){10}}}
\matrixput(90,40)(10,0){1}(0,10){2}{\textcolor{blue}{\line(1,0){10}}}
\matrixput(90,40)(10,0){2}(0,10){1}{\textcolor{blue}{\line(0,1){10}}}
\matrixput(90,40)(10,0){1}(0,10){1}{\textcolor{blue}{\line(1,1){10}}}
\matrixput(90,50)(10,0){1}(0,10){1}{\textcolor{blue}{\line(1,-1){10}}}
\matrixput(200,0)(10,0){2}(0,10){6}{\textcolor{blue}{\line(1,0){10}}}
\matrixput(200,0)(10,0){3}(0,10){5}{\textcolor{blue}{\line(0,1){10}}}
\matrixput(200,0)(10,0){2}(0,10){5}{\textcolor{blue}{\line(1,1){10}}}
\matrixput(200,10)(10,0){2}(0,10){5}{\textcolor{blue}{\line(1,-1){10}}}
\matrixput(180,10)(10,0){3}(0,10){5}{\textcolor{blue}{\line(1,0){10}}}
\matrixput(180,10)(10,0){4}(0,10){4}{\textcolor{blue}{\line(0,1){10}}}
\matrixput(180,10)(10,0){3}(0,10){4}{\textcolor{blue}{\line(1,1){10}}}
\matrixput(180,20)(10,0){3}(0,10){4}{\textcolor{blue}{\line(1,-1){10}}}
\matrixput(150,20)(10,0){3}(0,10){4}{\textcolor{blue}{\line(1,0){10}}}
\matrixput(150,20)(10,0){4}(0,10){3}{\textcolor{blue}{\line(0,1){10}}}
\matrixput(150,20)(10,0){3}(0,10){3}{\textcolor{blue}{\line(1,1){10}}}
\matrixput(150,30)(10,0){3}(0,10){3}{\textcolor{blue}{\line(1,-1){10}}}
\matrixput(130,30)(10,0){3}(0,10){3}{\textcolor{blue}{\line(1,0){10}}}
\matrixput(130,30)(10,0){4}(0,10){2}{\textcolor{blue}{\line(0,1){10}}}
\matrixput(130,30)(10,0){3}(0,10){2}{\textcolor{blue}{\line(1,1){10}}}
\matrixput(130,40)(10,0){3}(0,10){2}{\textcolor{blue}{\line(1,-1){10}}}
\matrixput(120,40)(10,0){2}(0,10){2}{\textcolor{blue}{\line(1,0){10}}}
\matrixput(120,40)(10,0){3}(0,10){1}{\textcolor{blue}{\line(0,1){10}}}
\matrixput(120,40)(10,0){2}(0,10){1}{\textcolor{blue}{\line(1,1){10}}}
\matrixput(120,50)(10,0){2}(0,10){1}{\textcolor{blue}{\line(1,-1){10}}}
\put(-10,1){\footnotesize 3}
\put(-10,11.7){\footnotesize 5}
\put(-10,22){\footnotesize 8}
\put(-10,32){\footnotesize 9}
\put(-13,42.5){\footnotesize 10}
\put(225,1){\footnotesize 2}
\put(225,11.7){\footnotesize 4}
\put(225,22){\footnotesize 7}
\put(225,32){\footnotesize 9}
\put(222,42.5){\footnotesize 10}
\end{picture}\,\, 
\end{align}
We refer to these configulations of Young diagram as {\em avalanches}. 
Therefore, we conclude: 
\begin{Proposition} [Avalanches]
The discrete solutions to the non-perturbative completion are 
labeled by configurations of avalanches in terms of Young diagram. 
\label{AvalanchesProposition}
\end{Proposition}
Note that one can also move some snow on the one side to the other side. 
It is also worth mentioning the following two transformations: 
\begin{itemize}
\item The first transformation is called {\em dual}, 
\begin{align}
&\bigl(n_1,n_2,\cdots,n_{m};
\widetilde n_1,\widetilde n_2,\cdots,\widetilde n_{m}\bigr)\nn\\
&\qquad \to \qquad \bigl(k-n_1,k-n_2,\cdots,k-n_{m};
k-\widetilde n_1,k-\widetilde n_2,\cdots,k-\widetilde n_{m}\bigr),
\label{DualTransformationYoung}
\end{align}
In the terminology of Young diagram, the dual transformation \eq{DualTransformationYoung} 
exchanges the sky and snow of the left and right Young diagrams simultaneously. 
In particular, the following diagram shows the action of the dual transformation on the left Young diagram:
\begin{align}
\begin{picture}(120,50)(-10,0)
\matrixput(0,0)(10,0){11}(0,10){6}{\line(1,0){10}}
\matrixput(0,0)(10,0){12}(0,10){5}{\line(0,1){10}}
\thicklines
\matrixput(0,0)(10,0){1}(0,10){6}{\textcolor{blue}{\line(1,0){10}}}
\matrixput(0,0)(10,0){2}(0,10){5}{\textcolor{blue}{\line(0,1){10}}}
\matrixput(0,0)(10,0){1}(0,10){5}{\textcolor{blue}{\line(1,1){10}}}
\matrixput(0,10)(10,0){1}(0,10){5}{\textcolor{blue}{\line(1,-1){10}}}
\matrixput(10,10)(10,0){1}(0,10){5}{\textcolor{blue}{\line(1,0){10}}}
\matrixput(10,10)(10,0){2}(0,10){4}{\textcolor{blue}{\line(0,1){10}}}
\matrixput(10,10)(10,0){1}(0,10){4}{\textcolor{blue}{\line(1,1){10}}}
\matrixput(10,20)(10,0){1}(0,10){4}{\textcolor{blue}{\line(1,-1){10}}}
\matrixput(20,20)(10,0){2}(0,10){4}{\textcolor{blue}{\line(1,0){10}}}
\matrixput(20,20)(10,0){3}(0,10){3}{\textcolor{blue}{\line(0,1){10}}}
\matrixput(20,20)(10,0){2}(0,10){3}{\textcolor{blue}{\line(1,1){10}}}
\matrixput(20,30)(10,0){2}(0,10){3}{\textcolor{blue}{\line(1,-1){10}}}
\matrixput(40,30)(10,0){2}(0,10){3}{\textcolor{blue}{\line(1,0){10}}}
\matrixput(40,30)(10,0){3}(0,10){2}{\textcolor{blue}{\line(0,1){10}}}
\matrixput(40,30)(10,0){2}(0,10){2}{\textcolor{blue}{\line(1,1){10}}}
\matrixput(40,40)(10,0){2}(0,10){2}{\textcolor{blue}{\line(1,-1){10}}}
\matrixput(60,40)(10,0){3}(0,10){2}{\textcolor{blue}{\line(1,0){10}}}
\matrixput(60,40)(10,0){4}(0,10){1}{\textcolor{blue}{\line(0,1){10}}}
\matrixput(60,40)(10,0){3}(0,10){1}{\textcolor{blue}{\line(1,1){10}}}
\matrixput(60,50)(10,0){3}(0,10){1}{\textcolor{blue}{\line(1,-1){10}}}
\put(-10,1){\footnotesize 1}
\put(-10,11.7){\footnotesize 2}
\put(-10,22){\footnotesize 4}
\put(-10,32){\footnotesize 6}
\put(-10,42.5){\footnotesize 9}
\end{picture}\qquad\to\qquad 
\begin{picture}(120,50)(-10,0)
\matrixput(0,0)(10,0){11}(0,10){6}{\line(1,0){10}}
\matrixput(0,0)(10,0){12}(0,10){5}{\line(0,1){10}}
\thicklines
\matrixput(0,0)(10,0){2}(0,10){6}{\textcolor{blue}{\line(1,0){10}}}
\matrixput(0,0)(10,0){3}(0,10){5}{\textcolor{blue}{\line(0,1){10}}}
\matrixput(0,0)(10,0){2}(0,10){5}{\textcolor{blue}{\line(1,1){10}}}
\matrixput(0,10)(10,0){2}(0,10){5}{\textcolor{blue}{\line(1,-1){10}}}
\matrixput(20,10)(10,0){3}(0,10){5}{\textcolor{blue}{\line(1,0){10}}}
\matrixput(20,10)(10,0){4}(0,10){4}{\textcolor{blue}{\line(0,1){10}}}
\matrixput(20,10)(10,0){3}(0,10){4}{\textcolor{blue}{\line(1,1){10}}}
\matrixput(20,20)(10,0){3}(0,10){4}{\textcolor{blue}{\line(1,-1){10}}}
\matrixput(50,20)(10,0){2}(0,10){4}{\textcolor{blue}{\line(1,0){10}}}
\matrixput(50,20)(10,0){3}(0,10){3}{\textcolor{blue}{\line(0,1){10}}}
\matrixput(50,20)(10,0){2}(0,10){3}{\textcolor{blue}{\line(1,1){10}}}
\matrixput(50,30)(10,0){2}(0,10){3}{\textcolor{blue}{\line(1,-1){10}}}
\matrixput(70,30)(10,0){2}(0,10){3}{\textcolor{blue}{\line(1,0){10}}}
\matrixput(70,30)(10,0){3}(0,10){2}{\textcolor{blue}{\line(0,1){10}}}
\matrixput(70,30)(10,0){2}(0,10){2}{\textcolor{blue}{\line(1,1){10}}}
\matrixput(70,40)(10,0){2}(0,10){2}{\textcolor{blue}{\line(1,-1){10}}}
\matrixput(90,40)(10,0){1}(0,10){2}{\textcolor{blue}{\line(1,0){10}}}
\matrixput(90,40)(10,0){2}(0,10){1}{\textcolor{blue}{\line(0,1){10}}}
\matrixput(90,40)(10,0){1}(0,10){1}{\textcolor{blue}{\line(1,1){10}}}
\matrixput(90,50)(10,0){1}(0,10){1}{\textcolor{blue}{\line(1,-1){10}}}
\put(-10,1){\footnotesize 2}
\put(-10,11.7){\footnotesize 5}
\put(-10,22){\footnotesize 7}
\put(-10,32){\footnotesize 9}
\put(-15,42.5){\footnotesize 10}
\end{picture}\,.
\end{align} 
\item The following is called {\em reflection}, 
which exchanges the snows of the left and right Young diagrams: 
\begin{align}
\bigl(n_1,n_2,\cdots,n_{m};
\widetilde n_1,\widetilde n_2,\cdots,\widetilde n_{m}\bigr)
\quad \to \quad 
\bigl(\widetilde n_1,\widetilde n_2,\cdots,\widetilde n_{m};
n_1,n_2,\cdots,n_{m}\bigr). 
\end{align}
\end{itemize}
Note that these two transformations also 
automorphisms which fix the condition \eq{YoungDiagramRule222}. 


\subsubsection{Continuum solutions}

\begin{Theorem} [Continuum Solution]
The following Stokes multipliers $s_{l,i,j}$ 
(written with $\theta_n,\widetilde \theta_n$ of Proposition \ref{ThetaParamProposition}) 
are solutions to the multi-cut boundary condition 
in the $\mathbb Z_k$ symmetric $(\hat p,\hat q)=(1,1)$ $k$-cut critical points 
($k=2m+1,\gamma=r=2$):
\begin{align}
\theta_n = {\sigma}_{n}
(\{-\omega^{n_j}\}_{j=1}^{m}),\qquad 
\widetilde \theta_n = \mathcal S_n \bigl(\{\theta_j\}_{j=1}^{m}\bigr) 
+\widetilde\theta_{m-n+1}^* \theta_{m}^*, 
\qquad \bigl(n=1,2,\cdots, m\bigr), \label{ContinuumThetaDefinition}
\end{align}
with the polynomial $\mathcal S_n(x)$ which are defined by 
the following recursion relation: 
\begin{align}
\mathcal S_{n}\bigl(\{x_j\}_{j\in \mathbb Z}\bigr) 
= -\sum_{i=1}^n x_i\,  \mathcal S_{n-i}\bigl(\{x_j\}_{j\in \mathbb Z}\bigr),\qquad 
S_0\bigl(\{x_j\}_{j\in \mathbb Z}\bigr) =1, \label{RecursionForPolySSS}
\end{align}
if and only if the integers 
$\bigl(n_1,n_2,\cdots, n_{\lfloor \frac{k}{2}\rfloor}\bigr)$ 
satisfy $n_i\not\equiv n_j \mod k \quad (i\neq j)$. 
\label{TheoremContinuumSolutions}
\end{Theorem}
A derivation of this solution is shown in Appendix \ref{ProofOfSolutionSectionAppendix}. 
The concrete expression of Eq.~\eq{ContinuumThetaDefinition} 
(and therefore the polynomials $\mathcal S_n(x)$) is given as 
\begin{align}
\widetilde \theta_1 &= \bigl(- \theta_1 \bigr)+ 
\widetilde\theta_{m}^* \theta_{m}^*, \nn\\
\widetilde \theta_2 &= \bigl(- \theta_2 + \theta_1^2\bigr)+
\widetilde\theta_{m-1}^* \theta_{m}^*, \nn\\
\widetilde \theta_3 &= \bigl(- \theta_3 + 2\theta_1\theta_2- \theta_1^3\bigr)+
\widetilde\theta_{m-2}^* \theta_{m}^*, \nn\\
\widetilde \theta_4 &= \bigl(- \theta_4 + \theta_2^2-3\theta_1^2\theta_2 + 2 \theta_1\theta_3
+ \theta_1^4\bigr)+
\widetilde\theta_{m-3}^* \theta_{m}^*, \nn\\
\widetilde \theta_5 &= \bigl(- \theta_5 + \theta_4\theta_1-3\theta_1^2\theta_3 + \theta_2\theta_3
+4 \theta_1^3\theta_2 - 3 \theta_1\theta_2^2 -\theta_1^5 \bigr)+
\widetilde\theta_{m-4}^* \theta_{m}^*. 
\end{align}
It is worth mentioning the relation to the Schur polynomials $P_n\bigl(\{x_j\}_{j\in \mathbb Z}\bigr)$:
\begin{align}
\mathcal S_n\bigl(\{x_j\}_{j\in \mathbb Z}\bigr)
= P_n\bigl(\{y_j\}_{j\in \mathbb Z}\bigr),
\qquad x_n=P_n\bigl(\{-y_j\}_{j\in \mathbb Z}\bigr),
\end{align}
where the Schur polynomials $P_n\bigl(\{x_j\}_{j\in \mathbb Z}\bigr)$ are defined as 
\begin{align}
\sum_{n=0}^\infty z^n P_n\bigl(\{x_j\}_{j\in \mathbb Z}\bigr) 
= \exp\Bigl[\sum_{n=1}^\infty z^n x_n\Bigr]. 
\end{align}
Note that these solutions includes $m \, (=\lfloor \frac{k}{2} \rfloor)$ real parameters. 
Sometimes, eigenvalues of the matrix $S_0^{(\rm sym)}\,\Gamma^{-1}$ 
of the discrete solutions are distinct. In this case, 
such a discrete solution is a special case of the continuum solution. 
However generally these solutions do not include the discrete solutions in Section 
\ref{AvalanchesSubSubSection}, 
since the discrete solutions generally include degeneracy of eigenvalues 
which cannot be resolved by these continuum parameters.

\section{Stability of perturbative backgrounds\label{RHapproachSection}}

In this section, we briefly review 
the Riemann-Hilbert approach and the Deift-Zhou method \cite{RHcite,RHPIIcite,DeiftZhou}, 
and also discuss its physical interpretations in non-critical string theory. 
In particular, we argue that this procedure implies 
{\em an additional physical requirement} about stability 
of classical (or perturbative) backgrounds. We will see that this constraint results in the proper Stokes multipliers 
expected in the two-cut $(1,2)$ critical point. Classical background here means the 
spectral curves which appear as semi-classical (large $N$) solutions of matrix models. 

The role of the Riemann-Hilbert approach is to obtain the $t$ dependence of physical 
amplitudes (for example, asymptotic expansion in $t$) 
by using an integration expression 
which can be derived from the ODE system in $\zeta$. 
The review article \cite{ItsBook} contains useful references of the Riemann-Hilbert approach. 

\begin{figure}[htbp]
\begin{center}
\begin{picture}(180,190)(0,0)
\end{picture}
\begin{picture}(5,0)(200,-89)
\put(70,70){\line(0,1){10}\line(1,0){10}}
\put(73,75){$\zeta$}
\put(0,0){$0$}
{
\thicklines
\put(10,10){\rotatebox{-90}{\textcolor{red}{\line(1,0){70}}}}
\put(10,10){\rotatebox{-60}{\textcolor{green}{\line(1,0){70}}}}
\put(10,10){\rotatebox{-30}{\textcolor{blue}{\line(1,0){70}}}}
\put(10,10){\rotatebox{0}{\textcolor{red}{\line(1,0){70}}}}
\put(9.5,10){\rotatebox{30}{\textcolor{green}{\line(1,0){70}}}}
\put(9.5,10){\rotatebox{60}{\textcolor{blue}{\line(1,0){70}}}}
\put(9.5,10){\rotatebox{90}{\textcolor{red}{\line(1,0){70}}}}
\put(-25.5,10){\rotatebox{120}{\textcolor{green}{\line(1,0){70}}}}
\put(-51,10){\rotatebox{150}{\textcolor{blue}{\line(1,0){70}}}}
\put(-60,9.5){\rotatebox{180}{\textcolor{red}{\line(1,0){70}}}}
\put(-50,10){\rotatebox{-150}{\textcolor{green}{\line(1,0){70}}}}
\put(-25,10){\rotatebox{-120}{\textcolor{blue}{\line(1,0){70}}}}
}
\put(-8,0){
\rotatebox{0}{
\begin{picture}(0,0)(0,0)
\put(12,18){\line(3,-3){15}}
\put(12,25){\line(3,-3){32}}
\put(12,32){\line(3,-3){49}}
\put(12,39){\line(3,-3){62}}
\put(12,46){\line(3,-3){62}}
\put(12,53){\line(3,-3){62}}
\put(12,60){\line(3,-3){62}}
\put(12,67){\line(3,-3){62}}
\put(12,74){\line(3,-3){62}}
\put(12,81){\line(3,-3){62}}
\put(19,81){\line(3,-3){55}}
\put(26,81){\line(3,-3){48}}
\put(33,81){\line(3,-3){41}}
\put(40,81){\line(3,-3){34}}
\put(47,81){\line(3,-3){27}}
\put(54,81){\line(3,-3){20}}
\put(61,81){\line(3,-3){13}}
\put(68,81){\line(3,-3){6}}
\thicklines
\put(13,11){\rotatebox{-30}{\line(1,0){70}}}
\put(12,11){\rotatebox{90}{\line(1,0){70}}}
\end{picture}}}
\put(15.5,-4){
\rotatebox{90}{
\begin{picture}(0,0)(0,0)
\put(12,18){\line(3,-3){15}}
\put(12,25){\line(3,-3){32}}
\put(12,32){\line(3,-3){49}}
\put(12,39){\line(3,-3){62}}
\put(12,46){\line(3,-3){62}}
\put(12,53){\line(3,-3){62}}
\put(12,60){\line(3,-3){62}}
\put(12,67){\line(3,-3){62}}
\put(12,74){\line(3,-3){62}}
\put(12,81){\line(3,-3){62}}
\put(19,81){\line(3,-3){55}}
\put(26,81){\line(3,-3){48}}
\put(33,81){\line(3,-3){41}}
\put(40,81){\line(3,-3){34}}
\put(47,81){\line(3,-3){27}}
\put(54,81){\line(3,-3){20}}
\put(61,81){\line(3,-3){13}}
\put(68,81){\line(3,-3){6}}
\thicklines
\put(12,11){\rotatebox{-30}{\line(1,0){70}}}
\put(12,11){\rotatebox{90}{\line(1,0){70}}}
\end{picture}}}
\put(16,19.5){
\rotatebox{180}{
\begin{picture}(0,0)(0,0)
\put(12,18){\line(3,-3){15}}
\put(12,25){\line(3,-3){32}}
\put(12,32){\line(3,-3){49}}
\put(12,39){\line(3,-3){62}}
\put(12,46){\line(3,-3){62}}
\put(12,53){\line(3,-3){62}}
\put(12,60){\line(3,-3){62}}
\put(12,67){\line(3,-3){62}}
\put(12,74){\line(3,-3){62}}
\put(12,81){\line(3,-3){62}}
\put(19,81){\line(3,-3){55}}
\put(26,81){\line(3,-3){48}}
\put(33,81){\line(3,-3){41}}
\put(40,81){\line(3,-3){34}}
\put(47,81){\line(3,-3){27}}
\put(54,81){\line(3,-3){20}}
\put(61,81){\line(3,-3){13}}
\put(68,81){\line(3,-3){6}}
\thicklines
\put(12,11){\rotatebox{-30}{\line(1,0){70}}}
\put(12,11){\rotatebox{90}{\line(1,0){70}}}
\end{picture}}}
\put(-3,24){
\rotatebox{-90}{
\begin{picture}(0,0)(0,0)
\put(12,18){\line(3,-3){15}}
\put(12,25){\line(3,-3){32}}
\put(12,32){\line(3,-3){49}}
\put(12,39){\line(3,-3){62}}
\put(12,46){\line(3,-3){62}}
\put(12,53){\line(3,-3){62}}
\put(12,60){\line(3,-3){62}}
\put(12,67){\line(3,-3){62}}
\put(12,74){\line(3,-3){62}}
\put(12,81){\line(3,-3){62}}
\put(19,81){\line(3,-3){55}}
\put(26,81){\line(3,-3){48}}
\put(33,81){\line(3,-3){41}}
\put(40,81){\line(3,-3){34}}
\put(47,81){\line(3,-3){27}}
\put(54,81){\line(3,-3){20}}
\put(61,81){\line(3,-3){13}}
\put(68,81){\line(3,-3){6}}
\thicklines
\put(12,11){\rotatebox{-30}{\line(1,0){70}}}
\put(12,11){\rotatebox{90}{\line(1,0){70}}}
\end{picture}}}
\put(3,9.5){
\rotatebox{0}{
\begin{picture}(0,0)(0,0)
\thicklines
\put(0,0){\rotatebox{-15}{\textcolor{magenta}{\dashline[50]{4}[0.7](1,0)(70,0)}}}
\put(0,0){\rotatebox{-15}{\qquad \quad \textcolor{magenta}{\vector(1,0){5}}}}
\end{picture}}}
\put(6,6){
\rotatebox{90}{
\begin{picture}(0,0)(0,0)
\thicklines
\put(0,0){\rotatebox{-15}{\textcolor{magenta}{\dashline[50]{4}[0.7](1,0)(70,0)}}}
\put(0,0){\rotatebox{-15}{\qquad \quad \textcolor{magenta}{\vector(1,0){5}}}}
\end{picture}}}
\put(6,14){
\rotatebox{-90}{
\begin{picture}(0,0)(0,0)
\thicklines
\put(0,0){\rotatebox{-15}{\textcolor{magenta}{\dashline[50]{4}[0.7](1,0)(70,0)}}}
\put(0,0){\rotatebox{-15}{\qquad \quad \textcolor{magenta}{\vector(1,0){5}}}}
\end{picture}}}
\put(6,10){
\rotatebox{180}{
\begin{picture}(0,0)(0,0)
\thicklines
\put(0,0){\rotatebox{-15}{\textcolor{magenta}{\dashline[50]{4}[0.7](1,0)(70,0)}}}
\put(0,0){\rotatebox{-15}{\qquad \quad \textcolor{magenta}{\vector(1,0){5}}}}
\end{picture}}}
\put(58,47){\footnotesize (1,2)}
\put(70,14){\footnotesize (2,3)}
\put(33,74){\footnotesize (3,1)}
\put(-33,74){\footnotesize (1,2)}
\put(-70,14){\footnotesize (2,3)}
\put(-58,47){\footnotesize (3,1)}
\put(33,-58){\footnotesize (1,2)}
\put(0,-67){\footnotesize (2,3)}
\put(58,-33){\footnotesize (3,1)}
\put(-58,-33){\footnotesize (1,2)}
\put(0,82){\footnotesize (2,3)}
\put(-33,-58){\footnotesize (3,1)}
\put(3,-93){(a)}
\end{picture}
\begin{picture}(5,0)(0,-89)
\put(70,70){\line(0,1){10}\line(1,0){10}}
\put(73,75){$\zeta$}
\put(0,0){$0$}
{
\thicklines
\put(10,10){\rotatebox{-90}{\textcolor{red}{\line(1,0){70}}}}
\put(10,10){\rotatebox{-60}{\textcolor{green}{\line(1,0){70}}}}
\put(10,10){\rotatebox{-30}{\textcolor{blue}{\line(1,0){70}}}}
\put(10,10){\rotatebox{0}{\textcolor{red}{\line(1,0){70}}}}
\put(9.5,10){\rotatebox{30}{\textcolor{green}{\line(1,0){70}}}}
\put(9.5,10){\rotatebox{60}{\textcolor{blue}{\line(1,0){70}}}}
\put(9.5,10){\rotatebox{90}{\textcolor{red}{\line(1,0){70}}}}
\put(-25.5,10){\rotatebox{120}{\textcolor{green}{\line(1,0){70}}}}
\put(-51,10){\rotatebox{150}{\textcolor{blue}{\line(1,0){70}}}}
\put(-60,9.5){\rotatebox{180}{\textcolor{red}{\line(1,0){70}}}}
\put(-50,10){\rotatebox{-150}{\textcolor{green}{\line(1,0){70}}}}
\put(-25,10){\rotatebox{-120}{\textcolor{blue}{\line(1,0){70}}}}
}
\put(3,-6.5){
\rotatebox{45}{
\begin{picture}(0,0)(0,0)
{
\thicklines
\put(10,10){\rotatebox{-90}{\textcolor{red}{\dashline[50]{4}[0.7](1,0)(70,0)}}}
\put(10,10){\rotatebox{-90}{\qquad \quad \textcolor{red}{\vector(1,0){5}}}}
\put(10,10){\rotatebox{-60}{\textcolor{green}{\dashline[50]{4}[0.7](1,0)(70,0)}}}
\put(10,10){\rotatebox{-60}{\qquad \quad \textcolor{green}{\vector(1,0){5}}}}
\put(10,10){\rotatebox{-30}{\textcolor{blue}{\dashline[50]{4}[0.7](1,0)(70,0)}}}
\put(10,10){\rotatebox{-30}{\qquad \quad \textcolor{blue}{\vector(1,0){5}}}}
\put(10,10){\rotatebox{0}{\textcolor{red}{\dashline[50]{4}[0.7](1,0)(70,0)}}}
\put(10,10){\rotatebox{0}{\qquad \quad \textcolor{red}{\vector(1,0){5}}}}
\put(6.5,10){\rotatebox{30}{\textcolor{green}{\dashline[50]{4}[0.7](1,0)(70,0)}
\quad\quad \, \,\textcolor{green}{\vector(1,0){5}}}}
\put(2,10){\rotatebox{60}{\textcolor{blue}{\dashline[50]{4}[0.7](1,0)(70,0)}\quad \quad\, \,\,\,\textcolor{blue}{\vector(1,0){5}}}}
\put(0.5,10){\rotatebox{90}{\textcolor{red}{\dashline[50]{4}[0.7](1,0)(70,0)}\qquad \quad \textcolor{red}{\vector(1,0){5}}}}
\put(-19,10){\rotatebox{120}{\textcolor{green}{\dashline[50]{4}[0.7](1,0)(70,0)}\qquad \quad \textcolor{green}{\vector(1,0){5}}}}
\put(-31,10){\rotatebox{150}{\textcolor{blue}{\dashline[50]{4}[0.7](1,0)(70,0)}\qquad \quad \textcolor{blue}{\vector(1,0){5}}}}
\put(-30,9.5){\rotatebox{180}{\textcolor{red}{\dashline[50]{4}[0.7](1,0)(70,0)}\qquad \quad \textcolor{red}{\vector(1,0){5}}}}
\put(-25,9.5){\rotatebox{-150}{\textcolor{green}{\dashline[50]{4}[0.7](1,0)(70,0)}\qquad \quad \textcolor{green}{\vector(1,0){5}}}}
\put(-9.5,10){\rotatebox{-120}{\textcolor{blue}{\dashline[50]{4}[0.7](1,0)(70,0)}\qquad \quad \textcolor{blue}{\vector(1,0){5}}}}
}
\end{picture}}}
\put(58,47){\footnotesize (1,2)}
\put(70,14){\footnotesize (2,3)}
\put(33,74){\footnotesize (3,1)}
\put(-33,74){\footnotesize (1,2)}
\put(-70,14){\footnotesize (2,3)}
\put(-58,47){\footnotesize (3,1)}
\put(33,-58){\footnotesize (1,2)}
\put(0,-67){\footnotesize (2,3)}
\put(58,-33){\footnotesize (3,1)}
\put(-58,-33){\footnotesize (1,2)}
\put(0,82){\footnotesize (2,3)}
\put(-33,-58){\footnotesize (3,1)}
\put(3,-93){(b)}
\end{picture}
\end{center}
 \caption{\footnotesize These are examples in the $3$-cut $(1,1)$ critical point. 
a) The coarse Stokes sectors (shadowed domains) and the discontinuity lines $\mathcal K$ 
(dashed lines). Basically, any lines in the intersections $D_{3n}\cap D_{3(n+1)}$ are allowed. 
b) The discontinuity lines $\mathcal K$ (dashed lines) with respect to the fine Stokes sectors. They are related 
to the lines in (a) by continuous deformations which do not cross any divergence in the 
Riemann-Hilbert integral \eq{TheSolutionOfTheRiemannHilbertProblem}. 
\label{FigureOfJumpLines}}
\end{figure}
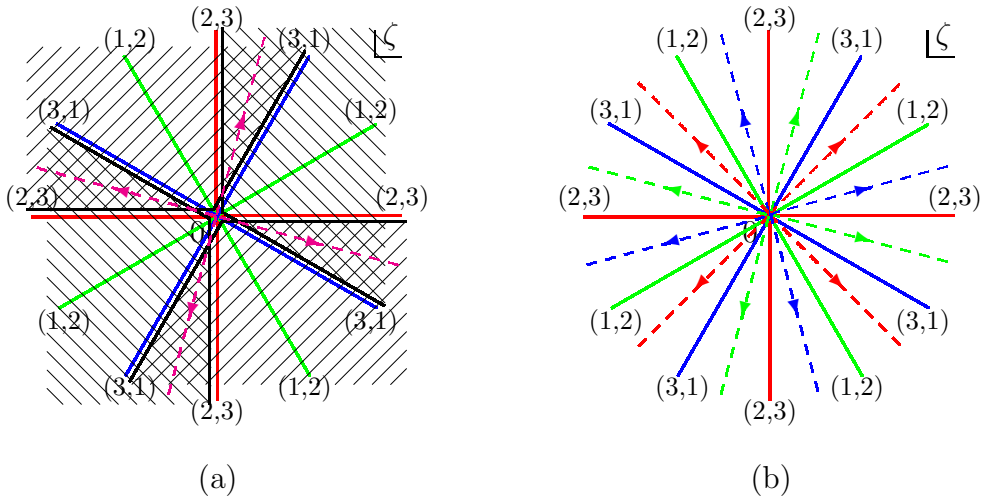

Roughly speaking, in the Riemann-Hilbert approach, we first discard the analytic continuity 
of the canonical solutions \eq{FirstAppearanceOfCanonicalSolutions} 
and keep the form of asymptotic expansion \eq{FirstAppearanceOfAsymptoticExpansion} 
in the complex plane $\mathbb C$. 
In practice, we introduce some Stokes sectors (here we consider fine Stokes sectors) $D_n$ 
and canonical solutions on them, $\widetilde \Psi_n(t;\zeta)$. 
As it has been reviewed in Section \ref{Section2StokesPheno}, these canonical solutions have 
the same asymptotic expansion in each Stokes sector \eq{FirstAppearanceOfCanonicalSolutions} 
and the difference of these canonical solutions is expressed by Stokes matrices 
\eq{FirstAppearanceOfStokesMatrix}. 
Therefore, inside the intersection of two Stokes sectors, 
we introduce a semi-infinite straight line from the origin, $\mathcal K_n$, 
\begin{align}
\mathcal K_n = \{\zeta = u \,e^{i\widetilde \chi_n};\, u\in \mathbb R_+\}\quad 
\text{with \quad ${}^\exists\widetilde \chi_n\in [0,2\pi )$ \quad  s.t. }
\quad \mathcal K_n \subset D_n \cap D_{n+1},
\end{align}
and define the following new function 
$\Psi_{\rm RH}(t;\zeta)$ which is analytic in $\zeta \in \mathbb C \setminus  \mathcal K$ with 
$\mathcal K\equiv \bigcup_n\mathcal K_n$:
\begin{align}
\Psi_{\rm RH}(t;\zeta)= \widetilde \Psi_n(t;\zeta) \qquad \zeta \in D(\widetilde \chi_{n-1},\widetilde \chi_{n}),\qquad (n=1,2,\cdots),
\end{align}
which has the following uniform asymptotic expansion in 
$\zeta \in \mathbb C \setminus  \bigcup_n\mathcal K_n$:
\begin{align}
\Psi_{\rm RH}(t;\zeta) \asymeq \widetilde \Psi_{\rm asym}(t;\zeta)
=\widetilde Y(t;\zeta)\, e^{\widetilde \varphi(t;\zeta)}, 
\qquad \zeta \to \infty \in \mathbb C \setminus  \bigcup_n\mathcal K_n. 
\label{PhiFunctionWithJump}
\end{align}
The lines $\mathcal K$ is referred to as discontinuity lines, and examples are 
shown in Fig.~\ref{FigureOfJumpLines}. 
Note that the function $\Psi_{\rm RH}(t;\zeta)$ 
has enough information to recover all the canonical solutions 
simply by analytically continuing the argument $\zeta$. 

An essence of the Deift-Zhou method for the Riemann-Hilbert problem \cite{DeiftZhou} 
is introduction of the following $k\times k$ function $g(t;\zeta)$ 
which we shall call {\em (off-shell) string background}:
\begin{align}
&g(t;\zeta)= \diag \bigl(g^{(1)}(t;\zeta),\cdots,g^{(k)}(t;\zeta)\bigr),\nn\\
&\qquad \text{with}\quad 
g^{(i)}(t;\zeta)\equiv \sum_{n=1}^{r} t_n^{(i)} \zeta^n +t_0^{(i)} \ln \zeta 
+ \sum_{n=0}^\infty \frac{1}{n}\,g_n^{(i)}\, \zeta^{-n}. 
\end{align}
If one focuses on the aspect of algebraic curves, 
the function $g(t;\zeta)$ is referred to as {\em (off-shell) background spectral curve}. 
We then obtain the following setting of the Riemann-Hilbert problem:
\begin{Lemma}
[Setting of the Riemann-Hilbert problem]
There exists the set of parameters $t_n^{(i)}\, (i=1,2,\cdots,k;\, n=1,2,\cdots,r)$ 
which satisfies 
\begin{align}
Z(t;\zeta)\equiv \Psi_{\rm RH}(t;\zeta) e^{-g(t;\zeta)}\to I_k, \qquad \bigl(\zeta \to \infty\in \mathbb C\setminus \bigcup_n \mathcal K_n\bigr). \label{RiemannHilbertBC1}
\end{align}
The $k\times k$ matrix function $Z(t;\zeta)$ then satisfies the following discontinuity relation: 
\begin{align}
Z_+(t;\zeta) =Z_-(t;\zeta)\, G_n(t;\zeta),\qquad
G_n(t;\zeta)\equiv e^{g(t;\zeta)} S_n e^{-g(t;\zeta)},
\qquad \text{along}\qquad  \zeta \in \mathcal K_n,
\label{RiemannHilbertBC2}
\end{align}
where $n=1,2,\cdots$ and 
we define $Z_\pm (t;\zeta)\equiv \lim_{a\to 0} Z(t;\zeta\pm a\epsilon)$ 
with a vector $\epsilon$ which directs towards the left hand side of the line $\mathcal K_n$. 
\end{Lemma}
In general, the parameters $t_n^{(i)}\, (i=1,2,\cdots,k;\, n=1,2,\cdots,r)$ 
are the integrable deformations of the $k$-component KP hierarchy \cite{kcKP}. 
These are then given by the Lax equations: 
\begin{align}
g_{\rm str} \frac{\del}{\del t_n^{(i)}} \widetilde \Psi(t;\zeta) 
&= \Bigl[\widetilde {\mathcal P}_{-n}^{(i)}\, \zeta^{n}
+ \widetilde {\mathcal P}_{-n+1}^{(i)}(t)\, \zeta^{n-1}+\cdots+ 
\widetilde {\mathcal P}_{0}^{(i)}\Bigr] \widetilde \Psi(t;\zeta) 
\equiv \widetilde {\mathcal P}^{(i)}(t;\zeta)\,\widetilde \Psi(t;\zeta). 
\end{align}
This information is understood as given information of the system 
and non-normalizable string moduli space which should not be minimized 
by the string dynamics \cite{SeibergShenker}. 
Note that the Stokes matrices are invariants of these 
integrable deformation: 
\begin{align}
\frac{\del S_m}{\del t_n^{(i)}}= 0,
\qquad \bigl(i=1,\cdots,k;\,n=1,2,\cdots,r;\,m=1,2,\cdots \bigr),
\end{align}
and therefore the multipliers are integration constants (initial conditions) of these deformations. 
In this sense, they are also understood as non-normalizable string moduli space 
of the dynamics in the strong-coupling region of string theory. 

Since the Stokes multipliers are integration constants of the system, 
we can uniquely obtain all the information by identifying the deformation parameters 
$t_n^{(i)}$ $(i=1,2,\cdots,k;\, n=1,2,\cdots,r)$
and the Stokes multipliers. The fact is given in the form of the following theorem:
\begin{Theorem} 
[The Riemann-Hilbert problem (see \cite{ItsBook})]
For a given analytic function $G(t;\zeta)$ on the discontinuity line $\zeta \in \mathcal K = \bigcup_n \mathcal K_n$, 
\begin{align}
G(t;\zeta) = G_n(t;\zeta) \equiv  e^{g(t;\zeta)} S_n e^{-g(t;\zeta)} \qquad \zeta \in \mathcal K_n 
\qquad (n=1,2,\cdots), 
\end{align}
there exists a unique holomorphic function $Z(t;\zeta)$ which satisfies Eqs.~\eq{RiemannHilbertBC1} and \eq{RiemannHilbertBC2}, and is given as 
\begin{align}
Z(t;\zeta)&= I_k+\int_{\mathcal K} \frac{d\lambda}{2\pi i} 
\frac{\rho(\lambda)(G(\lambda)-I_k)}{\lambda-\zeta} \nn\\
&= I_k + \sum_{n,i,j} s_{n,i,j}\int_{\mathcal K_n} 
\frac{d\lambda}{2\pi i} 
\frac{\rho(\lambda) E_{i,j}}{\lambda-\zeta}
e^{g^{(i)}(t;\lambda)-g^{(j)}(t;\lambda)}, \label{TheSolutionOfTheRiemannHilbertProblem}
\end{align}
with $\rho(\zeta) \equiv Z_-(\zeta)$ on $\zeta \in \mathcal K=\bigcup_n \mathcal K_n$.  
\label{TheoremRiemannHilbertProblem}
\end{Theorem}
By using the Riemann-Hilbert solution \eq{TheSolutionOfTheRiemannHilbertProblem}, 
one can obtain the canonical solutions to the ODE system 
(defined in \eq{PhiFunctionWithJump}) as a function of $t$:
\begin{align}
\Psi_{\rm RH}(t;\zeta) = Z(t;\zeta) \, e^{g(t;\zeta)} 
\asymeq \widetilde \Psi_{\rm asym}(t;\zeta)=\widetilde Y(t;\zeta)  \, e^{\widetilde \varphi(t;\zeta)},
\qquad \zeta \to \infty \in \mathbb C \setminus \mathcal K. \label{SolutionOfCanonicalSolutions}
\end{align}
Note that the ``density function $\rho(\zeta)$'' is given by $Z(t;\zeta)$ itself, 
and then the function $\rho(\zeta)$ satisfies the following integral equation:
\begin{align}
\rho(\zeta) = I_k + \int_{\mathcal K} \frac{d\lambda}{2\pi i} \frac{\rho(\lambda) (G(\lambda)-I_k)}{\lambda-\zeta + \epsilon},\qquad \zeta \in \mathcal K. 
\end{align}
Therefore, one can recursively solve it and the solution is given as the following infinite sum of integrals: 
\begin{align}
Z(t;\zeta) = I_k + \sum_{n=1}^\infty \prod_{i=1}^n 
\biggl[ \int_{\mathcal K} \frac{d\lambda_i}{2\pi i}\biggr]
\prod_{j=2}^{n} \biggl[\frac{G(\lambda_j)-I_k}{\lambda_j-\lambda_{j-1}+\epsilon}\biggr] \frac{G(\lambda_1)-I_k}{\lambda_1-\zeta},
\end{align}
with the assumption that
\begin{align}
\int_{\mathcal K} \frac{d\lambda}{2\pi i } \bigl(G(\lambda)-I_k\bigr),
\end{align}
is sufficiently small. 
Note that we use the following multiplication rule of matrices: 
$\prod_{j=1}^n A_j\equiv A_n A_{n-1}\cdots A_1$. 
In terms of componets, this is expressed as 
\begin{align}
&Z(t;\zeta)  = I_k + \sum_{n,i,j} s_{n,i,j} E_{i,j} \int_{\mathcal K_n} 
\frac{d\lambda_1}{2\pi i} 
\frac{e^{g^{(i)}(t;\lambda_1)-g^{(j)}(t;\lambda_1)}}{\lambda_1-\zeta}+\nn\\
&\quad + \sum_{n_1,n_2,i,j,l } s_{n_2,i,l}s_{n_1,l,j} E_{i,j} \int_{\mathcal K_{n_1}} 
\frac{d\lambda_1}{2\pi i} \int_{\mathcal K_{n_2}} 
\frac{d\lambda_2}{2\pi i} 
\frac{e^{g^{(i)}(t;\lambda_2)-g^{(l)}(t;\lambda_2)+g^{(l)}(t;\lambda_1)-g^{(j)}(t;\lambda_1)}}{(\lambda_2-\lambda_1-\epsilon)(\lambda_1-\zeta)}+\cdots. 
\label{RHProblemRecursionIntegral}
\end{align}
This expression is formally convergent if the subsequent integral are small enough. 
In this case, one can evaluate the leading contribution by truncating higher terms 
(the so-called Born approximation). It is worth mentioning that this integral is quite 
similar to the D-instanton operator formalism in the free-fermion formulation \cite{fy12,fy3} 
by interpreting $g^{(i)}(t;\zeta)$ as the free boson operator $\varphi^{(i)}_0(\zeta)$ in the system.  

An important point here is that, in the Riemann-Hilbert approach, 
the string background $g(t;\zeta)$ is {\em arbitrary} except for the parameters 
$t_n^{(i)}\, (i=1,2,\cdots,k;\, n=1,2,\cdots,r)$, and then 
generally is different from 
the semi-classical resolvent amplitudes $\widetilde \varphi(t;\zeta)$ of 
Eq.~\eq{VerPhiDefinition} 
which is obtained as a solution to the equation of motion (or loop equations) 
in the large $N$ limit of the matrix models. 
As one can see in Theorem \ref{TheoremRiemannHilbertProblem}, 
the role of the string background $g(t;\zeta)$ is 
a reference background in the Riemann-Hilbert problem. 
Therefore, from the string-theory viewpoints, 
the string backgrounds $g(t;\zeta)$ are 
generally understood as {\em off-shell backgrounds of string theory} 
and in this sense the Riemann-Hilbert approach realizes 
{\em an off-shell background independent formulation of string theory}. 

In order to understand $g(t;\zeta)$ as off-shell backgrounds of string theory, 
it is worth mentioning the interpretation of {\em the position of cuts}. 
Taking into account the consideration 
given around Eq.~\eq{GeneralStokeslinesForCuts}, 
we can define the cuts on the off-shell background as a combination of general Stokes lines: 
\begin{align}
{\rm Re}\Bigl(g^{(i)}(t;\zeta)-g^{(j)}(t;\zeta)\Bigr) = 0,
\end{align}
which is obtained by an analytic deformation of the matrix contour 
$\omega^{1/2}\mathcal C^{(k)}$ (so that it realizes the multi-cut boundary condition 
around $\zeta\to \infty$). 
Note that this consideration is possible after imposing proper Stokes phenomena 
which solve the multi-cut boundary condition, 
as it is carried out in Section \ref{MultiCutBCSection}.

This viewpoint also provides the following consideration: 
If one chooses $g(t;\zeta)$ as a semi-classical resolvent function 
$\widetilde \varphi(t;\zeta)$, then the evaluation of Eqs.~\eq{SolutionOfCanonicalSolutions} 
and \eq{TheSolutionOfTheRiemannHilbertProblem} in $g_{\rm str}\to 0$, 
\begin{align}
\Psi_{\rm RH}(t,\zeta)= Z(t;\zeta)\, e^{g(t;\zeta)} = \Bigl[I_k + \cdots \Bigr]\,e^{g(t;\zeta)}, 
\end{align}
is a calculation of quantum corrections from the background spectral curve 
$g(t;\zeta)$ which is given by the semi-classical resolvent.
Therefore, {\em if the resolvent background is a stable vacuum of this system, 
the non-perturbative corrections should be exponentially small.} This is the additional constraint 
for the Stokes multipliers and is referred to as {\em small-instanton condition}. 

\subsection{The small-instanton condition for the $2$-cut critical point \label{TwoCutSectionSmallInstantonCond}}

Here we consider the small-instanton condition in the $2$-cut $(1,2)$ critical point. 
Mathematically, the Riemann-Hilbert problem in this case has been evaluated in 
\cite{RHPIIcite,DeiftZhou,Kap12,Novok,IK,BI} 
in the larger classes of Stokes multipliers (See the review \cite{ItsBook}). 
In particular, according to the Deift-Zhou procedure \cite{DeiftZhou}, 
one first deforms the discontinuity lines $\mathcal K$ to {\em anti-Stokes lines}.
The concept of anti-Stokes lines depends on saddle points of the string background $g(t;\zeta)$:
\begin{align}
\text{saddle points $\zeta_*=\zeta_{i,j}^{(n)}$:}\qquad 
\frac{\del }{\del \zeta}\Bigl[g^{(i)}(t;\zeta)-g^{(j)}(t;\zeta)\Bigr]\bigg|_{\zeta = \zeta_*} = 0,
\qquad (i,j=1,2,\cdots,k). 
\end{align}

\begin{Definition} [Anti-Stokes lines] 
Anti-Stokes lines ${\rm ASL}_{i,j}^{(n)}$ 
are defined for each pair of $(i,j)$ as 
\begin{align}
{\rm ASL}_{i,j}^{(n)}=\biggl\{ \zeta \in \mathbb C;\, {\rm Im} \Bigl[g^{(i,j)}(t;\zeta)\Bigr] = {\rm Im}\Bigl[g^{(i,j)}(t;\zeta_{i,j}^{(n)})\Bigr] \biggr\}, 
\end{align}
where $\zeta_{i,j}^{(n)}$ is a saddle point of the function $g^{(i,j)}(t;\zeta)\equiv g^{(i)}(t;\zeta)-g^{(j)}(t;\zeta)$. 
\end{Definition}
In the procedure of the Deift-Zhou method, one can choose the string background $g(t;\zeta)$, 
however, we know that the $2$-cut $(1,2)$ critical point has two phases with respect to 
the sign of $t$ cosmological constant \cite{UniCom}. 
Therefore, we {\em choose} the string background 
according to the actual phase appearing in the two-cut matrix model:%
\footnote{Note that we are here imposing {\em a physical requirement}, 
by taking into account the Deift-Zhou method \cite{DeiftZhou}. In the Deift-Zhou procedure, 
one considers an arbitrary Stokes multipliers, and the function $g(t;\zeta)$ 
is a function which we choose so that there is no divergence in the RH calculation. 
In this way, we can obtain the asymptotic form in $t$ for these arbitrary Stokes multipliers. 
In this section, on the other hand, we impose a physical constraint in which 
the physical background $g(t;\zeta)$ obtained from the matrix models 
is stable perturbative background with small non-perturbative effects. 
Therefore, this constraint picks up the special and physical Stokes multipliers. }
\begin{align}
g(t;\zeta)= \sigma_3 \Bigl[\frac{1}{3}\zeta^3 + t \zeta + \cdots \Bigr] = \left\{
\begin{array}{ll} 
\ds \sigma_3 \Bigl[\frac{1}{3}\bigl(\zeta^2+2 t\bigr)^{3/2}\Bigr] &: \text{two-cut phase $(t>0)$} \cr
\ds \sigma_3\Bigl[\frac{1}{3}\zeta^3+t \zeta \Bigr] &: \text{one-cut phase $(t<0)$}
\end{array}
\right.. \label{PerturbativeVacuaInTwoCutCriticalPointsRH}
\end{align}
Since we know that these curves are realized in the critical point as its stable vacua, 
these perturbative vacua should satisfy the small-instanton condition. 
Below we consider each case separately. 
We skip the calculation which is the same as that in \cite{ItsBook}. 

\begin{figure}[htbp]
\begin{center}
\begin{picture}(130,160)(0,0)
\begin{picture}(80,0)(85,-80)
\put(0,0){\vector(1,0){60}}
\put(0,0){\vector(-1,0){60}}
\thicklines
\put(0,0){\rotatebox{30}{\dottedline{2}(0,0)(60,0)}}
\put(0,0){\dottedline{2}(0,0)(0,60)}
\put(-52,0){\rotatebox{60}{\dottedline{2}(0,0)(0,60)}}
\put(0,0){\rotatebox{-30}{\dottedline{2}(0,0)(60,0)}}
\put(0,0){\dottedline{2}(0,0)(0,-60)}
\put(-52,-30){\rotatebox{-60}{\dottedline{2}(0,0)(0,60)}}
\put(0,0){
\rotatebox{0}{
\put(-11.5,0){
\rotatebox{0}{
\begin{picture}(50,0)(0,0)
\thicklines
\put(0,0){\rotatebox{0}{\textcolor{red}{\line(1,0){70}}}}
\put(0,0){\qquad \quad \textcolor{red}{\vector(1,0){5}}}
\end{picture}}}
\put(-10,-4){
\rotatebox{60}{
\begin{picture}(0,0)(0,0)
\thicklines
\put(0,0){\rotatebox{0}{\textcolor{blue}{\line(1,0){70}}}}
\put(0,0){\qquad \quad \textcolor{blue}{\vector(1,0){5}}}
\end{picture}}}
\put(-10,4){
\rotatebox{-60}{
\begin{picture}(0,0)(0,0)
\thicklines
\put(0,0){\rotatebox{0}{\textcolor{blue}{\line(1,0){70}}}}
\put(0,0){\qquad \quad \textcolor{blue}{\vector(1,0){5}}}
\end{picture}}}
\put(-8,-2.5){
\rotatebox{120}{
\begin{picture}(0,0)(0,0)
\thicklines
\put(0,0){\rotatebox{0}{\textcolor{red}{\line(1,0){70}}}}
\put(0,0){\qquad \quad \textcolor{red}{\vector(1,0){5}}}
\end{picture}}}
\put(-8,3){
\rotatebox{-120}{
\begin{picture}(0,0)(0,0)
\thicklines
\put(0,0){\rotatebox{0}{\textcolor{red}{\line(1,0){70}}}}
\put(0,0){\qquad \quad \textcolor{red}{\vector(1,0){5}}}
\end{picture}}}
\put(-8,-0.262){
\rotatebox{180}{
\begin{picture}(0,0)(0,0)
\thicklines
\put(0,0){\rotatebox{0}{\textcolor{blue}{\line(1,0){70}}}}
\put(0,0){\qquad \quad \textcolor{blue}{\vector(1,0){5}}}
\end{picture}}}
}}
\put(-7,-80){(a)}
\put(60,5){\footnotesize $\mathcal K_0$}
\put(37,55){\footnotesize $\mathcal K_1$}
\put(-50,55){\footnotesize $\mathcal K_2$}
\put(-70,5){\footnotesize $\mathcal K_3$}
\put(-50,-55){\footnotesize $\mathcal K_4$}
\put(37,-55){\footnotesize $\mathcal K_5$}
\put(3.5,0){\begin{picture}(0,0)(0,0)
\put(-6.6,-3){$\bullet$}
\end{picture}}
\end{picture}
\begin{picture}(90,0)(13,-80)
\put(0,0){\vector(1,0){60}}
\put(0,0){\vector(-1,0){60}}
{\thicklines
\put(0,0){\rotatebox{30}{\dottedline{2}(0,0)(60,0)}}
\put(0,0){\dottedline{2}(0,0)(0,60)}
\put(-52,0){\rotatebox{60}{\dottedline{2}(0,0)(0,60)}}
\put(0,0){\rotatebox{-30}{\dottedline{2}(0,0)(60,0)}}
\put(0,0){\dottedline{2}(0,0)(0,-60)}
\put(-52,-30){\rotatebox{-60}{\dottedline{2}(0,0)(0,60)}}
}
\put(0,0){
\rotatebox{0}{
\put(-11.5,0){
\rotatebox{0}{
\begin{picture}(0,0)(0,0)
\thicklines
\put(0,0){\rotatebox{0}{\textcolor{red}{\line(1,0){70}}}}
\put(0,0){\qquad \quad \textcolor{red}{\vector(1,0){5}}}
\end{picture}}}
\put(-10,-4){
\rotatebox{60}{
\begin{picture}(0,0)(0,0)
\put(0,0){\rotatebox{0}{\textcolor{blue}{\dottedline{2}(0,0)(70,0)}}}
\end{picture}}}
\put(-10,4){
\rotatebox{-60}{
\begin{picture}(0,0)(0,0)
\put(0,0){\rotatebox{0}{\textcolor{blue}{\dottedline{2}(0,0)(70,0)}}}
\end{picture}}}
\put(-8,-2.5){
\rotatebox{120}{
\begin{picture}(0,0)(0,0)
\put(0,0){\rotatebox{0}{\textcolor{red}{\dottedline{2}(0,0)(70,0)}}}
\end{picture}}}
\put(-8,3){
\rotatebox{-120}{
\begin{picture}(0,0)(0,0)
\put(0,0){\rotatebox{0}{\textcolor{red}{\dottedline{2}(0,0)(70,0)}}}
\end{picture}}}
\put(-8,-0.262){
\rotatebox{180}{
\begin{picture}(0,0)(0,0)
\thicklines
\put(0,0){\rotatebox{0}{\textcolor{blue}{\line(1,0){70}}}}
\put(0,0){\qquad \quad \textcolor{blue}{\vector(1,0){5}}}
\end{picture}}}
\put(-8,-0.262){
\rotatebox{0}{
\begin{picture}(0,0)(0,0)
\thicklines
\put(-3.7,0){\rotatebox{0}{\textcolor{green}{\line(0,1){30}\vector(0,1){20}}}}
\put(-3.7,0){\rotatebox{0}{\textcolor{green}{\line(0,-1){30}\vector(0,-1){20}}}}
\end{picture}}}
}}
\put(0,0){
\begin{picture}(0,0)(0,0)
\thicklines
\put(-3.7,0){\textcolor{red}{\qbezier(-33,60)(-17,40)(0,30)}}
\put(-20,41.5){\textcolor{red}{\vector(-1,1){5}}}
\put(-3.7,0){\textcolor{blue}{\qbezier(33,60)(17,40)(0,30)}}
\put(12,41.5){\textcolor{blue}{\vector(1,1){5}}}
\put(-3.7,0){\textcolor{red}{\qbezier(-33,-60)(-17,-40)(0,-30)}}
\put(-20,-41.5){\textcolor{red}{\vector(-1,-1){5}}}
\put(-3.7,0){\textcolor{blue}{\qbezier(33,-60)(17,-40)(0,-30)}}
\put(12,-41.5){\textcolor{blue}{\vector(1,-1){5}}}
\end{picture}}
\put(-7,-80){(b)}
\put(60,5){\footnotesize $\mathcal K_0$}
\put(37,55){\footnotesize $\widetilde{\mathcal K}_1$}
\put(-50,55){ \footnotesize $\widetilde{\mathcal K}_2$}
\put(-70,5){\footnotesize $\mathcal K_3$}
\put(-50,-55){\footnotesize $\widetilde{\mathcal K}_4$}
\put(37,-55){\footnotesize $\widetilde{\mathcal K}_5$}
\put(-17,15){\footnotesize $\widetilde{\mathcal K}_U$}
\put(-17,-20){\footnotesize $\widetilde{\mathcal K}_D$}
\put(2,2){\footnotesize $0$}
\put(2,23){\footnotesize $i\sqrt{2t}$}
\put(2,-27){\footnotesize $-i\sqrt{2t}$}
\put(3.5,0){\begin{picture}(0,0)(0,0)
\put(-6.6,-3){$\bullet$}
\end{picture}}
\put(3.5,30){\begin{picture}(0,0)(0,0)
\put(-6.6,-3){$\bullet$}
\end{picture}}
\put(3.5,-30){\begin{picture}(0,0)(0,0)
\put(-6.6,-3){$\bullet$}
\end{picture}}
\end{picture}
\begin{picture}(90,0)(-38,-80)
\put(0,0){\vector(1,0){60}}
\put(0,0){\vector(-1,0){60}}
{\thicklines
\put(0,0){\rotatebox{30}{\dottedline{2}(0,0)(60,0)}}
\put(0,0){\dottedline{2}(0,0)(0,60)}
\put(-52,0){\rotatebox{60}{\dottedline{2}(0,0)(0,60)}}
\put(0,0){\rotatebox{-30}{\dottedline{2}(0,0)(60,0)}}
\put(0,0){\dottedline{2}(0,0)(0,-60)}
\put(-52,-30){\rotatebox{-60}{\dottedline{2}(0,0)(0,60)}}
}
\put(0,0){
\rotatebox{0}{
\put(-11.5,0){
\rotatebox{0}{
\begin{picture}(0,0)(0,0)
\thicklines
\end{picture}}}
\put(-10,-4){
\rotatebox{60}{
\begin{picture}(0,0)(0,0)
\put(0,0){\rotatebox{0}{\textcolor{blue}{\dottedline{2}(0,0)(70,0)}}}
\end{picture}}}
\put(-10,4){
\rotatebox{-60}{
\begin{picture}(0,0)(0,0)
\put(0,0){\rotatebox{0}{\textcolor{blue}{\dottedline{2}(0,0)(70,0)}}}
\end{picture}}}
\put(-8,-2.5){
\rotatebox{120}{
\begin{picture}(0,0)(0,0)
\put(0,0){\rotatebox{0}{\textcolor{red}{\dottedline{2}(0,0)(70,0)}}}
\end{picture}}}
\put(-8,3){
\rotatebox{-120}{
\begin{picture}(0,0)(0,0)
\put(0,0){\rotatebox{0}{\textcolor{red}{\dottedline{2}(0,0)(70,0)}}}
\end{picture}}}
\put(-8,-0.262){
\rotatebox{180}{
\begin{picture}(0,0)(0,0)
\thicklines
\end{picture}}}
\put(-8,-0.262){
\rotatebox{0}{
\begin{picture}(0,0)(0,0)
\thicklines
\put(-3.7,0){\rotatebox{0}{\textcolor{green}{\line(0,1){30}\vector(0,1){20}}}}
\put(-3.7,0){\rotatebox{0}{\textcolor{green}{\line(0,-1){30}\vector(0,-1){20}}}}
\end{picture}}}
}}
\put(0,0){
\begin{picture}(0,0)(0,0)
\thicklines
\put(-3.7,0){\textcolor{red}{\qbezier(-33,60)(-17,40)(0,30)}}
\put(-20,41.5){\textcolor{red}{\vector(-1,1){5}}}
\put(-3.7,0){\textcolor{blue}{\qbezier(33,60)(17,40)(0,30)}}
\put(12,41.5){\textcolor{blue}{\vector(1,1){5}}}
\put(-3.7,0){\textcolor{red}{\qbezier(-33,-60)(-17,-40)(0,-30)}}
\put(-20,-41.5){\textcolor{red}{\vector(-1,-1){5}}}
\put(-3.7,0){\textcolor{blue}{\qbezier(33,-60)(17,-40)(0,-30)}}
\put(12,-41.5){\textcolor{blue}{\vector(1,-1){5}}}
\put(-3.7,0){\textcolor{red}{\qbezier(0,30)(0,0)(60,1)}}
\put(0,16.5){\textcolor{red}{\vector(1,-1){5}}}
\put(-3.7,0){\textcolor{red}{\qbezier(0,-30)(0,0)(60,-1)}}
\put(0,-16.5){\textcolor{red}{\vector(1,1){5}}}
\end{picture}}
\put(-7,-80){(c)}
\put(50,5){\footnotesize $\widetilde{\mathcal K}_{0+}$}
\put(50,-14){\footnotesize $\widetilde{\mathcal K}_{0-}$}
\put(37,55){\footnotesize $\widetilde{\mathcal K}_1$}
\put(-50,55){ \footnotesize $\widetilde{\mathcal K}_2$}
\put(-50,-55){\footnotesize $\widetilde{\mathcal K}_4$}
\put(37,-55){\footnotesize $\widetilde{\mathcal K}_5$}
\put(-17,15){\footnotesize $\widetilde{\mathcal K}_U$}
\put(-17,-20){\footnotesize $\widetilde{\mathcal K}_D$}
\put(2,2){\footnotesize $0$}
\put(2,23){\footnotesize $i\sqrt{2t}$}
\put(2,-27){\footnotesize $-i\sqrt{2t}$}
\put(0,0){
\begin{picture}(0,0)(0,0)
\put(-6.6,-3){$\bullet$}
\end{picture}}
\put(3.5,30){\begin{picture}(0,0)(0,0)
\put(-6.6,-3){$\bullet$}
\end{picture}}
\put(3.5,-30){\begin{picture}(0,0)(0,0)
\put(-6.6,-3){$\bullet$}
\end{picture}}
\end{picture}
\end{picture}
\end{center}
\caption{\footnotesize The discontinuity lines and the DZ curves 
in the two-cut $(1,2)$ critical point of the two-cut phase. 
a) The discontinuity lines $\mathcal K$. There are two kinds of lines: the one kind is 
the lines $\mathcal K_{2n+1}$ on which the integral \eq{RHProblemRecursionIntegral} 
only includes the contributions from
the exponent $e^{g^{(1,2)}(\zeta)}$. The other kind is the lines $\mathcal K_{2n}$ 
on which the integral \eq{RHProblemRecursionIntegral} only includes the contributions from 
the exponent $e^{g^{(2,1)}(\zeta)}$.
b) The DZ curves which are obtained from analytic deformation 
of the original lines $\mathcal K$. A large D-instanton effect appears 
around the origin on the line $\mathcal K_3$. Therefore, we require $\alpha=0$ 
so that this large instanton vanishes. 
c) The resulting DZ lines with $\alpha=0$. 
Two lines along the real axes $\widetilde {\mathcal K}_{0\pm}$ come from 
the Stokes matrices 
on the lines $\widetilde {\mathcal K}_{U}$ and $\widetilde {\mathcal K}_{D}$. 
Saddle point approximation on each line gives ZZ branes in the Liouville theory, 
however contributions from these lines are the same and canceled 
by the $\mathbb Z_2$ symmetry. 
\label{TwoCutDZcurvesFigure}}
\end{figure}
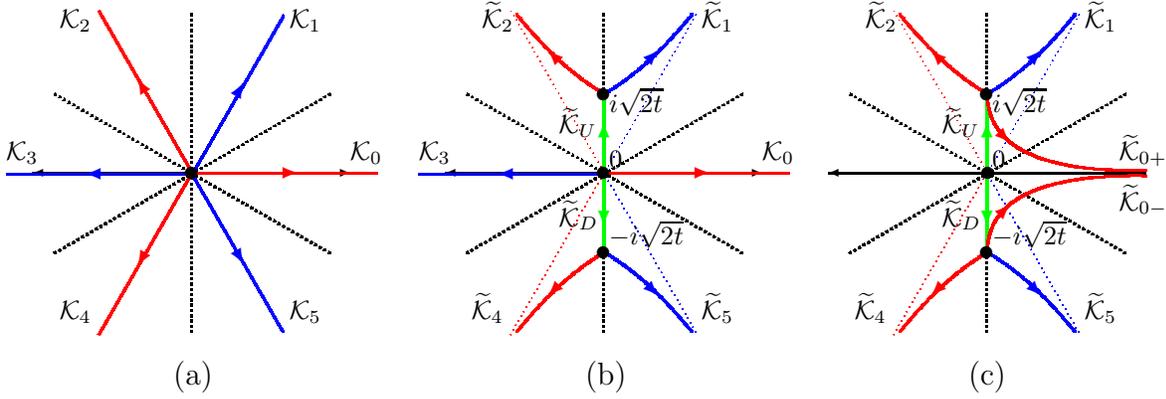

\paragraph{The two-cut phase $(t>0)$}
There are three saddle points of the function $g^{(1,2)}(t;\zeta)\equiv g^{(1)}(t;\zeta)-g^{(2)}(t;\zeta)$:
\begin{align}
\zeta= \zeta_{1,2}^{(n)}:\qquad 
\zeta_{1,2}^{(0)}=0,\qquad \zeta_{1,2}^{(\pm1)}=\pm i\sqrt{2t},
\end{align}
and the values of the function at these saddle points are 
\begin{align}
g^{(1,2)}(t;\zeta_{1,2}^{(0)})= \frac{2}{3}\bigl(2t\bigr)^{3/2},\qquad 
g^{(1,2)}(t;\zeta_{1,2}^{(\pm1)})= 0. 
\end{align}
Note the saddle-point value of the function $g^{(2,1)}(t;\zeta)=-g^{(1,2)}(t;\zeta)$. 
They are understood as instanton actions for the saddle points. 
The deformation of discrete lines $\mathcal K$ to the DZ curves 
is given in Fig.~\ref{TwoCutDZcurvesFigure}. 

On the DZ curves, we then evaluate the integral \eq{RHProblemRecursionIntegral} 
at saddle points \cite{ItsBook}. The small-instanton condition becomes relevant when 
the saddle point $\zeta_{1,2}^{(0)}=0$ of $g^{(1,2)}(t;\zeta)$ contributes in the 
Riemann-Hilbert integral \eq{RHProblemRecursionIntegral}. 
This happens in the integral on the curve $\mathcal K_3$. The relevant part is given as 
\begin{align}
Z(t;\zeta) = \alpha E_{1,2} \int_{\mathcal K_3} \frac{d \lambda}{2\pi i}
\frac{e^{g^{(1,2)}(t;\zeta)}}{\lambda - \zeta}+ \cdots.
\end{align}
The parameter $\alpha$ is the Stokes multipliers of this system \eq{DefAlfBetGam}. 
Therefore, the small-instanton condition requires 
\begin{align}
\alpha =s_0=s_3= 0,
\end{align}
otherwise this perturbative vacuum 
\eq{PerturbativeVacuaInTwoCutCriticalPointsRH} breaks down 
(or decays into some stable vacuum) 
by the large non-perturbative effects. 
Consequently, the solutions to the non-perturbative completion are finally fixed to be
\begin{align}
\alpha= 0,\qquad \beta=\pm1 = -\gamma, \label{SolutionToSmallInstantonTwoCut}
\end{align}
which is known as the Hastings-McLeod solution 
in the Painlev\'e II equation \cite{HastingsMcLeod}. 
As it has been calculated in \cite{HastingsMcLeod}, the final result is given as%
\footnote{In this calculation, we use the local Riemann-Hilbert problems. 
Since the evaluation of the Riemann-Hilbert problem is not our purpose, 
we here skip the calculation. See the review \cite{ItsBook}. 
An intuitive reason for vanishing the D-instanton effects 
(or physical interpretation of the mathematical result) is cancellation 
due to the $\mathbb Z_2$ symmetry of the system. For example, if one introduces 
the formal monodromy (as mentioned around \eq{MonodromyFreeTwoCutInTermsOfPsi}, 
i.e.~adding D0-brane charges in the background) then the instanton effect from the origin $\zeta=0$ appears. }
\begin{align}
f(t)=-2\beta \sqrt{2t}+\cdots\qquad \text{with}\quad \beta=\pm 1, 
\end{align}
especially, the instanton effect which comes 
from a single ZZ-brane at the origin $\zeta=0$ vanishes in this phase.

\paragraph{The one-cut phase $(t<0)$}

\begin{figure}[htbp]
\begin{center}
\begin{picture}(130,160)(0,0)
\begin{picture}(80,0)(85,-80)
\put(0,0){\vector(1,0){60}}
\put(0,0){\vector(-1,0){60}}
\thicklines
\put(0,0){\rotatebox{30}{\dottedline{2}(0,0)(60,0)}}
\put(0,0){\dottedline{2}(0,0)(0,60)}
\put(-52,0){\rotatebox{60}{\dottedline{2}(0,0)(0,60)}}
\put(0,0){\rotatebox{-30}{\dottedline{2}(0,0)(60,0)}}
\put(0,0){\dottedline{2}(0,0)(0,-60)}
\put(-52,-30){\rotatebox{-60}{\dottedline{2}(0,0)(0,60)}}
\put(0,0){
\rotatebox{0}{
\put(-11.5,0){
\rotatebox{0}{
\begin{picture}(50,0)(0,0)
\thicklines
\put(0,0){\rotatebox{0}{\textcolor{red}{\line(1,0){70}}}}
\put(0,0){\qquad \quad \textcolor{red}{\vector(1,0){5}}}
\end{picture}}}
\put(-10,-4){
\rotatebox{60}{
\begin{picture}(0,0)(0,0)
\thicklines
\put(0,0){\rotatebox{0}{\textcolor{blue}{\line(1,0){70}}}}
\put(0,0){\qquad \quad \textcolor{blue}{\vector(1,0){5}}}
\end{picture}}}
\put(-10,4){
\rotatebox{-60}{
\begin{picture}(0,0)(0,0)
\thicklines
\put(0,0){\rotatebox{0}{\textcolor{blue}{\line(1,0){70}}}}
\put(0,0){\qquad \quad \textcolor{blue}{\vector(1,0){5}}}
\end{picture}}}
\put(-8,-2.5){
\rotatebox{120}{
\begin{picture}(0,0)(0,0)
\thicklines
\put(0,0){\rotatebox{0}{\textcolor{red}{\line(1,0){70}}}}
\put(0,0){\qquad \quad \textcolor{red}{\vector(1,0){5}}}
\end{picture}}}
\put(-8,3){
\rotatebox{-120}{
\begin{picture}(0,0)(0,0)
\thicklines
\put(0,0){\rotatebox{0}{\textcolor{red}{\line(1,0){70}}}}
\put(0,0){\qquad \quad \textcolor{red}{\vector(1,0){5}}}
\end{picture}}}
\put(-8,-0.262){
\rotatebox{180}{
\begin{picture}(0,0)(0,0)
\thicklines
\put(0,0){\rotatebox{0}{\textcolor{blue}{\line(1,0){70}}}}
\put(0,0){\qquad \quad \textcolor{blue}{\vector(1,0){5}}}
\end{picture}}}
}}
\put(-7,-80){(a)}
\put(60,5){\footnotesize $\mathcal K_0$}
\put(37,55){\footnotesize $\mathcal K_1$}
\put(-50,55){\footnotesize $\mathcal K_2$}
\put(-70,5){\footnotesize $\mathcal K_3$}
\put(-50,-55){\footnotesize $\mathcal K_4$}
\put(37,-55){\footnotesize $\mathcal K_5$}
\put(3.5,0){\begin{picture}(0,0)(0,0)
\put(-6.6,-3){$\bullet$}
\end{picture}}
\end{picture}
\begin{picture}(90,0)(13,-80)
\put(0,0){\vector(1,0){60}}
\put(0,0){\vector(-1,0){60}}
{\thicklines
\put(0,0){\rotatebox{30}{\dottedline{2}(0,0)(60,0)}}
\put(0,0){\dottedline{2}(0,0)(0,60)}
\put(-52,0){\rotatebox{60}{\dottedline{2}(0,0)(0,60)}}
\put(0,0){\rotatebox{-30}{\dottedline{2}(0,0)(60,0)}}
\put(0,0){\dottedline{2}(0,0)(0,-60)}
\put(-52,-30){\rotatebox{-60}{\dottedline{2}(0,0)(0,60)}}
}
\put(0,0){
\rotatebox{0}{
\put(-11.5,0){
\rotatebox{0}{
\begin{picture}(0,0)(0,0)
\thicklines
\put(0,0){\rotatebox{0}{\textcolor{red}{\line(1,0){70}}}}
\put(0,0){\qquad \quad \textcolor{red}{\vector(1,0){5}}}
\end{picture}}}
\put(-10,-4){
\rotatebox{60}{
\begin{picture}(0,0)(0,0)
\put(0,0){\rotatebox{0}{\textcolor{blue}{\dottedline{2}(0,0)(70,0)}}}
\end{picture}}}
\put(-10,4){
\rotatebox{-60}{
\begin{picture}(0,0)(0,0)
\put(0,0){\rotatebox{0}{\textcolor{blue}{\dottedline{2}(0,0)(70,0)}}}
\end{picture}}}
\put(-8,-2.5){
\rotatebox{120}{
\begin{picture}(0,0)(0,0)
\put(0,0){\rotatebox{0}{\textcolor{red}{\dottedline{2}(0,0)(70,0)}}}
\end{picture}}}
\put(-8,3){
\rotatebox{-120}{
\begin{picture}(0,0)(0,0)
\put(0,0){\rotatebox{0}{\textcolor{red}{\dottedline{2}(0,0)(70,0)}}}
\end{picture}}}
\put(-8,-0.262){
\rotatebox{180}{
\begin{picture}(0,0)(0,0)
\thicklines
\put(0,0){\rotatebox{0}{\textcolor{blue}{\line(1,0){70}}}}
\put(0,0){\qquad \quad \textcolor{blue}{\vector(1,0){5}}}
\end{picture}}}
\put(-8,-0.262){
\rotatebox{0}{
\begin{picture}(0,0)(0,0)
\thicklines
\put(-21,0.5){\rotatebox{0}{\textcolor{green}{\line(1,0){35}}}}
\put(-21,0){\rotatebox{0}{\textcolor{green}{\line(1,0){35}}}}
\put(-21,0.5){\rotatebox{0}{\textcolor{green}{\vector(1,0){20}}}}
\end{picture}}}
}}
\put(0,0){
\begin{picture}(0,0)(0,0)
\thicklines
\put(-3.7,0){\textcolor{red}{\qbezier(-35,60)(0,0)(-35,-60)}}
\put(-27,36){\textcolor{red}{\vector(-1,2){3}}}
\put(-27,-36){\textcolor{red}{\vector(-1,-2){3}}}
\put(-3.7,0){\textcolor{blue}{\qbezier(35,60)(0,0)(35,-60)}}
\put(19.5,-36){\textcolor{blue}{\vector(1,-2){3}}}
\put(19.5,36){\textcolor{blue}{\vector(1,2){3}}}
\end{picture}}
\put(-7,-80){(b)}
\put(60,5){\footnotesize $\mathcal K_0$}
\put(37,55){\footnotesize $\widetilde{\mathcal K}_1$}
\put(-50,55){ \footnotesize $\widetilde{\mathcal K}_2$}
\put(-70,5){\footnotesize $\mathcal K_3$}
\put(-50,-55){\footnotesize $\widetilde{\mathcal K}_4$}
\put(37,-55){\footnotesize $\widetilde{\mathcal K}_5$}
\put(-10,-13){\footnotesize $\widetilde{\mathcal K}_C$}
\put(2,2){\footnotesize $0$}
\put(20,4){\footnotesize $\sqrt{t}$}
\put(-42,4){\footnotesize $-\sqrt{t}$}
\put(-14,0){\begin{picture}(0,0)(0,0)
\put(-6.6,-3){$\bullet$}
\end{picture}}
\put(21.2,0){\begin{picture}(0,0)(0,0)
\put(-6.6,-3){$\bullet$}
\end{picture}}
\end{picture}
\begin{picture}(90,0)(-40,-80)
\put(0,0){\vector(1,0){60}}
\put(0,0){\vector(-1,0){60}}
{\thicklines
\put(0,0){\rotatebox{30}{\dottedline{2}(0,0)(60,0)}}
\put(0,0){\dottedline{2}(0,0)(0,60)}
\put(-52,0){\rotatebox{60}{\dottedline{2}(0,0)(0,60)}}
\put(0,0){\rotatebox{-30}{\dottedline{2}(0,0)(60,0)}}
\put(0,0){\dottedline{2}(0,0)(0,-60)}
\put(-52,-30){\rotatebox{-60}{\dottedline{2}(0,0)(0,60)}}
}
\put(0,0){
\rotatebox{0}{
\put(-10,-4){
\rotatebox{60}{
\begin{picture}(0,0)(0,0)
\put(0,0){\rotatebox{0}{\textcolor{blue}{\dottedline{2}(0,0)(70,0)}}}
\end{picture}}}
\put(-10,4){
\rotatebox{-60}{
\begin{picture}(0,0)(0,0)
\put(0,0){\rotatebox{0}{\textcolor{blue}{\dottedline{2}(0,0)(70,0)}}}
\end{picture}}}
\put(-8,-2.5){
\rotatebox{120}{
\begin{picture}(0,0)(0,0)
\put(0,0){\rotatebox{0}{\textcolor{red}{\dottedline{2}(0,0)(70,0)}}}
\end{picture}}}
\put(-8,3){
\rotatebox{-120}{
\begin{picture}(0,0)(0,0)
\put(0,0){\rotatebox{0}{\textcolor{red}{\dottedline{2}(0,0)(70,0)}}}
\end{picture}}}
}}
\put(0,0){
\begin{picture}(0,0)(0,0)
\thicklines
\put(-3.7,0){\textcolor{red}{\qbezier(-35,60)(0,0)(-35,-60)}}
\put(-27,36){\textcolor{red}{\vector(-1,2){3}}}
\put(-27,-36){\textcolor{red}{\vector(-1,-2){3}}}
\put(-3.7,0){\textcolor{blue}{\qbezier(35,60)(0,0)(35,-60)}}
\put(19.5,-36){\textcolor{blue}{\vector(1,-2){3}}}
\put(19.5,36){\textcolor{blue}{\vector(1,2){3}}}
\end{picture}}
\put(-7,-80){(c)}
\put(37,55){\footnotesize $\widetilde{\mathcal K}_1$}
\put(-50,55){ \footnotesize $\widetilde{\mathcal K}_2$}
\put(-50,-55){\footnotesize $\widetilde{\mathcal K}_4$}
\put(37,-55){\footnotesize $\widetilde{\mathcal K}_5$}
\put(2,2){\footnotesize $0$}
\put(20,4){\footnotesize $\sqrt{t}$}
\put(-42,4){\footnotesize $-\sqrt{t}$}
\put(-13,0){\begin{picture}(0,0)(0,0)
\put(-6.6,-3){$\bullet$}
\end{picture}}
\put(21,0){\begin{picture}(0,0)(0,0)
\put(-6.6,-3){$\bullet$}
\end{picture}}
\end{picture}
\end{picture}
\end{center}
\caption{\footnotesize The discontinuity lines and the DZ curves 
in the two-cut $(1,2)$ critical point of the one-cut phase. 
a) The discontinuity lines $\mathcal K$ which is the same as two-cut phase. 
b) The DZ curves which are obtained from analytic deformation 
of the original lines $\mathcal K$. A large D-instanton effect appears 
around the saddle point $\zeta=+\sqrt{t}$ on the line $\mathcal K_0$, 
and around the saddle point $\zeta=-\sqrt{t}$ on the line $\mathcal K_3$. 
Therefore, we require $\alpha=0$ so that these large instantons vanishes. 
c) The resulting DZ lines with $\alpha=0$. 
By taking into account the sign of the Stokes multipliers, 
one observes that the integral \eq{RHProblemRecursionIntegral} 
along connected lines $\widetilde{\mathcal K}_2$ 
and $\widetilde{\mathcal K}_4$ 
(and also $\widetilde{\mathcal K}_1$ and $\widetilde{\mathcal K}_5$ 
in the same way) 
can be considered as an integral on the single contour. 
Saddle point approximation on each line gives ZZ branes in the Liouville theory of the one-cut phase. 
\label{OneCutDZcurvesFigure}}
\end{figure}
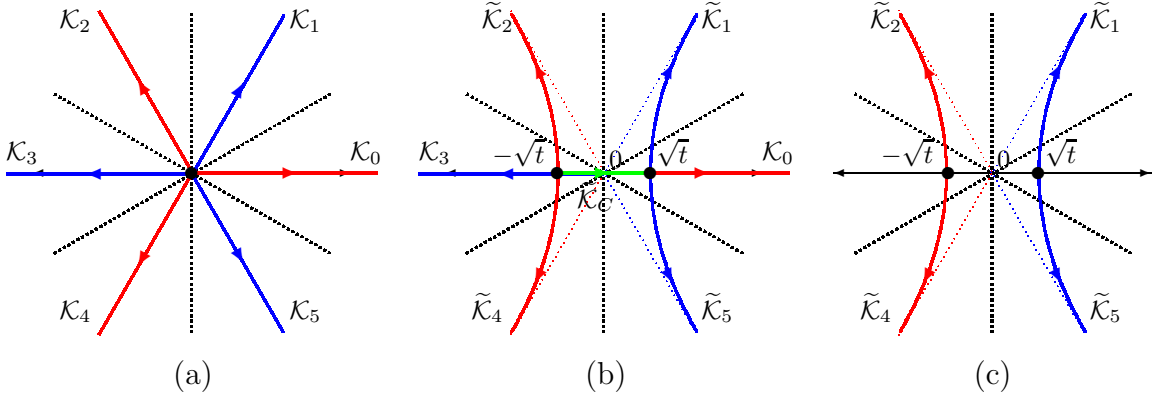

There are two saddle points of the function $g^{(1,2)}(t;\zeta)\equiv g^{(1)}(t;\zeta)-g^{(2)}(t;\zeta)$:
\begin{align}
\zeta= \zeta_{1,2}^{(n)}:\qquad 
\zeta_{1,2}^{(\pm1)}=\pm \sqrt{-t},
\end{align}
and the values of the function at these saddle points are 
\begin{align}
g^{(1,2)}(t;\zeta_{1,2}^{(\pm1)})= \mp \frac{4}{3} (-t)^{3/2}. 
\end{align}
The deformation of discrete lines $\mathcal K$ to the DZ curves 
is given in Fig.~\ref{OneCutDZcurvesFigure}. Note that 
existence of this phase also requires the same constraint $\alpha=0$. 
By taking into account the solution to the non-perturbative completion 
\eq{SolutionToSmallInstantonTwoCut}, 
the Riemann-Hilbert integral \eq{RHProblemRecursionIntegral} 
becomes the following simple contour integrals:
\begin{align}
Z(t;\zeta) &= I_k 
+ \beta E_{1,2}\int_{\mathcal K_{1,2}} \frac{d\lambda}{2\pi i }
\frac{e^{g^{(1,2)}(t;\lambda)}}{\lambda-\zeta} 
- \beta E_{2,1}\int_{\mathcal K_{2,1}} \frac{d\lambda}{2\pi i }
\frac{e^{g^{(2,1)}(t;\lambda)}}{\lambda-\zeta}+\cdots, \nn\\
&= I_k + \frac{\beta}{2\pi i }\Bigl[ i \sqrt{\frac{\pi}{2 \sqrt{-t}}} 
\frac{E_{1,2}}{\sqrt{-t}-\zeta}
-i \sqrt{\frac{\pi}{2 \sqrt{-t}}} 
\frac{E_{2,1}}{-\sqrt{-t}-\zeta}\Bigr]e^{-\frac{4}{3} (-t)^{3/2}}+\cdots, 
\end{align}
therefore the asymptotic expression of $f(t)$ is given as 
\begin{align}
f(t)= -\frac{\beta}{\sqrt{2\pi \sqrt{-t}}}e^{-\frac{4}{3} (-t)^{3/2}}+\cdots \qquad 
\text{with}\quad \beta=\pm 1.
\end{align}
See Eq.~\eq{AsymTwoCutIIIIIIIII}. 
It is worth mentioning that a similar expression 
was found in the $2$-cut $(1,2)$ critical points \cite{fi2} which comes from 
an explicit expression of fermion state within the free-fermion formulation \cite{fkn,fy12,fy3,fi1}, 
although the expression there is given by an infinite sum of super-matrix integrals. 

\subsection{The small-instanton condition for the $k$-cut critical points}

Here we consider the small-instanton constraint in the $k$-cut $(1,1)$ critical points. 
Since we here focus on the additional constraint, 
we only study the saddle point actions for the semi-classical string background 
and evaluation of the Riemann-Hilbert integrals is remained for future investigation. 
The classical backgrounds in these cases are calculated in \cite{CISY1} and given 
in terms of parameter $z$ as 
\begin{align}
&g(t;\zeta)= \diag\bigl(g^{(1)}(t;\zeta),\cdots,g^{(k)}(t;\zeta)\bigr),\qquad g^{(j)}(t;\zeta)= \int^{\omega^{-(j-1)}\zeta} y(z) \, dx(z),\nn\\
&\quad \text{with}\qquad 
x(z)= t \sqrt[k]{\bigl(z-c\bigr)^{l}\bigl(z-b\bigr)^{k-l}},\qquad 
y(z)= t \sqrt[k]{\bigl(z-c\bigr)^{k-l}\bigl(z-b\bigr)^{l}}, \label{ClassicalSolutionsInKcutCases}
\end{align}
with $0=c\,l+b\,(k-l)$. The index $l\, (=0,1,2,\cdots,k-1)$ labels 
generally different solutions. The classical background $g(t;\zeta)$ is then expressed as
\begin{align}
g^{(j)}(t;\zeta)=g^{(1)}\bigl(t;\omega^{-(j-1)}\zeta\bigr),\qquad 
g^{(1)}\bigl(t;x\bigr)=\frac{1}{2} \bigl(z(x)\bigr)^2 - (c+b)\, z(x). \label{InTermsOfZForG}
\end{align}
Here $z(x)$ is the inverse of the function $x(z)$ in Eq.~\eq{ClassicalSolutionsInKcutCases}. 
The saddle points for $g^{(i,j)}(t;\zeta)=g^{(i)}(t;\zeta)-g^{(j)}(t;\zeta)$ are given as 
\begin{align}
\frac{d}{d\zeta} g^{(i,j)}(t;\zeta)=0\qquad 
\Leftrightarrow\qquad \omega^{i-1} x(z) = \omega^{j-1} x(z'),\quad 
\omega^{-(i-1)} y(z)= \omega^{-(j-1)} y(z'), 
\end{align}
and then this can be solved as 
\begin{align}
z'=z^{(n)}_{i,j}\equiv  \biggl(\frac{b\, e^{\frac{i}{2}\chi_{i,j}^{(n)}}  + c\, e^{-\frac{i}{2}\chi_{i,j}^{(n)}} }{2\cos\bigl(\chi_{i,j}^{(n)}/2\bigr)}\biggr),
\qquad 
z=z^{(-n)}_{j,i}\equiv \biggl(\frac{b\, e^{-\frac{i}{2}\chi_{i,j}^{(n)}}  + c\, e^{\frac{i}{2}\chi_{i,j}^{(n)}} }{2\cos\bigl(\chi_{i,j}^{(n)}/2\bigr)}\biggr),
\end{align}
with $\chi_{i,j}^{(n)} \equiv 2\pi \frac{(i-j)+nk}{k-2l},$  $(n=1,2,\cdots)$. 
Substituting these values in Eq.~\eq{InTermsOfZForG}, we obtain the saddle point action: 
\begin{align}
g^{(i,j)}\bigl(t;\zeta_{i,j}^{(n)}\bigr) 
&= \frac{1}{2}\Bigl((z_{i,j}^{(n)})^2-(z_{j,i}^{(-n)})^2\Bigr)
-(b+c)\Bigl(z_{i,j}^{(n)}-z_{j,i}^{(-n)}\Bigr) \nn\\
&= i\frac{c^2-b^2}{2}\tan\bigl(\frac{\chi_{i,j}^{(n)}}{2}\bigr) \in i\,\mathbb R. 
\end{align}
This means that the saddle point action always contributes order 
$\mathcal O(g_{\rm str}^0)$ and then identified as perturbative corrections 
(not as instantons). Therefore, 
there is no additional (small-instanton) constraints on the solutions obtained
in Section \ref{MultiCutBCSection}. 

\section{Conclusion and discussions \label{SectionConclusionDiscussion}}

In this paper, we give concrete solutions to the non-perturbative completion in 
the $k$-cut two-matrix models by a quantitative study of Stokes phenomena. 
The non-perturbative completion problem consists of 
the multi-cut boundary condition for the orthonormal polynomial systems 
and the non-perturbative stability condition for the semi-classical spectral curves in the large $N$ limit. 
By carrying out these procedures, we demonstrated two classes of solutions, 
which are referred to as discrete and continuum solutions. 
Interestingly, the solutions possess kind of ``charges'' in terms of Young diagram representation. 

We note that the continuum solutions to the non-perturbative completion still include 
continuous free parameters, although the two-cut cases have been completely fixed. 
It is conceivable that we might need to rely on further independent physical arguments to reduce 
these degrees of freedom, here we would like to interpret these free parameters as physical moduli parameters 
in the non-perturbative region of the string theory. 
Since the strong-coupling dual theory of the multi-cut matrix models seems 
to be non-critical M theory \cite{CIY1}, these continuous parameters would correspond to 
the non-perturbative (non-normalizable) moduli space of M theory, 
$\mathcal M_{\text{M-theory}}^{(\rm non-norm.)}$ which is a distinct parameter space 
from the string-theory moduli space, $\mathcal M_{\text{string}}^{(\rm non-norm.)}$ and $\mathcal M_{\text{string}}^{(\rm norm.)}$. 
Below we provide a list of issues which deserve further exploration. 

\begin{itemize}
\item In this paper, we have solved Stokes phenomena 
in $\mathbb Z_k$-symmetric critical points. It is also interesting to consider 
similar program in the fractional-superstring critical points \cite{irie2}. 
In particular,we would like to see the emergence of the non-critical M theory 
from the $k\to \infty$ limit \cite{CIY1}. 
\item Our procedure is directly related to Riemann-Hilbert calculus. 
It is useful to examine higher order instanton sectors and generalize the results in \cite{GKM}. 
\item In this paper, we focus on the cases with $\hat p=1$ and small $\hat q$. 
In order to extend this procedure to the general $\hat q$ cases, one should resolve 
several complexities as shown in Eq.~\eq{GeneralrDiagram}. It is of great interest 
to obtain the Stokes multipliers in higher $(\hat p,\hat q)$ critical points. 
In particular, evaluation in the bosonic cases would clarify 
the issue raised in \cite{HHIKKMT}. Also we have to take into account the smoothing 
of the cuts as shown in \cite{MMSS} (also see Appendix \ref{AppendixAiryFunction}). 
\item It is interesting to investigate {\em whether 
the Riemann-Hilbert representation can be written in language of matrix models? }
This resembles the supermatrix models \cite{fi2} which appear by evaluating 
tau-function in terms of free fermions. 
Also it is interesting to compare it with Kontsevich type matrix models \cite{KontsevichMatrixModel}
and also with the non-perturbative topological string-theory block recently proposed in \cite{CDV}. 
\item The Riemann-Hilbert representation is a background independent formulation, 
which allows us to introduce general off-shell background in string theory. 
Therefore, it is interesting to study physics in off-shell backgrounds 
and general concept of background independence in matrix models/string theory. 
\item In the multi-cut matrix models, there are two kinds of perturbative string vacua \cite{CIY1}: 
One is perturbatively isolated sectors (perturbative superselection sectors) 
which are decoupled with other sectors in all-order perturbation theory.   
This phenomenon is an origin of the extra-dimension in M theory. 
The other is perturbative vacua in the string-theory moduli space. 
For survey for the second vacua, the Riemann-Hilbert representation is even more powerful, 
since the off-shell moduli space is understand as the space of 
off-shell string-theory backgrounds. Furthermore, the $\mathbb Z_k$ symmetric critical 
points in the multi-cut matrix models 
have several perturbative vacua which satisfy loop equations. 
Therefore, it is interesting to study non-perturbative string-theory landscape 
from the Riemann-Hilbert approach. In particular, it might be possible to identify 
which observables are suitable for a discription of {\em a potential picture in the moduli space}. 
\item We obtained several solutions to Stokes phenomena which are characterized 
by several charges carried by Young diagrams. 
What is the physical meaning of our solutions? Any relation to W-symmetry or WZNW? 
\item Our solutions are natural generalizations of the Hastings-McLeod solution 
in the Painlev\'e II equation. The Hastings-McLeod solution is known to have 
several special features, for instance analyticity of the solution (See also \cite{ItsBook}).  
Therefore, it is mathematically interesting to understand the analyticity of the solutions in $t$ and to identify the standing point of our solutions 
in general solutions of the string equations. 
\item As is well-known, the integrable deformations in the usual integrable system 
correspond to the moduli space of worldsheet conformal field theory. On the other hand, 
non-trivial deformations of our solutions can be interpreted as 
non-perturbative integrable deformations in {\em physical solutions of string equations}. 
Therefore, these deformations are related to the moduli space of 
the dynamical degree of freedom in the strong coupling region, 
i.e.~degree of freedom in non-critical M theory. 
It is interesting if there is a comprehensive understanding of 
these non-perturbative integrable deformations. 

\end{itemize}

\vspace{1cm}
\noindent
{\bf \large Acknowledgment}  
\vspace{0.2cm}

\noindent
The authors would like to thank Hiroyuki Fuji, Masafumi Fukuma, Kazuyuki Furuuchi, 
Martin Guest, Yasuaki Hikida, Pei-Ming Ho, Alexander R.~Its, Hsien-Chung Kao, Hikaru Kawai, 
Tsunehide Kuroki, Chang-Shou Lin, Feng-Li Lin, Yoshinori Matsuo, Toshio Nakatsu, 
So Okada, Ricardo Schiappa and Dan Tomino for useful discussions and comments. 
C.-T. Chan and H. Irie are supported by National Science Council of Taiwan under the contract 
No.~96-2112-M-021-002-MY3 (C.-T. Chan), No.~99-2112-M-029-001-MY3 (C.-T. Chan) 
and No.~100-2119-M-007-001 (H. Irie). The authors are also supported 
by String Theory Focus Group in National Center for Theoretical Science 
under NSC No.~98-2119-M-002-001 and No.~99-2119-M-007-001.

\appendix

\section{Stokes phenomenon in the Airy function \label{AppendixAiryFunction}}
The non-perturbative relations between the resolvent and the orthonormal polynomials 
are first studied in \cite{MMSS} 
in the $(2,1)$ critical point of bosonic minimal string. 
Since this study also uncovers another aspect of cuts in the resolvent curves 
for the cases of $\hat p\geq 2$, 
we here briefly review the results and summarize the key points. 

In the bosonic $(2,1)$ critical point, 
the orthonormal polynomials satisfy the following differential equation: 
\begin{align}
\zeta \Psi_{\rm orth}(t;\zeta) 
&= \bigl(\del^2 + u(t) \bigr)\, \Psi_{\rm orth}(t;\zeta),\label{AiryBA1} \\ 
g_{\rm str}\frac{\del }{\del \zeta } \Psi_{\rm orth}(t;\zeta) 
&= \del \, \Psi_{\rm orth}(t;\zeta). \label{AiryBA2}
\end{align}
By taking into account the definition $\del \equiv g_{\rm str} \del_t$, 
one can show that the orthonormal polynomial is given as Airy function:
\begin{align}
0 = \Bigl(g_{\rm str}^2\, \frac{\del^2}{\del \zeta^2 } -\zeta-t \Bigr)\, 
\Psi_{\rm orth}(t;\zeta), \qquad 
\Psi_{\rm orth}(t;\zeta) = {\rm Ai}(\zeta+t). 
\end{align}
Here we have concluded $u(t)=-t$ by imposing the integrability condition of \eq{AiryBA1} 
and \eq{AiryBA2}, and also have chosen the damping solution (Airy function) 
as the physical solution \cite{MMSS}: 
\begin{align}
\Psi_{\rm orth}(t;\zeta) \to 0, \qquad \zeta \to \infty.  \label{BCInAiryFunction}
\end{align}
As it is well-known, the asymptotic behavior of the orthonormal polynomial 
$\Psi_{\rm orth}(t;\zeta)$ (i.e.~the Airy function) around the real axes, $\zeta \to \pm \infty$, is given as 
\begin{align}
\Psi_{\rm orth}(t;\zeta) &\asymeq
\Bigl(\frac{g_{\rm str} \pi }{(\zeta + t)^{1/2}}\Bigr)^{1/2}\,
e^{-\frac{2}{3 g_{\rm str}} (\zeta + t)^{3/2}}+\cdots,
\end{align}
in $\zeta \to \infty $ with the angle, $|\arg(\zeta)|<\pi$, and 
\begin{align}
\Psi_{\rm orth}(t;\zeta) &\asymeq
\Bigl(\frac{g_{\rm str} \pi }{(\zeta + t)^{1/2}}\Bigr)^{1/2}\,
\Bigl[
e^{-\frac{2}{3 g_{\rm str}} (\zeta + t)^{3/2}}
+i
e^{\frac{2}{3 g_{\rm str}} (\zeta + t)^{3/2}}
\Bigr]+\cdots,
\end{align}
in $\zeta \to e^{\pi i} \times \infty$ with the angle, $|\arg(-\zeta)|<2\pi/3$. 
Note that both two expressions in the intersections, 
$\pi/3 < |\arg(\zeta)|<\pi$, have common asymptotic expansions, 
and therefore, the appearance/disappearance of different exponents 
in different asymptotic regions is understood as 
the Stokes phenomenon. 
As a consequence, the resolvent in the weak coupling 
limit $g_{\rm str}\to 0$ is 
smooth in $\zeta$ with $\arg(\zeta)<\pi$, and 
the discontinuity 
only appears along $\zeta \in (-\infty,-t)$, that is,
\begin{align}
\lim_{\epsilon \to \pm 0}
\biggl[\lim_{g_{\rm str}\to 0}\Psi_{\rm orth}(t;\zeta +i\epsilon)
\biggr] \sim 
e^{\mp \frac{2}{3 g_{\rm str}} (\zeta + t)^{3/2}},\qquad \zeta \in (-\infty,-t). 
\end{align}
An important point in \cite{MMSS} is that the resolvent curve itself 
has a cut around $\zeta\to \infty$. However the explicit cuts are smeared by 
the superposition of the exponents $e^{(\zeta + t)^{3/2}}$ and $e^{-(\zeta + t)^{3/2}}$. 
Note that the solution corresponding to matrix models can be chosen by the single condition 
Eq.~\eq{BCInAiryFunction}. This is due to the simplicity of Airy system. In more general system, however, 
one needs the multi-cut boundary condition proposed in this paper. 

\section{Lax operators in the multi-cut matrix models \label{AppendixLaxOP}}

Here we summarize the Lax operators used in this paper. 

\subsection{The $\mathbb Z_k$ symmetric $(1,1)$ critical points}

This class of critical points are characterized by the following 
Lax operators:
\begin{align}
\bP(t;\del)&= \Gamma \del + H(t), \nn\\
\qquad \bQ(t;\del)&= \bigl(\bGamma^{-2}(t;\del)\, \bP(t;\del) \bigr)_{+} 
-\mu \bigl(\bGamma^{-1}(t;\del)\bigr)_+ \nn\\
&= \Gamma^{-1} \del -\Gamma^{-1} H \Gamma^{-1} - \mu \Gamma^{-1}. 
\end{align}
Note that the $\mathbb Z_k$ symmetry requires 
\begin{align}
H(t) =
\begin{pmatrix}
0 & * &  \cr
   & 0 & * & \cr
   &    &  \ddots & \ddots & \cr
   &    &             &       0     &  * \cr
* &    &             &              &  0
\end{pmatrix}, 
\end{align}
and the Lax operator $\bGamma(t;\del)$ is defined as 
\begin{align}
\bGamma(t;\del) = \Gamma + \sum_{n=1}^\infty S_n(t)\, \del^{-n},
\qquad \bigl(\bGamma(t;\del)\bigr)^k =I_k,\qquad \bigl[\bGamma(t;\del),\bP(t;\del)\bigr]=0. 
\end{align}
From these operators, one can calculate 
the operator $\mathcal Q(t;\zeta)$ 
(see Eq.~\eq{GeneralODEQ}) which is given as 
\begin{align}
\mathcal Q(t;\zeta)= \Gamma^{-2} \zeta - \Gamma^{-1}\bigl(\{\Gamma^{-1},H\}+\mu\bigr). 
\end{align}
The coefficients of the asymptotic expansion \eq{FormalExpansionAsymptoticExpansionsMultiCut} are then calculated as 
\begin{align}
\varphi(t;\zeta) &= \frac{(\Gamma^{-1}\zeta)^2}{2} - \mu\, \Gamma^{-1} \zeta + \mathcal O(1/\zeta), \nn\\
Y(t;\zeta) &= I_k + \frac{1}{\zeta}{\rm adj}^{-1}(\Gamma^{-2})\bigl[\Gamma^{-1}\{\Gamma^{-1},H(t)\}\bigr]+ \mathcal O(1/\zeta^2),
\end{align}
where ${\rm adj}^{-1}$ is the inverse operator of ${\rm adj}(A)[B]=AB-BA$. 
In the $k=3$ case, by using the formula, ${\rm adj}^{-1}(\Gamma^{-1})[X] = [\Gamma^{-1},X]/3$, 
one can show 
\begin{align}
Y_1(t) = \frac{1}{3}\Bigl(H(t)- \Gamma^{-1} H(t) \Gamma \Bigr).
\end{align}
Here we have checked that $\varphi_0(t)=0$ is true for first few cases $k=3,4,5$ 
and this is consistent with our solutions.

\subsection{Fractional-superstring $(\hat p,\hat q)=(1,2)$ critical points ($r=3$) \label{FSSTLaxOperatorAppendixxx}}
In this case, we only study $k=2$ case, but generally 
one can calculate as follows: 
The Lax operators in these cases are 
\begin{align}
\bP(t;\del) = \Gamma \del + H(t),
\qquad \bQ(t;\del)&= \bigl(\bGamma^{-1}(t;\del)\, \bP^2(t;\del) \bigr)_{+} 
-\mu \bigl(\bGamma^{-1}(t;\del)\bigr)_+ \nn\\
&= \Gamma\, \del^2 +H(t)\,\del -S_2(t) -\mu \Gamma^{-1}
\end{align}
Therefore, the operator $\mathcal Q(t;\zeta)$ is given as 
\begin{align}
\mathcal Q(t;\zeta)= \Gamma^{-1}\, \zeta^2 
- \Gamma^{-1} H(t)\, \zeta - \del H(t) - S_2(t) - \mu \Gamma^{-1},
\end{align}
or 
\begin{align}
\mathcal Q_{-3}(t) = \Gamma^{-1},
\quad \mathcal Q_{-2}(t) =-\Gamma^{-1} H(t),
\quad \mathcal Q_{-1}(t) = -\del H(t) -S_2(t) - \mu \Gamma^{-1}.
\end{align}
Here $S_2(t)$ satisfies%
\footnote{We define the symmetric product $\{A_1 ,A_2 ,\cdots, A_k\}_k$ as
\begin{align}
\Bigl(\sum_{i} a_i\Bigr)^k \equiv \sum_{i_1,i_2,\cdots,i_k} \bigl\{a_{i_1},a_{i_2},\cdots,a_{i_k}\bigr\}_k. 
\end{align}
}
\begin{align}
\bigl[\Gamma,S_2(t)\bigr] + \Gamma \del H = 0,
\qquad \{\Gamma,\cdots,\Gamma,S_2(t)\}_k + \{\Gamma,\cdots,\Gamma, H(t),H(t)\}_k= 0. 
\end{align}
In the $k=2$ case, $S_2(t)$ is given as 
\begin{align}
S_2(t) = \frac{1}{2}\Bigl(\sigma_1 f^2(t) - i \sigma_2 \,\del f(t)\Bigr),\qquad 
H(t) = i \sigma_2 f(t). 
\end{align}
The coefficients of the asymptotic expansion are given as 
\begin{align}
\varphi(\zeta) &= \sigma_1\bigl(\frac{\zeta^3}{3} - \mu  \zeta \bigr) + \mathcal O(1/\zeta), \nn\\
Y(\zeta) &= I_2 +\frac{1}{2}i\sigma_2 \frac{ f(t)}{\zeta}- \frac{1}{4}\sigma_3 \frac{\del f(t)}{\zeta^2} 
+\frac{1}{8}i\sigma_2 \frac{f^3(t)-4\mu f(t)}{\zeta^3}+
\mathcal O(1/\zeta^4). 
\end{align}

\section{Supplements to Theorem \ref{ThmMCBCRec} 
and Theorem \ref{ThmConplementaryBCRec} \label{ExamplesProfilesSections}}

In this appendix, we first show the explicit form of the linear expressions Eq.~\eq{LinearEqForYBC} and Eq.~\eq{LinearEqForYBCComp}, 
and then show some examples. Before we focus on these cases, we summarize the general properties of 
these systems by introducing the 
following four categories of the indices $i$ of $y_{n,i}$: 
\begin{align}
&\text{(I)}\quad 1 \leq i\leq \Big\lfloor \frac{k+3}{4} \Big\rfloor =:A,
&\text{(II)}\quad B:=\Big\lfloor \frac{k+3}{4} \Big\rfloor +1
\leq i\leq \frac{k+1}{2},\nn\\
&\text{(III)}\quad \frac{k+1}{2}+1 \leq i\leq \Big\lfloor \frac{3k+3}{4} \Big\rfloor=:C,
&\text{(IV)}\quad D:= \Big\lfloor \frac{3k+3}{4} \Big\rfloor +1
\leq i\leq k, \label{RegionABCDIV}
\end{align}
which is closely related to the multi-cut boundary conditions Eq.~\eq{BCforXYinMatrix} 
and Eq.~\eq{BCforXYinMatrixComp}. 
On the other hand, we show this division in the profile $\mathcal J_{k,2}^{(\rm sym)}$. 
Here we show the categories of I and III with {\it italic font} and the categories of II and IV with {\bf bold font}: 
\begin{align}
&\underline{\text{$k=4k_0+1,\, (k_0 \in \mathbb N):$}} \nn\\
&\mathcal J_{k,2}^{(\rm sym)}= \nn\\
&
\left[
\begin{array}{c|c|c|c|c|c|c|c|c|c}
\bf  B & (\bf D & \bf D+1) & (\it A & \bf B+1) & (\it C & \bf D+2) & (\it A-1 &\bf  B+2) &  \cdots \cr
\hline
 (\bf D &\bf  B) & (\it A & \bf D+1) & (\it C & \bf B+1) & (\it A-1 & \bf D+2) & (\it C-1 & \cdots \cr
\hline
\bf  D & (\it A & \bf B) & (\it C & \bf D+1) & (\it A-1 &\bf  B+1) & (\it C-1 &\bf  D+2) & \cdots \cr
\hline
 (\it A & \bf D) & (\it C &\bf  B) & (\it A-1 & \bf D+1) & (\it C-1 & \bf B+1) & (\it A-2 & \cdots \cr
\end{array}
\right. \nn\\
&\qquad \left.
\begin{array}{c|c|c|c|c|c|c|c|c|c|c}
...   & (\it  4 &\bf  \frac{k-1}{2}) & ( \frac{k+7}{2} & \bf k) & (\it 3 & \bf \frac{k+1}{2}) & ( \it \frac{k+5}{2} &\it 1) & ( \it 2 & \it \frac{k+3}{2}) \cr
\hline
...   & \bf k-1) & ( \frac{k+7}{2} & \bf \frac{k-1}{2}) & (\it 3 & \bf k) & (\it \frac{k+5}{2} &\bf \frac{k+1}{2}) & (\it 2 &\it  1) &\it  \frac{k+3}{2} \cr
\hline
...   & (\frac{k+7}{2} & \bf k-1) & (\it 3 & \bf \frac{k-1}{2}) & ( \it \frac{k+5}{2} &\bf k) & (\it  2 & \bf \frac{k+1}{2}) &(\it  \frac{k+3}{2} &\it  1) \cr
\hline
...    & \bf \frac{k-3}{2}) & (\it 3 & \bf k-1) & (\it \frac{k+5}{2} & \bf \frac{k-1}{2}) & ( \it 2 & \bf k) & (\it \frac{k+3}{2} &  \bf \frac{k+1}{2}) &\it  1 \cr
\end{array}
\right]\!\!\!\!
\begin{array}{l}
:J_{3} \cr 
:J_2 \cr 
:J_1 \cr 
:J_0
\end{array}
\label{DPgeneralKAppendix1}
\end{align}
and some concrete examples ($k=9$ and $13$) are 
\begin{align}
&\mathcal J_{9,2}^{(\rm sym)}= 
\left[
\begin{array}{c|c|c|c|c|c|c|c|c}
\bf 4 & (\bf 8 & \bf 9) & (\it 3 & \bf 5) & (\it 7 & \it 1) & (\it 2 & \it 6) \cr
\hline
(\bf 8 & \bf 4) & (\it 3 & \bf 9) & (\it 7 & \bf 5) & (\it 2 & \it 1) & \it 6 \cr
\hline
\bf 8 & (\it 3 & \bf 4)&(\it 7 & \bf 9)&(\it 2 & \bf 5)&(\it 6 & \it 1) \cr
\hline
(\it 3 & \bf 8)&(\it 7 &\bf  4)&(\it 2 & \bf 9)&(\it 6 & \bf 5) & \it 1 \cr
\end{array}
\right]\!\!\!\!
\begin{array}{l}
:J_{3} \cr 
:J_2 \cr 
:J_1 \cr 
:J_0
\end{array}, \nn\\
\nn\\
&\mathcal J_{13,2}^{(\rm sym)}= 
\left[
\begin{array}{c|c|c|c|c|c|c|c|c|c|c|c|c}
\bf 5 & (\bf 11 & \bf 12) & (\it 4 & \bf 6) & (\it 10 & \bf 13) & (\it 3 & \bf 7) & (\it 9 & \it 1) & (\it 2 & \it 8) \cr
\hline
(\bf 11 & \bf 5) & (\it 4 & \bf 12) & (\it 10 & \bf 6) & (\it 3 & \bf 13) & (\it 9 & \bf 7) & (\it 2 & \it 1) & \it 8 \cr
\hline
\bf 11 & (\it 4 & \bf 5) & (\it 10 & \bf 12) & (\it 3 & \bf 6) & (\it 9 & \bf 13) & (\it 2 & \bf 7) & (\it 8 & \it 1) \cr
\hline
(\it 4 & \bf 11) & (\it 10 & \bf 5) & (\it 3 & \bf 12) & (\it 9 & \bf 6) & (\it 2 & \bf 13) & (\it 8 & \bf 7) & \it 1 \cr
\end{array}
\right]\!\!\!\!
\begin{array}{l}
:J_{3} \cr 
:J_2 \cr 
:J_1 \cr 
:J_0
\end{array}. 
\end{align}
\begin{align}
&\underline{\text{$k=4k_0+3,\, (k_0\in \mathbb N):$}} \nn\\
&\mathcal J_{k,2}^{(\rm sym)}= \nn\\
&
\left[
\begin{array}{c|c|c|c|c|c|c|c|c|c}
 \bf D & (\bf B & \bf B+1) & (\it C & \bf D+1) & (\it A & \bf B+2) & (\it C-1 & \bf D+2) & \cdots \cr
\hline
 (\bf B & \bf D) & (\it C & \bf B+1) & (\it A & \bf D+1) & (\it C-1 & \bf B+2) & (\it A-1 &  \cdots \cr
\hline
 \bf B & (\it C & \bf D) & (\it A & \bf B+1) & (\it C-1 & \bf D+1) & (\it A-1 & \bf B+2) &  \cdots \cr
\hline
 (\it C & \bf B) & (\it A & \bf D) & (\it C-1 & \bf B+1) & (\it A-1 & \bf D+1) & (\it C-2 &  \cdots \cr
\end{array}
\right. \nn\\
&\qquad \left.
\begin{array}{c|c|c|c|c|c|c|c|c|c|c}
...   & (\it  4 &\bf  \frac{k-1}{2}) & ( \frac{k+7}{2} & \bf k) & (\it 3 & \bf \frac{k+1}{2}) & ( \it \frac{k+5}{2} &\it 1) & ( \it 2 & \it \frac{k+3}{2}) \cr
\hline
...   & \bf k-1) & ( \frac{k+7}{2} & \bf \frac{k-1}{2}) & (\it 3 & \bf k) & (\it \frac{k+5}{2} &\bf \frac{k+1}{2}) & (\it 2 &\it  1) &\it  \frac{k+3}{2} \cr
\hline
...   & (\frac{k+7}{2} & \bf k-1) & (\it 3 & \bf \frac{k-1}{2}) & ( \it \frac{k+5}{2} &\bf k) & (\it  2 & \bf \frac{k+1}{2}) &(\it  \frac{k+3}{2} &\it  1) \cr
\hline
...    & \bf \frac{k-3}{2}) & (\it 3 & \bf k-1) & (\it \frac{k+5}{2} & \bf \frac{k-1}{2}) & ( \it 2 & \bf k) & (\it \frac{k+3}{2} &  \bf \frac{k+1}{2}) &\it  1 \cr
\end{array}
\right]\!\!\!\!
\begin{array}{l}
:J_{3} \cr 
:J_2 \cr 
:J_1 \cr 
:J_0
\end{array}, 
\label{DPgeneralKAppendix2}
\end{align}
and some concrete examples ($k=7$ and $11$) are 
\begin{align}
&\mathcal J_{7,2}^{(\rm sym)}= 
\left[
\begin{array}{c|c|c|c|c|c|c}
\bf 7&(\bf 3&\bf 4) & (\it 6 & \it 1) & (\it 2 & \it 5)\cr
\hline
(\bf 3&\bf 7) & (\it 6 & \bf 4)& (\it 2 & \it 1) & \it 5 \cr
\hline
\bf 3&(\it 6&\bf 7) & (\it 2 & \bf 4)& (\it 5&\it 1) \cr
\hline
(\it 6& \bf 3)&(\it 2 & \bf 7)&(\it 5 & \bf 4) &\it  1 \cr
\end{array}
\right]\!\!\!\!
\begin{array}{l}
:J_{3} \cr 
:J_2 \cr 
:J_1 \cr 
:J_0
\end{array}, \nn\\
&\mathcal J_{11,2}^{(\rm sym)}= 
\left[
\begin{array}{c|c|c|c|c|c|c|c|c|c|c}
\bf 10 & (\bf 4 &\bf  5) & (\it 9 & \bf 11) & (\it 3 & \bf 6) & (\it 8 & \it 1) & (\it 2 & \it 7) \cr
\hline
(\bf 4 & \bf 10) & (\it 9 & \bf 5) & (\it 3 & \bf 11) & (\it 8 & \bf 6) & (\it 2 & \it 1) & \it 7 \cr
\hline
\bf 4 & (\it 9 & \bf 10) & (\it 3 & \bf 5) & (\it 8 & \bf 11) & (\it 2 & \bf 6) & (\it 7 & \it 1) \cr
\hline
(\it 9 & \bf 4) & (\it 3 & \bf 10) & (\it 8 & \bf 5) & (\it 2 &\bf  11) & (\it 7 & \bf 6) & \it 1 \cr
\end{array}
\right]\!\!\!\!
\begin{array}{l}
:J_{3} \cr 
:J_2 \cr 
:J_1 \cr 
:J_0
\end{array}. 
\end{align}
If one follows trajectories of numbers, 
One may notice that the trajectories of the numbers in the region I and III (written by {\it italic font}) 
almost form slash shape, ``$\diagup$'', 
and the trajectories of the numbers in the region II and IV (written by {\bf bold font}) 
almost form backslash shape, ``$\diagdown$''. 
In the left-hand and right-hand ends of the profile $\mathcal J_{k,2}^{(\rm sym)}$, 
there are a few exceptions which form curved shape as ``$<$'' 
and so on. 
From this property, we can see the following facts: 
\begin{itemize}
\item Stokes multipliers $s_{l,i,j} \leftrightarrow (j|i)_l$ 
are almost given by $i \in {\rm II, IV}$ ({\bf bold font}) and $j \in {\rm I, III}$ ({\it italic font}). 
We emphasize this fact by writing $s_{l,{\bf i},j} \leftrightarrow (j|{\bf i})_l$. 
Then there are only a few exceptions, $s_{*,{\bf i},{\bf j}} \, ({\bf i},{\bf j} \in {\rm II,IV})$ and $s_{*,i,j} \, (i,j \in {\rm I,III})$, which appear in the left-hand and right-hand ends of the profile $\mathcal J_{k,2}^{(\rm sym)}$. 
Interestingly, there is no Stokes multipliers of the type $s_{l,i,{\bf j}}$ with 
$i \in {\rm I, III}$ and ${\bf j} \in {\rm II, IV}$ in this $r=2$ case. 
\item 
From this fact, one can show that 
the relation between the symmetric Stokes multipliers $s_{0,i,j}^{(\rm sym)}$ and 
the fine Stokes multipliers $s_{l,i,j}$ (Eq.~\eq{MultiSymStokes}) are almost trivial: 
$s_{0,i,j}^{(\rm sym)}= s_{{}^\exists l,i,j}$ for almost all $(j|i)_l \in \mathcal J_{k,2}^{(\rm sym)}$, 
and that only the following 
few multipliers are the exceptions: 
\begin{align}
s_{0,{\it 1},{\it 2}}^{(\rm sym)} =s_{2,{\it 1},{\it 2}}
+ s_{1,{\it 1},{\it \frac{k+3}{2}}} \,s_{3,{\it \frac{k+3}{2}},{\it 2}},
\qquad s_{0,{\bf \frac{k+1}{2}},{\it 2}}^{(\rm sym)} =s_{1,{\bf \frac{k+1}{2}},{\it 2}}
+ s_{0,{\bf \frac{k+1}{2}},{\it \frac{k+3}{2}}} \,s_{3,{\it \frac{k+3}{2}},{\it 2}},
\label{FineSymTranslationReverse}
\end{align}
and 
\begin{align}
&\underline{\text{$k=4k_0+1:$}} \nn\\
&\quad s_{0,{\bf B},{\it A}}^{(\rm sym)} =s_{1,{\bf B},{\it A}}
+ s_{2,{\bf B},{\bf D}} \,s_{0,{\bf D},{\it A}},
\qquad s_{0,{\bf D+1},{\it A}}^{(\rm sym)} =s_{2,{\bf D+1},{\it A}}
+ s_{3,{\bf D+1},{\bf D}} \,s_{0,{\bf D},{\it A}}; \\
&\underline{\text{$k=4k_0+3:$}} \nn\\
&\quad s_{0,{\bf D},{\it C}}^{(\rm sym)} =s_{1,{\bf D},{\it C}}
+ s_{2,{\bf D},{\bf B}} \,s_{0,{\bf B},{\it C}},
\qquad s_{0,{\bf B+1},{\it C}}^{(\rm sym)} =s_{2,{\bf B+1},{\it C}}
+ s_{3,{\bf B+1},{\bf B}} \,s_{0,{\bf B},{\it C}}. 
\label{FineSymTranslationReverseComp1}
\end{align}
\item These relations are important not only because they are used in deriving the results in this appendix, 
but also because they provide a concrete example which guarantees the claim shown in Eq.~\eq{SymFineRelationCol}. In particular, we expect that one can extend this discussion to the general systems of $(k,r;\gamma)$ which 
are controlled by the method developed in Section \ref{MultiCutStokesSection}. 
\end{itemize}

\subsection{The explicit form of Eq.~\eq{LinearEqForYBC} and Eq.~\eq{LinearEqForYBCComp}}

Below is the explicit form of the linear expressions Eq.~\eq{LinearEqForYBC} and Eq.~\eq{LinearEqForYBCComp}. 
Note the function $\epsilon(k)$ appears in these formulas is given as 
\begin{align}
\epsilon (k)=
\left\{
\begin{array}{ll}
0 & (k=4k_0+1)\cr
1 & (k=4k_0+3)
\end{array}
\right..
\end{align}

\paragraph{The linear expression of Eq.~\eq{LinearEqForYBC}}

\begin{align}
&\underline{\text{Region I:}} \nn\\
&\qquad y_{n,i} \bigl( \{y_{m,1}\}_{m\in\mathbb Z} \bigr) = y_{n+i-1,1} \neq 0,\qquad i \in {\rm (I)},   \label{ShiftRegionI} \\
&\underline{\text{Region II:}} \nn\\
&\qquad y_{n,B+j} \bigl( \{y_{m,1}\}_{m\in\mathbb Z} \bigr) \equiv y_{n+A+j,1} 
 + \sum_{a=0}^{j-\epsilon(k)} s^{(\rm sym)}_{0,B+j-a, A-j+a+\epsilon(k)} \times y_{n+A-j-1+2a+\epsilon(k),1} +\nn\\
&\qquad\qquad\qquad\qquad\qquad +\sum_{a=0}^{j-1} s^{(\rm sym)}_{0,B+j-a, A-j+a+1+\epsilon(k)} \times y_{n+A-j+2a+\epsilon(k),1}, \label{YinTermsOfYn1ForRegionII}  \\
&\underline{\text{Region III:}} \nn\\
&\qquad y_{n,i} \bigl( \{y_{m,1}\}_{m\in\mathbb Z} \bigr) = 0,\qquad i \in {\rm (III)}. \\
&\underline{\text{Region IV:}} \nn\\
&\qquad y_{n,D+j} \bigl( \{y_{m,1}\}_{m\in\mathbb Z} \bigr) 
\equiv \sum_{a=0}^{j} s^{(\rm sym)}_{0,D+j-a, A-j+a} 
\times y_{n+A-j-1+2a,1} \nn\\
&\qquad \qquad\qquad\qquad\qquad + \sum_{a=0}^{j-1+\epsilon(k)} s^{(\rm sym)}_{0,D+j-a, A-j+a+1} 
\times y_{n+A-j+2a,1}. \label{sdfjsafjsflsdee}
\end{align}

\paragraph{The linear expression of Eq.~\eq{LinearEqForYBCComp}}
\begin{align}
&\underline{\text{Region I:}} \nn\\
&\qquad \widetilde y_{n,i} \bigl( \{\widetilde y_{m,\frac{k+3}{2}}\}_{m\in\mathbb Z} \bigr) = 0,\qquad i \in {\rm (I)}, \\
&\underline{\text{Region II:}} \nn\\
&\qquad \widetilde y_{n,B+j} \bigl( \{\widetilde y_{m,\frac{k+3}{2}}\}_{m\in\mathbb Z} \bigr) \equiv 
\sum_{a=0}^{j} s^{(\rm sym)}_{0,B+j-a, C-j+a} \times \widetilde y_{n+C-j-\frac{k+3}{2}+2a,\frac{k+3}{2}} +\nn\\
&\qquad \qquad\qquad\qquad\qquad +\sum_{a=0}^{j-\epsilon(k)} s^{(\rm sym)}_{0,B+j-a, C-j+a+1} \times \widetilde y_{n+C-j+2a+1-\frac{k+3}{2},\frac{k+3}{2}},\label{YinTermsOfYn1ForRegionIIAAAAA} \\
&\underline{\text{Region III:}} \nn\\
&\qquad \widetilde y_{n,i}\bigl(\{\widetilde y_{m,\frac{k+3}{2}}\}_{m\in\mathbb Z}\bigr) 
= \widetilde y_{n+i-\frac{k+3}{2},\frac{k+3}{2}} \neq 0, \qquad i \in {\rm (III)}, \\
&\underline{\text{Region IV:}} \nn\\
&\qquad \widetilde y_{n,D+j} \bigl( \{\widetilde y_{m,\frac{k+3}{2}}\}_{m\in\mathbb Z} \bigr) 
\equiv \widetilde y_{n+D+j-\frac{k+3}{2},\frac{k+3}{2}}+ \nn\\
&\qquad \qquad\qquad\qquad\qquad +\sum_{a=0}^{j-1} s^{(\rm sym)}_{0,D+j-a, C-j+a+1-\epsilon(k)} 
\times \widetilde y_{n+C-j+a+1-\epsilon(k)-\frac{k+3}{2},\frac{k+3}{2}} +\nn\\
&\qquad \qquad\qquad\qquad\qquad + \sum_{a=0}^{j-1+\epsilon(k)} s^{(\rm sym)}_{0,D+j-a, C-j+a+2-\epsilon(k)} 
\times \widetilde y_{n+C-j+a+2-\epsilon(k)-\frac{k+3}{2},\frac{k+3}{2}}. \label{sdfjsafjsflsdeeAAAAA}
\end{align}

\subsection{Some examples of Theorem \ref{ThmMCBCRec} }
Below we show the concrete expressions of the equations in Theorem \ref{ThmMCBCRec} 
for some special cases ($k=5,7,9$ and $11$). 
First of all, the vectors are expressed only by using $\{y_{n,1}\}_{n\in\mathbb Z}$: 
\begin{align}
&\underline{\text{$k=5$}}:
\quad Y^{(n)}=
\begin{pmatrix}
y_{n,1} \cr y_{n+1,1} \cr
\hline
-s_{0,4,2}^{(\rm sym)} \,y_{n,1} \cr
\hline
0 \cr
\hline
s_{0,5,2}^{(\rm sym)} \,y_{n+1,1}
\end{pmatrix}, 
\qquad\,\underline{\text{$k=7$}}:\quad
Y^{(n)}=
\begin{pmatrix}
y_{n,1} \cr y_{n+1,1} \cr 
\hline
y_{n+2,1} \cr -s_{0,5,2}^{(\rm sym)}\, y_{n,1} \cr
\hline
0 \cr 0 \cr
\hline
s_{0,7,2}^{(\rm sym)} \,y_{n+1,1}+s_{0,7,3}^{(\rm sym)}\, y_{n+2,1}
\end{pmatrix}, \nn\\
&\underline{\text{$k=9$}}: \qquad Y^{(n)}=
\begin{pmatrix}
y_{n,1} \cr y_{n+1,1} \cr y_{n+2,1} \cr
\hline
y_{n+3,1} + s_{0,4,3}^{(\rm sym)}\,
y_{n+2,1}\cr 
-s_{0,6,2}^{(\rm sym)}\, y_{n,1} \cr
\hline
0 \cr 0 \cr
\hline
s_{0,8,3}^{(\rm sym)} \,y_{n+2,1} \cr
s_{0,8,3}^{(\rm sym)} \,y_{n+3,1}
+ s_{0,9,3}^{(\rm sym)} \,y_{n+2,1}
+s_{0,9,2}^{(\rm sym)} \,y_{n+1,1}
\end{pmatrix}, \nn\\
&\underline{\text{$k=11$}}: \qquad Y^{(n)}=
\begin{pmatrix}
y_{n,1} \cr y_{n+1,1} \cr y_{n+2,1} \cr
\hline
y_{n+3,1} \cr
y_{n+4,1} + s_{0,5,3}^{(\rm sym)}\,
y_{n+2,1} + s_{0,5,4}^{(\rm sym)}\,
y_{n+3,1}\cr 
-s_{0,7,2}^{(\rm sym)}\, y_{n,1} \cr
\hline
0 \cr 0 \cr 0 \cr
\hline
s_{0,10,3}^{(\rm sym)} \,y_{n+2,1} +
s_{0,10,4}^{(\rm sym)} \,y_{n+3,1} \cr
s_{0,10,3}^{(\rm sym)} \,y_{n+3,1} +
s_{0,10,4}^{(\rm sym)} \,y_{n+4,1}
+ s_{0,11,2}^{(\rm sym)} \,y_{n+1,1}
+s_{0,11,3}^{(\rm sym)} \,y_{n+2,1}
\end{pmatrix}. 
\end{align}
Secondly, the multi-cut BC recursions are expressed as 
\begin{align}
&\underline{\text{$k=5$}} \nn\\
&\quad \mathcal F_5[y_{n,1}] \equiv y_{n+2,1} + s_{1,3,2}\, y_{n+1,1} + s_{3,4,2} \,y_{n,1}= 0, \nn\\
&\quad \mathcal G_5[y_{n,1}] \equiv s_{0,5,2}\, y_{n+2,1} + s_{2,1,2}\, y_{n+1,1} -y_{n,1}= 0, \nn\\
&\underline{\text{$k=7$}} \nn\\
&\quad \mathcal F_7[y_{n,1}] \equiv y_{n+3,1} + s_{3,4,3}\, y_{n+2,1} + s_{1,4,2}\, y_{n+1,1} + s_{3,5,2}\, y_{n,1}= 0, \nn\\
&\quad \mathcal G_7[y_{n,1}] \equiv s_{2,7,3}\, y_{n+3,1} + s_{0,7,2}\, y_{n+2,1} + s_{2,1,2}\,y_{n+1,1} - y_{n,1}= 0, \nn\\
&\underline{\text{$k=9$}} \nn\\
&\quad \mathcal F_9[y_{n,1}] \equiv y_{n+4,1} + s_{1,4,3}\, y_{n+3,1} + s_{3,5,3}\, y_{n+2,1} + s_{1,5,2}\, y_{n+1,1}+ s_{3,6,2}\, y_{n,1}= 0, \nn\\
&\quad \mathcal G_9[y_{n,1}] \equiv s_{0,8,3}\, y_{n+4,1} + s_{2,9,3}\, y_{n+3,1} + s_{0,9,2}\,y_{n+2,1} + s_{2,1,2}\,y_{n+1,1}- y_{n,1}= 0, \nn\\
&\underline{\text{$k=11$}} \nn\\
&\quad \mathcal F_{11}[y_{n,1}] \equiv y_{n+5,1} + s_{3,5,4}\, y_{n+4,1} + s_{1,5,3}\, y_{n+3,1} + s_{3,6,3}\, y_{n+2,1}+ s_{1,6,2}\, y_{n+1,1}+s_{3,7,2}\, y_{n,1}= 0, \nn\\
&\quad \mathcal G_{11}[y_{n,1}] \equiv s_{2,10,4}\, y_{n+5,1} + s_{0,10,3}\, y_{n+4,1} + s_{2,11,3}\,y_{n+3,1} + s_{0,11,2}\,y_{n+2,1}+s_{2,1,2}\,y_{n+1,1}- y_{n,1}= 0. 
\end{align}

\section{Derivation of the continuum solutions \label{ProofOfSolutionSectionAppendix}}

In this subsection, we derive the continuum solutions 
of Theorem \ref{TheoremContinuumSolutions}. 
According to Lemma \ref{MFcond3}, 
the monodromy free condition can be solved if the matrix $S_0^{(\rm sym)}\,\Gamma^{-1}$
is diagonalizable. For the continuum solutions, we solve this problem by requiring that 
{\em all the eigenvalues of the matrix $S_0^{(\rm sym)}\,\Gamma^{-1}$ are distinct}. 
This means that we require the characteristic polynomial $\mathcal H(x)$ 
of the matrix $S_0^{(\rm sym)}\,\Gamma^{-1}$ satisfy 
\begin{align}
\mathcal H(x)\equiv \det\Bigl( x I_k-S_0^{(\rm sym)}\,\Gamma^{-1}\Bigr) = x^k-1. 
\label{OurAnsatz4}
\end{align}
The coefficients of the characteristic polynomial are 
related to the Stokes multipliers and here are several examples:
\begin{align}
&\underline{\text{The $5$-cut case:}} \nn\\
&\mathcal H(x)=-1+x^5+x \left(-s_{1,3,2}+s_{0,5,2} s_{2,3,5}-s_{3,1,5}\right) +\nn\\
&\quad+x^2 \left(s_{0,3,4} s_{0,5,2}-s_{1,1,4}+s_{2,1,2} s_{2,3,5}-s_{1,3,2} s_{3,1,5}-s_{3,4,2}\right)+ \nn\\
&\quad+x^3 \left(-s_{0,5,2}-s_{1,1,4} s_{1,3,2}+s_{0,3,4} s_{2,1,2}-s_{2,3,5}-s_{3,1,5} s_{3,4,2}\right) +\nn\\
&\quad+x^4 \left(-s_{0,3,4}-s_{2,1,2}-s_{1,1,4} s_{3,4,2}\right), \nn\\
&\underline{\text{The $7$-cut case:}} \nn\\
&\mathcal H(x)=-1+x^7+x \left(-s_{1,7,6}+s_{0,3,6} s_{2,7,3}-s_{3,4,3}\right)+\nn\\
&\quad+x^2 \left(s_{0,3,6} s_{0,7,2}-s_{1,4,2}+s_{2,4,6} s_{2,7,3}-s_{3,1,6}-s_{1,7,6} s_{3,4,3}\right)+ \nn\\
&\quad+x^3 \left(-s_{1,1,5}-s_{1,4,2} s_{1,7,6}+s_{0,3,6} s_{2,1,2}+s_{0,7,2} s_{2,4,6}+s_{0,4,5} s_{2,7,3}-s_{3,1,6} s_{3,4,3}-s_{3,5,2}\right)+\nn\\
&\quad+x^4 \left(-s_{0,3,6}+s_{0,4,5} s_{0,7,2}+s_{2,1,2} s_{2,4,6}-s_{2,7,3}-s_{1,4,2} s_{3,1,6}-s_{1,1,5} s_{3,4,3}-s_{1,7,6} s_{3,5,2}\right)+\nn\\
&\quad+x^5 \left(-s_{0,7,2}-s_{1,1,5} s_{1,4,2}+s_{0,4,5} s_{2,1,2}-s_{2,4,6}-s_{3,1,6} s_{3,5,2}\right)+ \nn\\
&\quad+x^6 \left(-s_{0,4,5}-s_{2,1,2}-s_{1,1,5} s_{3,5,2}\right). 
\end{align}
These equations become simpler if one uses the notation given 
in Eqs.~\eq{ThetaNotation1} and \eq{ThetaNotation2}. 
One can read the general formula for $k=2m+1$: 
\begin{align}
\mathcal H(x) = x^k-1 &+ \sum_{n=1}^{m} x^n 
\Bigl[
\sum_{i=1}^{n} \theta_{m+1-i}^*
\widetilde \theta_{m-n+i}^*
-\sum_{i=0}^n \theta_{i} \widetilde \theta_{n-i}
\Bigr]+\nn\\
&+\sum_{n=1}^{m} x^{k-n}
\Bigl[
\sum_{i=0}^{n} \theta_{i}^* \widetilde \theta_{n-i}^*
-\sum_{i=1}^n \theta_{m+1-i}
\widetilde\theta_{m-n+i}
\Bigr], \label{ContinuumConstraintEquationsThetaTheta}
\end{align}
where we have introduced $\theta_0 \equiv 1$. 
Therefore, by comparing both sides of Eq.~\eq{OurAnsatz4}, 
we obtain constraints on the Stokes multipliers:
\begin{align}
0&= \theta_{m}^*
\widetilde \theta_{m}^*
- \theta_{1} -\widetilde \theta_{1}, \nn\\
0&= \theta_{m}^* 
\widetilde \theta_{m -1}^*
+\theta_{m-1}^* 
\widetilde \theta_{m }^*
- \theta_{2} -\theta_{1} \widetilde \theta_{1} -\widetilde \theta_{2}, \nn\\
0&= \theta_{m}^* 
\widetilde \theta_{m -2}^*
+\theta_{m -1}^* 
\widetilde \theta_{m -1}^*
+\theta_{m-2}^* 
\widetilde \theta_{m }^*
- \theta_{3} -\theta_{2} \widetilde \theta_{1}
-\theta_{1} \widetilde \theta_{2} -\widetilde \theta_{3}, \nn\\
0&= \theta_{m}^* 
\widetilde \theta_{m -3}^*
+\theta_{m -1}^* 
\widetilde \theta_{m -2}^*
+\theta_{m -2}^* 
\widetilde \theta_{m -1}^*
+\theta_{m -3}^* 
\widetilde \theta_{m}^*
- \theta_{4} 
-\theta_{3} \widetilde \theta_{1}
-\theta_{2} \widetilde \theta_{2}
-\theta_{1} \widetilde \theta_{3} 
-\widetilde \theta_{4}, \nn\\
& \qquad \cdots. 
\end{align}
Since a half of the Stokes multipliers $\{\theta_n\}_{n=1}^{m}$ 
are given as 
\begin{align}
\theta_n = {\sigma}_{n}
(\{-\omega^{n_j}\}_{j=1}^{m}),
\end{align}
we fix all the other Stokes multipliers
$\{\widetilde \theta_n\}_{n=1}^{m}$ from these constraints. 
Note that, since all the eigenvalues are distinct, the indices 
$(n_1,n_2,\cdots,n_{m})$ for $\{\theta_n\}$ are 
also $m \, (=\lfloor \frac{k}{2} \rfloor)$ distinct integers. Here we can freely choose the ordering: 
\begin{align}
(n_1,n_2,\cdots,n_{m}): \qquad 
1\leq n_1< n_2< \cdots < n_{m}\leq k. 
\end{align}
With noting the following relation:
\begin{align}
\theta_{m}^* \theta_n = \theta_{m-n}^*, \qquad m =\Bigl\lfloor \frac{k}{2} \Bigr\rfloor, 
\end{align}
and recursively rewriting the constraint for the continuum solutions, Eqs.~\eq{OurAnsatz4} 
and \eq{ContinuumConstraintEquationsThetaTheta}, 
we obtain the following simple form:
\begin{align}
\widetilde \theta_n = \mathcal S_n \bigl(\{\theta_j\}_{j\in \mathbb Z}\bigr) 
+\widetilde\theta_{m-n+1}^* \theta_{m}^*, 
\qquad \bigl(n=1,2,\cdots, m\bigr), \label{FormulaForContinuumSolutionAAAAAp}
\end{align}
with the polynomials $\mathcal S_n(x)$ (defined by Eq.~\eq{RecursionForPolySSS}). 
This results in Theorem \ref{TheoremContinuumSolutions}. 

\section{Calculation in the 3-cut $(1,1)$ critical point $(r=2)$\label{CalculationInConcreteSystems}}

The specialty of the $3$-cut $(1,1)$ critical point is that the symmetric Stokes sectors $D_{4n}$ 
(see Eq.~\eq{MultiCutDefinitionOfSymStokesSectors}) 
do not cover the whole plane $\mathbb C$. 
Therefore, we consider doubling of the sectors
\begin{align}
D_{2n},\qquad S_{2n}^{(\rm sym)}\equiv S_{2n}S_{2n+1},\qquad (n=0,1,\cdots,5),
\end{align}
and express 
the boundary condition \eq{BCforXYinMatrix} as follows:
\begin{align}
Y^{(4n)}&=
\begin{pmatrix}
y_{4n,1} \cr y_{4n,2} \cr y_{4n,3}
\end{pmatrix}
\equiv \Gamma^{n}X^{(4n)}= 
\begin{pmatrix}
x^{(4n)}_{n+1} \neq 0\cr x^{(4n)}_{n+2} \cr x^{(4n)}_{n+3}= 0
\end{pmatrix},\nn\\
Y^{(4n+2)}&=
\begin{pmatrix}
y_{4n+2,1} \cr y_{4n+2,2} \cr y_{4n+2,3}
\end{pmatrix}
\equiv \Gamma^{n}X^{(4n+2)}= 
\begin{pmatrix}
x^{(4n+2)}_{n+1} \cr x^{(4n+2)}_{n+2} \neq 0\cr x^{(4n+2)}_{n+3}= 0
\end{pmatrix},
\end{align}
with
\begin{align}
X^{(2n)} = S_{2n}^{(\rm sym)} X^{(2n+2)},\qquad (n=0,1,\cdots,5). 
\end{align}
This is then written as 
\begin{align}
& Y^{(4n)} = S_0^{(\rm sym)} Y^{(4n+2)},\qquad 
Y^{(4n+2)} = \bigl(S_2^{(\rm sym)} \Gamma^{-1}\bigr) Y^{(4n+4)}, \nn\\
\Leftrightarrow& \qquad
\left\{
\begin{array}{l}
y_{4n,i} = y_{4n+2,i} +\sum_{j} \Bigl[s_{0,i,j}^{(\rm sym)}\times  y_{4n+2,j}\Bigr], \cr 
y_{4n+2,i} = y_{4n+4,i-1} +\sum_{j} \Bigl[s_{2,i,j}^{(\rm sym)} \times y_{4n+4,j-1}\bigr]. 
\end{array}
\right.
\end{align}
These recursion relations are expressed as 
\begin{align}
&y_{4n,3}=y_{4n+2,3}=0, 
\qquad y_{4n,1} = y_{4n+2,1}\neq 0, \nn\\
&y_{4n,2}=y_{4n+2,2}\neq 0, \qquad 
y_{4n+2,2} = y_{4n+4,1} \neq 0,
\end{align}
and the following two recursion equation for $y_{4n,1}$
\begin{align}
y_{4n,1} = s_{2,1,2}\times y_{4n+4,1},\qquad
y_{4n+4,1} = -s_{3,3,2}\times y_{4n,1}.
\end{align}
As one may notice, this equation itself is the same 
as Eq.~\eq{MultiCutBC}.
The solutions (labeled by $l$) to this boundary condition is easily solved as 
\begin{align}
y_{4n,1}^{(l)}=\omega^{nl},\qquad s_{3,3,2}^{(l)}=-\omega^l,\qquad 
s_{2,1,2}^{(l)}= \omega^{-l},\qquad (l=0,1,2), \label{SolutionToThreeCutAAA}
\end{align}
and the general solution is given as 
\begin{align}
s_{0,2,3}^{(l)}=-\omega^{-l} + \omega^l (s_{1,1,3}^{(l)})^*,
\end{align}
with Eq.~\eq{SolutionToThreeCutAAA}. 
This provides the first case of the continuum solution \eq{FormulaForContinuumSolutionAAAAAp}. 

\section{Calculation in the 4-cut $(1,1)$ critical point $(r=2)$ \label{CalculationInConcreteSystems2}}

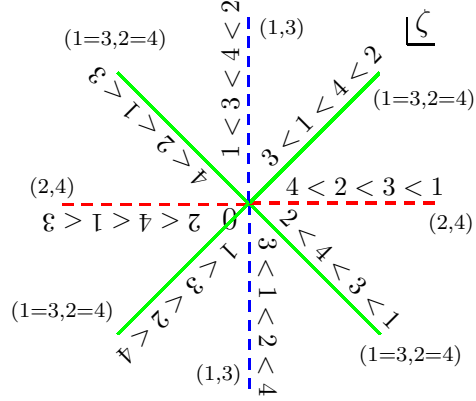
\begin{figure}[htbp]
\begin{center}
\begin{picture}(180,170)(0,0)
\end{picture}
\begin{picture}(5,0)(100,-60)
\put(70,70){\line(0,1){10}\line(1,0){10}}
\put(73,75){$\zeta$}
\put(0,0){$0$}
{
\thicklines
\put(10,10){\rotatebox{-90}{\textcolor{blue}{\dashline[50]{4}[0.7](1,0)(70,0)}}}
\put(10,10){\rotatebox{-45}{\textcolor{green}{\line(1,0){70}}}}
\put(10,10){\rotatebox{0}{\textcolor{red}{\dashline[50]{4}[0.7](1,0)(70,0)}}}
\put(9.5,10){\rotatebox{45}{\textcolor{green}{\line(1,0){70}}}}
\put(9.5,10){\rotatebox{90}{\textcolor{blue}{\dashline[50]{4}[0.7](1,0)(70,0)}}}
\put(-40,10){\rotatebox{135}{\textcolor{green}{\line(1,0){70}}}}
\put(10,9.5){\rotatebox{180}{\textcolor{red}{\dashline[50]{4}[0.7](1,0)(70,0)}}}
\put(-40,10){\rotatebox{-135}{\textcolor{green}{\line(1,0){70}}}}
}
\put(45,47){\quad\scriptsize (1=3,2=4)}
\put(70,0){\,\, \scriptsize (2,4)}
\put(10,74){\,\,\scriptsize (1,3)}
\put(-3,0){\rotatebox{0}{\footnotesize\put(27,13){$4<2<3<1$}}}
\put(-5,-5){\rotatebox{45}{\footnotesize\put(27,13){$3<1<4<2$}}}
\put(0,0){\rotatebox{90}{\footnotesize\put(27,13){$1<3<4<2$}}}
\put(10,10){\rotatebox{135}{\footnotesize\put(27,13){$4<2<1<3$}}}
\put(19,19){\rotatebox{180}{\footnotesize\put(27,13){$2<4<1<3$}}}
\put(12,24){\rotatebox{225}{\footnotesize\put(27,13){$1<3<2<4$}}}
\put(1,24){\rotatebox{270}{\footnotesize\put(27,13){$3<1<2<4$}}}
\put(-7,14){\rotatebox{315}{\footnotesize\put(27,13){$2<4<3<1$}}}
\put(-73,14){\scriptsize (2,4)}
\put(-60,70){\scriptsize (1=3,2=4)}
\put(-10,-57){\scriptsize (1,3)}
\put(40,-50){\quad \scriptsize (1=3,2=4)}
\put(-80,-33){\scriptsize (1=3,2=4)}
\end{picture}
\end{center}
\caption{\footnotesize 
The dominance profile in the $4$-cut $(1,1)$ case in terms of $\zeta$. The bold lines express 
the leading Stokes lines with degeneracy $\varphi^{(1)}\sim \varphi^{(3)}$ 
and $\varphi^{(2)}\sim \varphi^{(4)}$. The dashed lines express the sub leading Stokes lines 
for $(1,3)$ and $(2,4)$. 
\label{4CutR2Figure}}
\end{figure}

Here we calculate the $4$-cut $(1,1)$ critical point as an example in which 
the coprime condition of Eq.~\eq{NonDegenerateConditionCoprimeKR} is violated:
\begin{align}
(k,r)=(4,2). 
\end{align}
In this case, the leading exponents are degenerate:
\begin{align}
\varphi^{(1)}(t;\zeta) \sim \varphi^{(3)}(t;\zeta),\qquad 
\varphi^{(2)}(t;\zeta) \sim \varphi^{(4)}(t;\zeta), 
\end{align} 
and we consider the subleading Stokes lines:
\begin{align}
{\rm Re}\Bigl[\bigl(\varphi^{(1)}_{-r+1}-\varphi^{(3)}_{-r+1}\bigr)\zeta^{r-1}\Bigr]=0,
\qquad {\rm Re}\Bigl[\bigl(\varphi^{(2)}_{-r+1}-\varphi^{(4)}_{-r+1}\bigr)\zeta^{r-1}\Bigr]=0. 
\end{align}
The dominance profile in the $\zeta$ plane is shown in Fig.~\ref{4CutR2Figure}. 

Here we use the fine Stokes sectors $D_n$ (calculated in the leading Stokes lines) which are defined as 
\begin{align}
D_n\equiv D\Bigl(\frac{(n-1)\pi}{4},\frac{n\pi}{4}\Bigr),\qquad 
n=0,1,2,3. 
\end{align}
All fine Stokes matrices can be expressed in terms of $S_0$ as 
\begin{align}
S_{n}= \Gamma^{-n} S_0 \Gamma^{n}, \qquad 
S_0 = 
\begin{pmatrix}
1 & \cr
\alpha & 1 & \beta & \cr
\epsilon & & 1 &\cr
 \gamma & & \delta & 1
\end{pmatrix}. 
\end{align}
Then the multi-cut boundary condition is given as
\begin{align}
Y^{(n)} \equiv  \Gamma^n X^{(n)} = 
\begin{pmatrix}
y_{n,1} \neq 0 \cr
y_{n,2}=0 \cr
y_{n,3}=0 \cr
y_{n,4} \neq 0
\end{pmatrix}.
\end{align}
The recursive equations are expressed as
\begin{align}
y_{n,1} = y_{n+1,4},\qquad 0=\epsilon \times y_{n+1,4},\qquad
y_{n+1,1} +\alpha\times y_{n,1}=0,\qquad \gamma \times y_{n+1,1}-y_{n,1}=0
\end{align}
and there are four solutions which are labeled by $l$ ($\alpha \to \alpha^{(l)})$: 
\begin{align}
\alpha^{(l)} =-\omega^l,\qquad \gamma^{(l)}=\omega^{-l},\qquad \epsilon^{(l)}=0,\qquad y_{n,1}^{(l)}=\omega^{nl}\qquad (l=0,1,2,3). 
\end{align}
By directly solving the monodromy free condition, the other Stokes multipliers ($\beta^{(l)}$ and $\gamma^{(l)}$) 
are also fixed and 
the solution is given as
\begin{align}
S_0 = 
\begin{pmatrix}
1 & \cr
-\omega^l & 1 & -\omega^{-l} & \cr
0 & & 1 &\cr
\omega^{-l} & & \omega^l  & 1
\end{pmatrix},\qquad (l=0,1,2,3). 
\end{align}

\end{document}